\documentclass[11pt,screen,pagebackref]{ibtesis}
\usepackage{ibextra}
\usepackage{inputenc}
\usepackage{bm}
\usepackage{slashed}
\usepackage{subfigure}
\usepackage{color}
\usepackage{amssymb}
\usepackage{bbm} 
\usepackage{soul} 
\usepackage{MnSymbol}

\title{Non-local quark models for description of dense matter in the cores of
neutron stars
}
\author{Germ\'an Malfatti}
\director{Dr. Milva Gabriela Orsaria}
\codirector{Dr. Gustavo An\'ibal Contrera, Dr. Fridolin Weber}
\carrera{PhD. Thesis}
\grado{MSc.}
\laboratorio{}
\jurado{Dr. Horacio Falomir, Dr. Marcelo Miller, Dr. Norberto Sc\'occola}  
\date{}

\graphicspath{{figs/}} % Lugar donde encontrar las figuras generales (se puede poner uno en cada cap{\'{\i}}tulo)
\begin{document}

% Dentro del environment 'preliminary' va:
% la dedicatoria, resumen, abstract, indices

\begin{preliminary}

% Escriba su dedicatoria
\dedicatoria{

}

%%% \'{I}ndices %%%%

%\begin{abreviaturas}
                                %Abreviaturas
%\end{abreviaturas}

\tableofcontents                %\'{I}ndice

\listoffigures                  %Figuras

\listoftables                   %Tablas

\begin{resumen}%

This thesis work focuses on studying the possible existence of phase transitions in the immediate compact remnants of core collapse supernova, neutron stars, and the theoretical models that describe the interior of dense matter.

Specifically, we are interested in analyzing the feasibility of a transition from hadronic matter to quark matter in the cores of these objects. The density of matter inside neutron stars is several times that of atomic nuclei, and the equation of state that describes such matter in such a regime is still unknown. In this context, it is known that the interaction between the constituents of nucleons, the quarks, weakens with increasing density due to the intrinsic property of the QCD known as {\ it asymptotic freedom}. Therefore, matter should either dissolve into a quark-free state at high densities, or else form a superconducting state of color. This superconducting phase of color would be energetically favorable, if it were present in a cold neutron star, since a system of fermions that interact weakly at low temperature is unstable with respect to the formation of Cooper pairs. Although it is impossible to know both theoretically and experimentally whether these phases exist in neutron stars, the interpolation of the resolvable part of QCD at high densities, together with the hadronic equations of state at low densities, suggest that they could appear in the interior of compact objects.
 For the phase transition we will use two different formalisms: the Maxwell formalism, in which an abrupt phase transition between hadronic and quark matter without mixed phase formation is assumed, and the Gibbs formalism, in which a mixed phase in which hadrons and quarks coexist. For the description of hadronic matter, we will use different parametrizations of the relativistic mean field model with density-dependent coupling constants. For the description of quark matter we will use an effective nonlocal Nambu Jona-Lasinio model of three flavors with vector interactions, in which we will include the possibility of formation of diquarks to model a superconducting phase of color in $SU(3)$, which we will call $2SC + s$. Phase diagrams and equations of state of quark matter at finite temperature are presented, and the influence of that type of matter on observables associated with neutron stars is investigated. Likewise, using hybrid equations of state, the simplified thermal evolution of compact stars during their formation is studied, from their state of proto-neutron stars to that of cold neutron stars, and the results obtained are compared with recent astrophysical observations. The parameterizations used in this work are adjusted to the most recent measurements of masses and coupling constants of the QCD, which imposes strong restrictions on the existence of quark matter in proto-stars, unlike what happens with less realistic models or with more free parameters. However, the results obtained indicate that even considering these restrictions, the occurrence of quark matter in the nuclei of these stars remains a promising possibility. The remaining free parameters of the models were adjusted taking into account the observational restrictions, coming from precise determinations of the pulsars masses of $\sim$ 2 M$_{\odot}$, and the event corresponding to the fusion of two neutron stars, known as GW170817. The fact that the use of more realistic models for the description of the dense matter in these objects indicates the presence of quark matter inside neutron stars, could be an answer to the question of the behavior of that type of matter and the determination of its corresponding equation of state.

\end{resumen}

\begin{seccion}{Associated publications}
  \begin{enumerate}
  \item {\bf{Quark-hadron Phase Transition in Proto-Neutron Stars Cores based on a Non-local NJL Model}}, G.Malfatti, M.G.Orsaria, G.A.Contrera, F.Weber. Feb 20, 2017. 5 pp, Int.\ J.\ Mod.\ Phys.\ Conf.\ Ser.\  {\bf 45}, 1760039 (2017),  doi:10.1142/S2010194517600394,  [arXiv:1702.06114 [astro-ph.HE]].
  \item {\bf{Effects of Hadron-Quark Phase Transitions in Hybrid Stars within the NJL Model}}, I.F.Ranea-Sandoval, M.G.Orsaria, G.Malfatti, D.Curin, M.Mariani, G.A.Contrera, O.M.Guilera. Mar 27, 2019. 18 pp. Symmetry {\bf 11}, no. 3, 425 (2019), doi:10.3390/sym11030425,  [arXiv:1903.11974 [nucl-th]].
  \item {\bf{Phases of Hadron-Quark Matter in (Proto) Neutron Stars}}, F.Weber, D.Farrell, W.M.Spinella, M.G.Orsaria, G.A.Contrera, I.Maloney. Jul 10, 2019. 16 pp. Universe {\bf 5}, no. 7, 169 (2019), doi:10.3390/universe5070169, [arXiv:1907.06591 [nucl-th]].
  \item {\bf{Hot quark matter and (proto-) neutron stars}}, G.Malfatti, M.G.Orsaria, G.A.Contrera, F.Weber, I.F.Ranea-Sandoval. Jul 5, 2019. 27 pp, Phys.\ Rev.\ C {\bf 100}, no. 1, 015803 (2019),  doi:10.1103/PhysRevC.100.015803, [arXiv:1907.06597 [nucl-th]].
  \end{enumerate}
\end{seccion}

\end{preliminary}

% Podemos usar cualquiera de los dos comandos: \input o \include para incluir el texto
\graphicspath{{Introduccion/}}
\chapter{\label{ch:introduccion}Introduction}

The collapse of supernovae cores constitutes one of the most energetic explosions observed in the Universe, being able to outshine an entire galaxy for a brief moment. His possible remains,
the so-called neutron stars (NS), represent the densest stable objects (besides black holes) known, with masses of $\sim$ 1.4 M$_{\odot}$ and radii $\sim$ 10 km. Many NSs are radiopulsars, emitting radio waves periodically and having very high rotational speeds, with periods on the order of milliseconds. Pulsars can also emit X-rays or $\gamma$ rays and possess extreme magnetic fields. Some ENs, known as magnetars, have surface magnetic fields of $10^{14} -10^{15}$ Gauss (\cite{JPGreview} and references mentioned there).

The description of matter under such extreme conditions is an unsolved problem for the astrophysics community. In the last ten years, the discoveries of EN with masses around 2 M $_{\odot}$ \citep{Demorest:2010bx, Antoniadis:2013pzd, Fonseca:2016tux} represented a strong challenge when constructing the equations of state (EoS) that these objects describe. The EoS that were used up to that moment failed to describe masses of that order, so it was necessary to start considering additional degrees of freedom to describe the matter inside the NSs. Likewise, the historical event known as GW170817, the first detection of gravitational waves from the merger of two NS on August 17, 2017 \cite{PhysRevLett.121.161101}, also established new theoretical challenges for the construction of the EoS of the matter inside these compact objects.

\begin{figure}[ht!]
\begin{center}
\includegraphics[scale = 0.3 ]{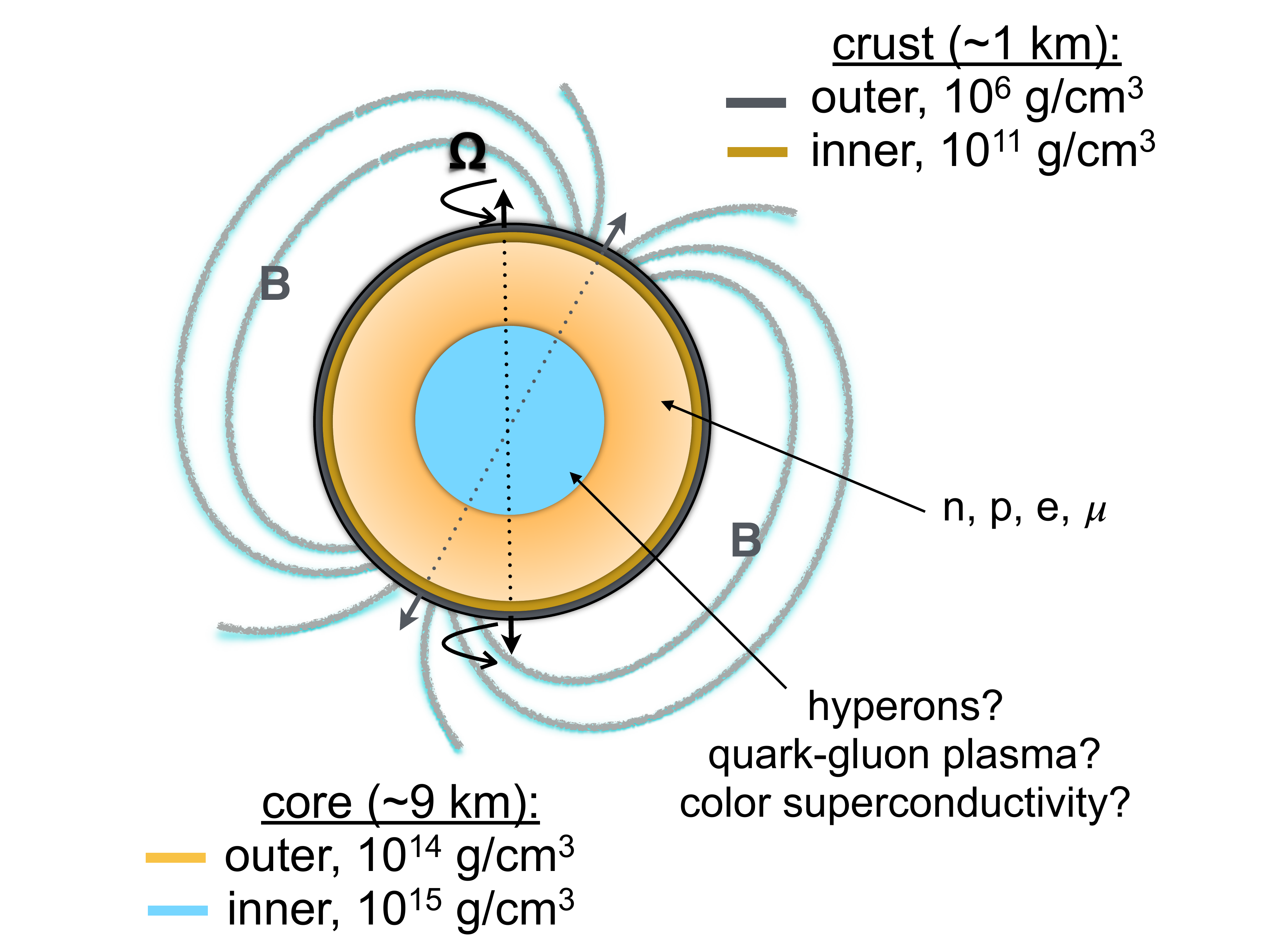}
\caption[Schematic structure of a neutron star] {Schematic structure of a neutron star. Figure adapted from the work of F. Weber \cite{WEBER2005}.}
\label{eneut}
\end{center}
\end{figure}

The densities in the inner nuclei of the NSs can be higher than the saturation density of the atomic nuclei ($\sim$ 2.3 $\times 10^{14}$ g/cm$^3$), and these conditions are not accessible in terrestrial laboratories. For this reason, the study of the composition of the internal cores of these objects is based, in most cases, on theoretical conjectures. The traditional image of an NS includes matter composed mainly of neutrons with a small fraction of protons and electrons. Even in this simple image, the internal structure of these objects is far from uniform: layers of protons and superconducting neutrons are expected to appear in the outer core, and uniform gas of nucleons between the outer core and the inner crust of the star. Other even more exotic configurations such as the formation of geometric structures known as {\it{nuclear pasta}} \cite{PhysRevC.88.065807} can appear in transition regions inside the star. In addition, at densities comparable to or greater than the density of nuclear matter, as we approach the inner core, exotic phases may appear that include hyperons, mesonic condensates, and quark matter. A schematic representation of an NS is shown in Figure {\ref{eneut}}. 
In particular, since the discovery of the substructure of nucleons formed by quarks \cite{PhysRevLett.30.1087}, many works have conjectured the appearance of a phase of deconfined quarks in the cores of these objects. The existence of quark matter inside NSs is of particular interest for this thesis work.

In this sense, the study and understanding of quantum chromodynamics (QCD), the theory that describes strong interactions, and their characteristics is essential to understand both the nature of NSs and the primordial Universe.

One of the main features of QCD is asymptotic freedom. Contrary to most fields of physics (for example, quantum electrodynamics), the interactions are weak at short distances ($ \ sim $ 0.4 fm)  between the quarks and they get stronger as the distance between them increases. Calculated at first order by Gross, Wilczek and Politzer in 1973 \citep{Gross_Wilczek, Politzer}, the experimental confirmation of asymptotic freedom made it possible for the QCD to be accepted as the theory that describes strong interactions. Therefore, in the short distance regime, QCD can be treated perturbatively. However, in the long-distance or, equivalently, low-energy regime ($ \lesssim $ 1 GeV), the theory becomes non-perturbative and the calculations are not mathematically feasible. In this context, the problems that QCD has in the non-perturbative regime are addressed either through \textit{ab initio} calculations using \textit {Lattice QCD} \cite{Ma:2017pxb}, or by building an effective model that shares, for a certain problem, some of the fundamental properties of the QCD while preserving a simpler mathematical structure. However, these two approaches have serious limitations. \textit {Lattice QCD} cannot describe the theory at finite chemical potentials due to the sign problem or the problem of the
  complex action \cite{Latt2003}, while most effective models are only capable of qualitatively reproducing QCD. Given that in this thesis we are interested in analyzing quantities and thermodynamic properties with finite chemical potential, we will use an effective non-local model for the description of quark matter and its different phases considered in the framework of NSs, which not only qualitatively reproduces properties of QCD, but also respects several fundamental symmetries of the theory.
 \begin{figure}[ht!]
\begin{center}
\includegraphics[scale = 0.3]{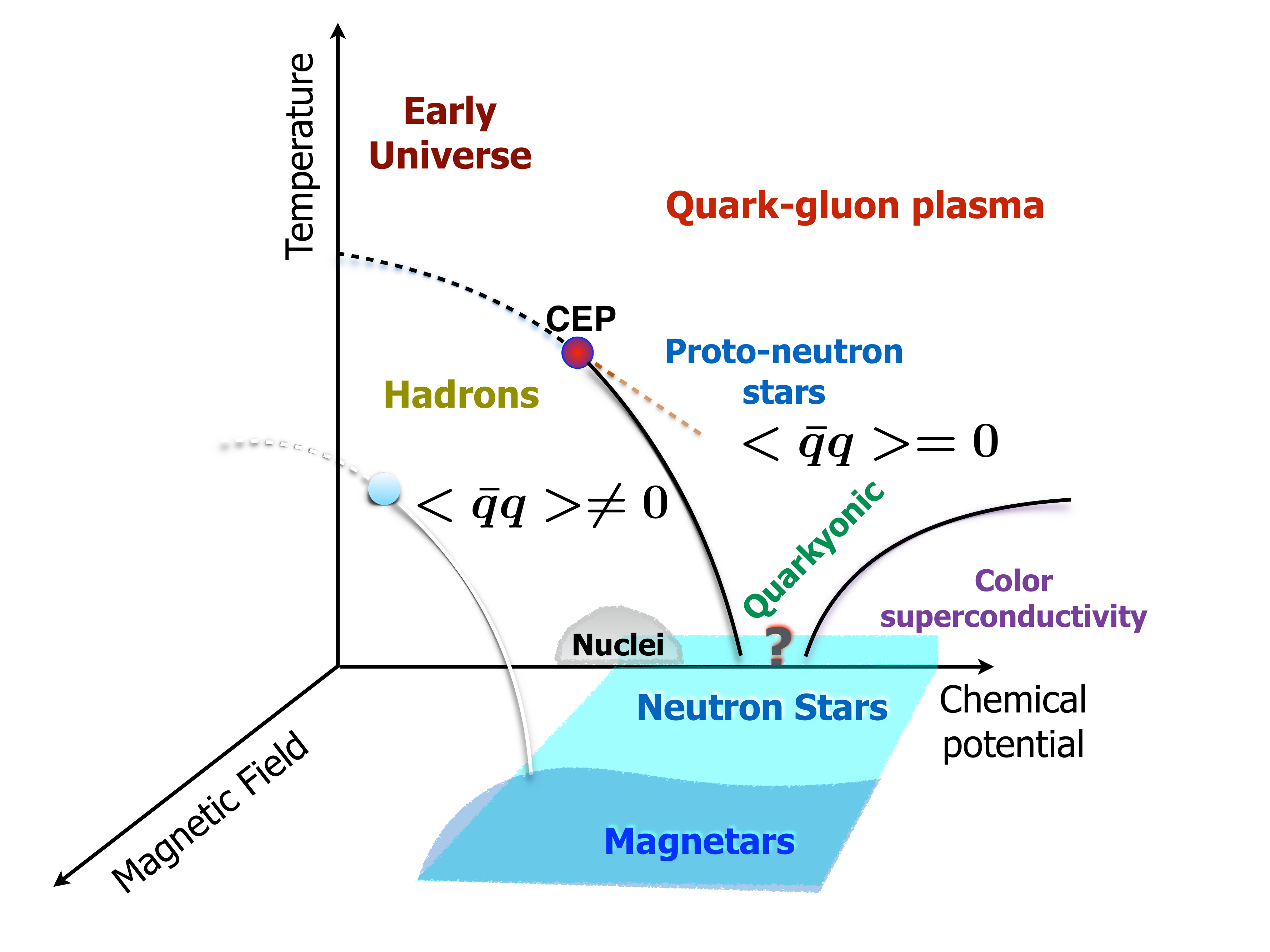}
\caption[Schematic phase diagram of QCD] {Schematic phase diagram of QCD. Figure adapted from the reference work \cite {JPGreview}.}
\label{QCDdiag}
\end{center}
\end{figure}

The schematic phase diagram of QCD in Figure \ref{QCDdiag} illustrates the different theoretical phases of quark matter at a given temperature and chemical potential. The \textit{ab-initio} calculations of \textit{Lattice QCD} indicate that at a small baryonic chemical potential and at high temperature, the phase transition of a hadron gas to the plasma phase of quarks and gluons is a soft transition, called crossover. On the other hand, a first order phase transition is expected in the region of high baryon chemical potentials. Both phases are separated by a main curve in which the transition is found, in which the confined and deconfined states are in equilibrium. The point indicated in the Figure with the acronym CEP (\textit{Critical End Point}), separates the crossover line from the first order line. The secondary curve separates the region of the phase diagram with the question mark from the color superconducting phase. This question sign refers to the fact that it is still unknown, from a theoretical point of view, what phase of quark matter would be found in that zone of low temperatures and high densities and/or chemical potentials. A superconducting color phase could be present inside a cold NS, since a system of fermions that interact weakly at low temperature is unstable with respect to the formation of Cooper pairs. The quarkyonic phase, in which only a partial restoration of the chiral symmetry would occur, is also shown as a possible alternative. There is great uncertainty in the theoretical calculations to determine what is the structure of the QCD phase diagram in the region of higher baryon densities. Finding the nature of the QCD phase transition and the corresponding critical point in said region are the main objectives of the Beam Energy Scan (BES) programs in the Relativistic Heavy-Ion Collider (RHIC) \cite{Aggarwal:2010cw} and in the Super-Proton Synchrotron (SPS) \cite {Abgrall:2014xwa}. Regarding the region of the Figure \ref{QCDdiag} with higher densities, where matter lies in conditions similar to those found inside ENs or magnetars, it is expected that the results of space missions such as NICER (\textit{Neutron star Interior Composition Explorer}) \footnote {https://www.nasa.gov/content/about-nicer}, or Strobe-X \footnote {https://gammaray.nsstc.nasa.gov/Strobe-X/} allow us to fully determine the equation of state of ultra-dense matter. In this thesis, we will study the thermodynamic properties of non-local models throughout the phase diagram and apply it to the study of possible phase transitions that may occur in the region where the the matter inside NSs would be located (Figure \ref {QCDdiag}).
 
With regard to non-local models, in the 90s, non-local interactions between quarks began to be included in the treatment of matter by using Nambu Jona-Lasinio (NJL) type QCD effective models \cite{Ripka:1997zb}. The NJL model \cite{Nambu:1961fr} was initially used to study the properties of hadrons and was later generalized to the study of quark matter. A natural way to introduce nonlocality into the quark-quark interaction is to use a phenomenologically modified gluon propagator (effective propagator) \cite{effectiveprop1994}. The fact of including non-local interactions, solves the divergences in the self-energies for the quarks and the mesonic masses that occur in the standard NJL model. Another characteristic feature of the standard NJL model is the absence of confinement. In order to describe both the spontaneous breaking of chiral symmetry and color confinement that occur in QCD, the Polyakov-Nambu-Jona-Lasinio (PNJL) models were proposed as an extension of the original NJL theory. The Polyakov loop, in a pure gauge theory, is an order parameter of the deconfining phase transition \cite{POLYAKOV1978477, SVETITSKY19861}. This peculiarity is related to the existence of a discrete symmetry, $Z_3$, of the pure gauge action, which breaks spontaneously when the deconfinement starts. The Polyakov loop disappears in the confined phase at low temperature but is different from zero in the deconfined phase, at high temperatures.

We have then, that the problems due to ultraviolet divergences and confinement, which arise naturally in NJL models, can be solved with the use of non-local interactions and the inclusion of the Polyakov loop. Non-local chiral quark models are able to provide a satisfactory description of the properties of hadrons at zero temperature \cite{Broniowski:2001cx}. However, for zero chemical potential, these models lead to a rather low critical temperature (about 120 MeV) for the chiral phase transition compared to the \textit {Lattice QCD} results. This problem is solved by incorporating the Polyakov loop, which significantly increases the chiral symmetry restoration temperature \cite{Contrera:2007wu}.

This thesis is organized as follows: in Chapter \ref{ch:Quarks_y_contexto_astrofisico}, we give an introduction to the astrophysical context in which our non-local quark models (nl-NJL) will be applied. The generalities of QCD and the theoretical formalism for the case of the three lightest flavors of quarks in the nl-NJ model are presented in Chapter \ref{ch:Nambu}. Chapter \ref{ch:PNJL} is dedicated to the extension of the model at finite temperature and to the incorporation of the Polyakov loop, nl-PNJL. The inclusion of color superconductivity at zero temperature and its extension at finite temperature is developed in Chapter \ref{ch:Superconductividad}. Chapter \ref{ch:Hadrones} describes the model used for the hadronic matter that will make up the outer core of NSs. The results of the astrophysical application of the effective models presented in this thesis are shown and discussed in the chapter \ref{ch:Resultados}. Finally, in the chapter \ref{ch:Conclusiones} we present the conclusions and future perspectives of the application of this type of models in the astrophysical context.

\graphicspath{{Quarks_y_contexto_astrofisico/}}
\chapter{\label{ch:Quarks_y_contexto_astrofisico}Astrophysical scenario and quark matter}

Neutron stars are the densest compact astrophysical objects in the universe ($10^6 \, g/cm^3 \, \lesssim \rho \lesssim 10^{15} \, g/cm^3)$ \cite{Glendenning:1997wn}, after black holes. Both arise as a result of supernova explosions from a massive star. These objects constitute natural laboratories that make it possible to investigate the influence of intense magnetic fields, superfluidity and superconductivity in matter subjected to extreme conditions, the properties of nuclear forces at high densities and the possible phase transitions of nuclear matter towards another, more exotic type of matter, such as unconfined quark matter or in a superconducting state of color.  Regarding deconfined quark matter, the Bodmer-Witten conjecture \cite{Bodmer:1971we, Witten:1984rs} arose in the 1980s, which suggests that a form of quark matter known as strange matter would be the fundamental state of hadronic interactions. From this conjecture, several studies of stars composed solely of quark matter arose \cite {Alcock:1986hz, Benvenuto:1989up, Chakrabarty:1995br}. Later, theoretical studies showed that the ground state of deconfined quark matter could be a color superconductor \cite {Alford:2001dt}. However, detailed studies in the framework of Nambu Jona-Lasinio models question the absolute stability of quark matter, whether in the form of quark-gluon plasma or in a superconducting phase of color \cite {Buballa:2005}. We will discuss this point in more detail in the \ref{quarks_stable} section.

Although gravity in NSs compresses matter to energy densities comparable to those achieved in collisions of heavy ions, in which the formation of a plasma of quarks and gluons is feasible, it is yet unknown if deconfined quark matter exists in their inner cores. This depends on whether the densities in the center of these objects are extreme enough to lead to the formation of a new deconfined state of matter. Although the last decade has seen notable advances in observations of NSs, what has hampered firm conclusions about the presence of quark matter within these objects is the lack of accurate QCD predictions for the properties of matter at high baryonic densities. What is clear is that both the theoretical and experimental studies of QCD indicate that at least two phases are clearly identifiable by the qualitative behavior of the equation of state of dense matter:  at low densities the QCD EoS follows the predictions of hadronic models and at high densities the description is compatible with perturbative models of QCD. This situation should not, in principle, be different in the context of a NS. Likewise, the existence of stars formed only by quark matter has not yet been ruled out \cite {Flores:2013yqa, Becerra-Vergara:2019uzm, Jimenez:2019iuc}. Another possibility is that NSs are formed by a core of deconfined quark matter, or by forming diquarks in a color superconducting phase, surrounded by several layers of hadronic matter. These types of stars are known as hybrid stars. There could also be a mixed phase where quarks and hadrons were present at the same time, since a sufficiently small surface tension ($\lesssim $ 70 MeV/fm$^2$) between the hadronic and quark interfaces would favor the formation of that type of phase \cite{SOTANI201337}. However, the surface tension value is still unknown and this has led to several studies showing conflicting results (see for example \cite{PhysRevC.88.045803} and references therein).

As we discussed in the introduction, the NSs internal structure is divided into several regions or layers. Therefore, the modeling of these objects from the moment they are born, in their proto-star stage, until they cool down, requires taking into account not only the physical transport processes that occur inside them, but also analyzing their structure and hydrostatic stability. Although most NSs are found in binary systems, isolated NSs have also been detected, such as those commonly known as ''the magnificent 7 '' (\textit {XINS: X-ray Isolated Neutron Stars}) \cite {Zampieri:2001ewa}. In this work we will consider isolated, static and spherically symmetric neutron stars in a process that can be described in a simple way through different stages of constant entropy per baryon, as we will describe later in section \ref{Estadios}.

\section{Structure equations for Neutron Stars (TOV)}
Since the discovery of pulsars in 1967 by Jocelyne Bell \cite{Hewish:1968bj} (which allowed its director, Antony Hewish to win the Nobel Prize in 1974), NSs have been the subject of intensive study, both from the point of view of of astronomical observations as from the theoretical one. The first theoretical speculation about the existence of an extremely dense object with a composition similar to that of atomic nuclei came from Lev Landau in 1931\cite{196560}. The discovery of neutrons the following year\cite{Chadwick:1932ma} provided additional support for Landau's conjecture, since an object composed of neutral particles would not suffer from the stability problem as a consequence of electrostatic repulsion. In 1934, astronomers Fritz Zwicky and Walter Baade proposed objects composed of pure neutron matter as a remnant of a supernova explosion \cite{PhysRev.46.76.2}. One of the crucial facts to understand the physics of NSs was the formulation of hydrostatic equilibrium equations of a self-gravitating fluid in the framework of general relativity. That formulation was completed independently by Oppenheimer and Volkoff \cite{Oppenheimer:1939ne} and by Tolman \cite{Tolman1939}. The corresponding equations are known as the Tolman-Oppenheimer-Volkof (TOV) equations. To solve them, the EoS of the matter that composes the mentioned fluid is needed.
 
\begin{figure}[ht!]
\begin{center}
\includegraphics[scale = 0.45]{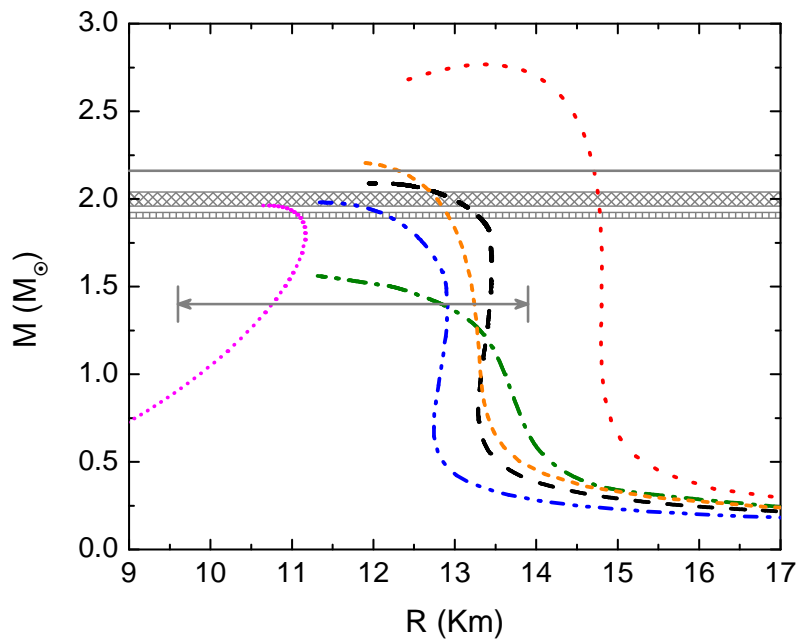}
\caption[Different types of solutions to the TOV equations in the gravitational mass-radius plane.] {Solutions of the TOV equations in the mass-radius plane for different EoS. The bars in gray are the restrictions imposed by the 2 $M_{\odot}$ pulsars. The line immediately above the bars indicates the restriction resulting from the analysis of the data obtained from the event GW170817, which imposes a limit for the maximum mass of ENs of 2.16 $M_{\odot} $ \cite {Most2018}. The lower line with double arrows indicates the restriction to the radii of the NSs from the data analysis of the same event from the merger of two NSs (see \cite{PhysRevLett.121.161101}, \cite{JPGreview} and references therein).}
\label{mradio}
\end{center}
\end{figure}

The relationship between the mass or energy density and the pressure inside the NS is what determines its EoS. In this way, for each EoS, it is possible to calculate a family of NSs with a well-determined mass and radius that will depend on the model or models chosen to describe the matter inside these objects. Combining the TOV equations and the EoS we can calculate the structure of each star that makes up the corresponding family.  The TOV equations are obtained by solving the equations of general relativity for a metric that describes a non-rotational static fluid with spherical symmetry. Starting from a metric that respects these conditions, and assuming that the matter within the sphere is isotropic and in gravitational equilibrium, we obtain

\begin{eqnarray}
\frac{dP}{dr} &=& -\frac{G m(r)\epsilon (r)}{r^2} \frac{\left[1 + \frac{P(r)}{\epsilon(r)}\right]\left[1 + 4\pi r^3 \frac{P(r)}{m(r)}\right]}{\left[1 - 2G \frac{m(r)}{r}\right]}, \label{TOveq1}\\
\frac{dm(r)}{dr} &=& 4\pi r^2 \epsilon(r),
\label{TOveq2}
\end{eqnarray}
where $r$ is the radius, $P(r)$ and $ \epsilon (r)$ are the pressure and energy density within a radius $r$, $ m (r) $ is the mass contained in that radius and $G$ is the universal gravitational constant. Usually these equations are solved with good precision using Runge Kutta methods of order 4. Giving a relation $\epsilon (p)$, the boundary conditions to satisfy in order to find mass-radius relations are: $ m(r=0) = 0 $ and $ P (r = R) = 0 $ for some value of $ R $ that will be the limit point of integration for which, if this condition is met, we will be on the edge of the star. Once these equations have been solved, we can arrive, depending on the EoS $\epsilon (p) $ from which we start, to mass-radius relationships like those in Figure \ref{mradio}.
Depending on the type of EoS provided, we can see that it is possible to obtain arbitrarily small radii (for the case of stars purely formed by strange quark matter) or mass-radio curves that go down to a lower limit of radii. In the latter case, the star that corresponds to the maximum mass point is the last stable star of the family, being those with smaller radii hydrostatically unstable, and possibly candidates to become black holes. However, there are works that propose that these instabilities could be overcome, depending on the nucleation time in relation to the radial oscillation frequency of the star, if an abrupt phase transition from hadronic matter to quark matter occurred in its interior \cite{Pereira:2017rmp}. It should be noted that although one can enter different equations of state into the TOV equations, there are general restrictions that have to be met, namely:

\begin{itemize}
 \item Causality: $\frac{dp}{d\epsilon} < c^2$, where $c$ is the light speed.
 \item Thermodynamic stability: $\frac{dp}{d\epsilon} > 0$.
\end{itemize}

In the next chapter we will develop in more detail the conditions that we will impose on the equations of state. For now, it is worth mentioning that the causality condition gives an upper limit in the mass-radius plane, which results in $ M \lesssim 0.35R $, which is less than the Schwarzchild limit $ M = 0.5R $ . That is, it is possible to construct EoS that violate causality before reaching the Schwarzchild limit.

\begin{figure}[ht!]
\begin{center}
 \includegraphics[scale = 0.45]{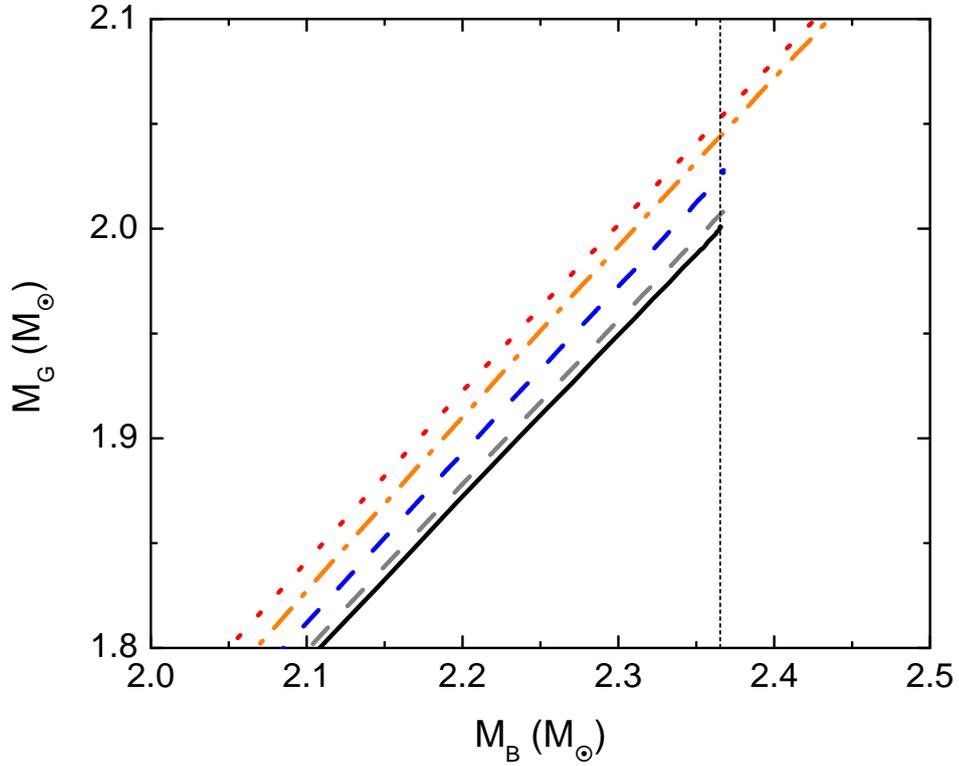}
\caption[Curves of gravitational mass versus baryonic mass as solutions of the TOV equations.] {Curves of gravitational mass as a function of baryonic mass at different temperatures, the solid black line corresponds to zero temperature, the vertical line indicates the value of baryonic mass that the stars that are candidates as predecessors of the cold star should have.}
\label{mgravsmbar}
\end{center}
\end{figure}

So far we have discussed the gravitational mass (which we will always express in terms of solar masses $ M_\odot $), which is defined as
\begin{equation}
 M_G = \int_0^R4\pi r^2 \epsilon(r) dr.
\end{equation}
It is useful, however, to define the baryon mass of the star:
\begin{equation}
 M_B = m_N \int_0^R \frac{4\pi r^2 n_B(r)}{\sqrt {1 - \frac{2G m(r)}{r}}} dr,
\end{equation}
where $m_N$ is the neutron's mass, and $ n_B (r) $ is the baryon density for a radius $ r $. This quantity can be calculated as follows: for each point $ \epsilon (p) $ we also enter $ n_B (p) $. Once the baryon mass has been calculated, it is possible to make curves in the $ (M_{G}, M_{B}) $ plane, like the one in Figure \ref{mgravsmbar}.
These curves will be useful for the following argument: if we suppose that the stars to be analyzed are isolated, their baryon mass has to be conserved during their evolution. Therefore, if we have stages of the star calculated at different temperatures, the condition to ask is that all the stars that we calculate fall on a line on a defined value of baryonic mass. For example, if our star to study at zero temperature has a baryon mass value of $ M_B = 2.4 M_\odot $ but the stars calculated at finite temperature never reach this value, they will not be possible candidates for predecessors of said star because they do not fulfil baryon mass conservation.

\section{Neutron stars stages and their relationship with the equations of state}
\label{Estadios}

So far we have presented the structure equations of neutron stars, where we said that we need as a primary ingredient an equation of state that gives us the relationship $ \epsilon (p) $ to solve the TOV equations. SHowever, depending on the type of compact object that we want to study and its internal composition, the EoS that we can use as an ingredient can be of different types: isothermal, isentropic, with and without neutrinos, with and without quarks, etc. To determine which EoS is appropriate, we must take into account the different stages in the evolution of the NSs and what restrictions these impose on the conditions of the matter that the EoSs represent. Therefore, below we will schematically describe the process of formation of the NSs.

\begin{figure}[ht!]
\begin{center}
 \includegraphics[scale = 0.3]{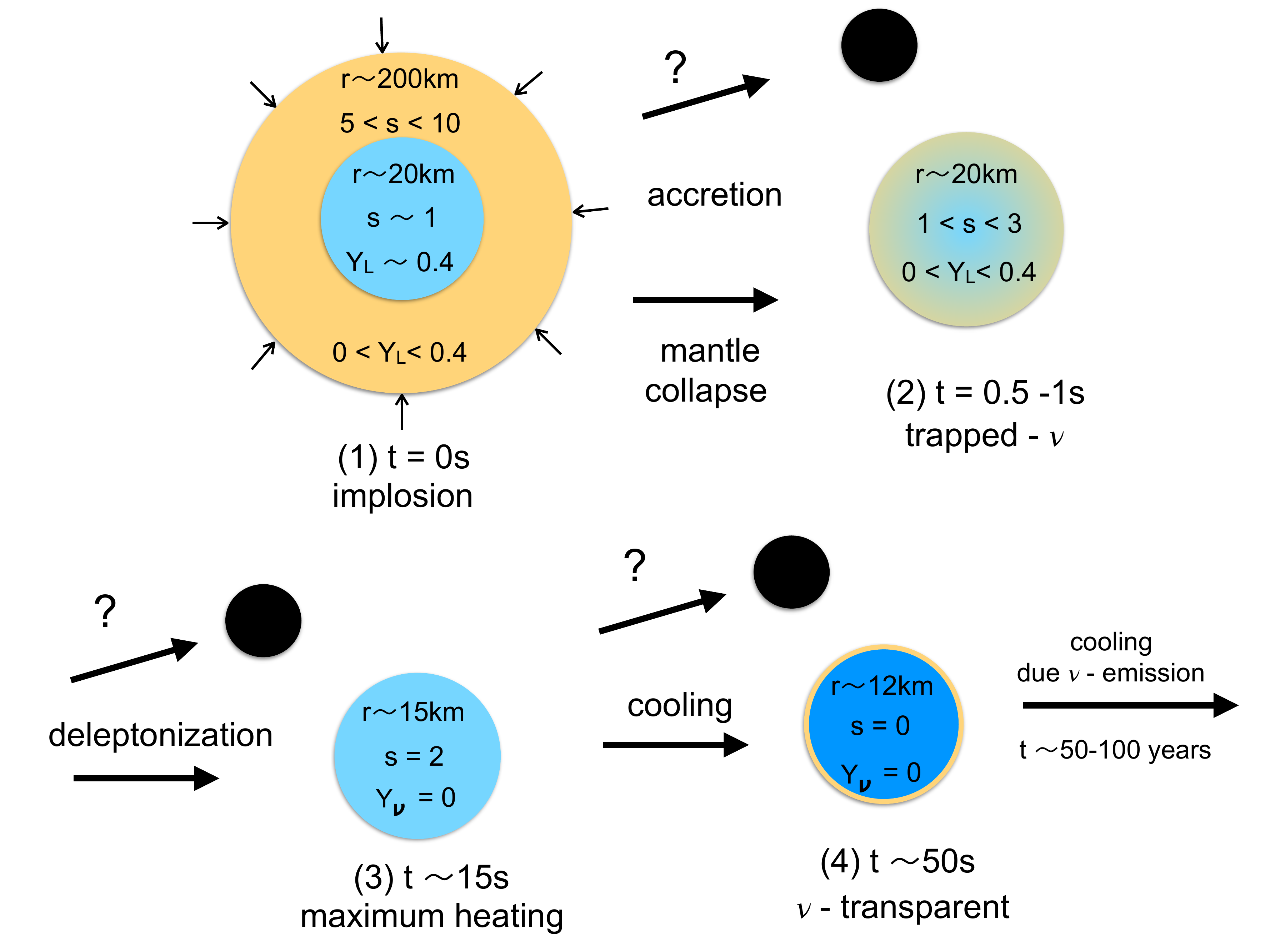}
\caption[Diagram of the evolution stages of a neutron star.] {Diagram of the evolution stages of a neutron star considered in this study (based on the reference \cite{Prakash:2000jr}). }
\label{figestadios}
\end{center}
\end{figure}

Once a star runs out of fuel to consume, it implodes as there is no longer pressure from nuclear reactions that can counteract the gravitational pressure. As the content that remains in the star are mostly fermions, the collapse reaches a point where there are no different states for each particle (by Pauli exclusion principle), and a rebound effect of the outer mantle of the star is generated, collapsing against the core. At this stage there is a low entropy core with trapped neutrinos (stage (1) in Figure \ref{figestadios}). The core is surrounded by a mantle of low density and high entropy, which is accreting matter from the outer iron core and at the same time is losing energy due to processes of electronic capture and emission of thermal neutrinos. This mantle extends for around 200 km and is approximately stable before an eventual explosion due to the aforementioned rebound.
After this stage, two things can happen: the explosion is not strong enough to deleptonize the outer mantle, and the matter becomes accreted again, giving way to the formation of a black hole, or the supernova explodes, generating a significant pressure loss in the disk by deleptonization, the mantle collapses, and the accretion becomes less important (stage (2)). The third stage occurs when the star rapidly begins to lose neutrinos. This leads to a reduction in its pressure by deleptonization, and here too it is possible that it will become a black hole if the gravitational pressure is large enough. If this does not happen, the star will finish deleptonizing as it is heated by the Joule effect of the escaping neutrinos and will reach stage 3. It is believed that once the neutrinos have finished escaping is when the maximum heating of the star occurs, but different analyzes of the results that we will see at the end of this thesis allow us to reconsider that statement. After the escape of neutrinos, the probability of occurrence of strange matter increases. In the next section we will analyze this hypothesis, but it is worth mentioning that if the production of strange matter is large enough, there is again the possibility of a black hole collapse. If this does not happen, unless there is accretion, this possibility is ruled out for the following stages. If the collapse does not occur, the star continues to cool and becomes transparent to neutrinos (stage (4)). For all the practical purposes of this work (remember that we will not analyze transport effects), the following stages correspond to the cold star, at zero entropy and temperature.

For the first stages, where there are trapped neutrinos, the matter is in beta equilibrium, which together with the conservation of charge implies that \cite {Steiner:2000bi}:
\begin{eqnarray}
 \mu_e - \mu_{\nu_e} &=& \mu_\mu - \mu_{\nu_\mu}, \\ 
 \mu_B &=& b_i \mu_N -q_i \mu_e + q_i \mu_{\nu_e}, 
\end{eqnarray}
where $ b_i $ and $ q_i $ are the baryon number and the electric charge for the corresponding hadron or quark and $ \mu_N $ is the neutron's chemical potential. We will ignore Coulomb and surface effects, so the leptons will be considered as free Fermi gases. Lepton fractions are defined as
\begin{eqnarray}
 Y_{L_e} &=& \frac{n_e + n_{\nu_e}}{n_B} = X, \\
 Y_{L_{\mu}} &=& \frac{n_\mu + n_{\nu_\mu}}{n_B} = 0, \\
\end{eqnarray}
where $ Y_L $ is the fraction of the particle species $ i $, $ n_i $ is the density and $ X $ is a number to be fixed depending on the stage to be considered. The fraction of muons is taken as null because it is suppressed while electron neutrinos are present, and it becomes non-zero after deleptonization, where for the same equilibrium reasons mentioned above it results $ \mu_\mu = \mu_e $.

In this work we will use the aforementioned stages as a first guide to establish what conditions we will ask of our EoS. In principle we would need to meet the following requirements:

\begin{itemize}
\item Fixed (constant) entropy per baryon or temperature.
\item Inclusion of leptons as free Fermi gas.
\item Fixed (constant) lepton fraction in the case of including neutrinos.
\item Description of hadronic matter and quark matter.
\item Possibility of phase transition between hadronic and quark matter (hybrid matter).
\end{itemize}

As we said before, as the density of the star increases, the probability for appearance of strange matter increases, which leads us to the need for EoS that contain $s$ quarks for quark matter, and that contain hyperons in the description of hadronic matter. Let us look at the reasons for this hypothesis in a little more detail in the next section.

\section{Quark matter in compact objects}
\label{quarks_stable}

\begin{figure}[ht!]
\begin{center}
\includegraphics[width=0.40\textwidth]{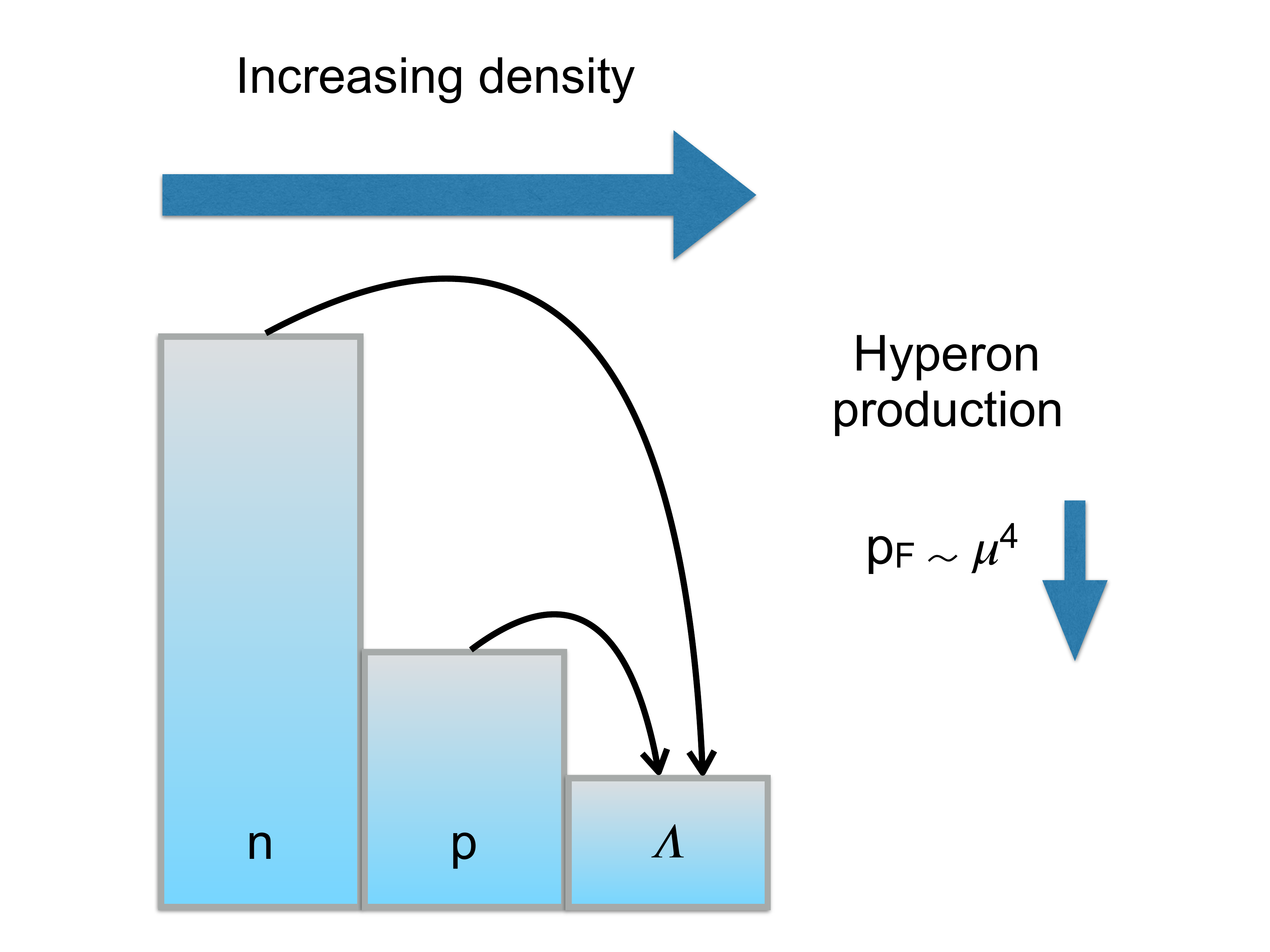}
\caption[Conversion scheme of nucleons to hyperons.] {Conversion scheme of nucleons (N = n, p) to hyperons ($ \Lambda$). As the density increases inside the NS, new degrees of freedom appear, such as hyperons, in addition to nucleons. Contrary to terrestrial conditions, where hyperons are unstable and decompose into nucleons through the weak interaction, equilibrium conditions in NSs can cause the reverse process to occur, so that the formation of hyperons becomes energetically favorable. As soon as the neutron's chemical potential $\mu_n$ becomes large enough, the more energetic neutrons (i.e. those on the Fermi surface) can break down through the weak interactions into hyperons $\Lambda$ and form a new Fermi sea for this hadronic species with $\mu_{\Lambda} = \mu_n$. In this way, the Fermi pressure, $p_F$, exerted by the baryons, decreases.}
\label{conversionhyp}
\end{center}
\end{figure}

The first models to describe matter inside ENs were based on considering protons, neutrons, and electrons as non-interacting particles in a Fermi gas. In 1939, Oppenheimer and Volkoff used the hydrostatic equilibrium equations to numerically calculate the models of NSs from a simple EoS, which did not take into account the interaction of neutrons. From the results of their calculations, they mistakenly concluded that the maximum gravitational mass of a stable static neutron star was M $ \sim $ 0.7 M$_{\odot} $. This limit is known as the Oppenheimer-Volkoff mass limit. In 1959, Cameron \cite{Cameron1959} emphasized the importance of taking into account the nuclear interaction when calculating the EoS of NSs and considering these interactions, he succeeded in extending the Oppenheimer-Volkoff mass limit of 0.7M$_{\odot} $ to 2M$_{\odot} $. Furthermore, Cameron was one of the first authors to suggest the existence of hyperons, that is, states of three quarks where at least one of them is a strange quark, in the cores of neutron stars. Then, as the nuclear models progressed, different tests were carried out leading to the model of the liquid drop, with its variants (such as asymmetry energy) as a possible model to describe the composition of these stars. These models gave fairly large mass predictions for stars, on the order of 2.8 $ M_\odot $. However, after the work of Ambartsumyan and Saakyan \cite{Ambartsumya}, the possibility that NSs also contain hyperons began to be considered more firmly. The reasoning for considering the existence of these degrees of freedom is that for sufficiently high densities, the Fermi moment of the nucleons (N) would be greater than the sum of the masses of, for example, the Kaon ($ K $) and the Lambda ($\Lambda$) hyperon, generating the reaction:

\begin{equation}
 N + N \to N + \Lambda + K,
\end{equation}
where then the Kaon decays in the usual way (photons, muons and neutrinos) and the $\Lambda $ hyperon is produced. This provokes a decrease in the system's energy, due to the leakage of the Kaon's decay products, neutrinos and photons. The hyperon production inside an NS occurs from a certain critical density in which new Fermi seas open for these degrees of freedom, producing a conversion of protons and neutrons into hyperons, as illustrated in Figure \ref{conversionhyp}. However, this makes matter more compressible and therefore makes it difficult for an equation of state containing hyperons to explain the masses of 2 $ M_\odot $ pulsars. This is known in the literature as {\it the hyperon puzzle} \cite{Bombaci:2016xzl}.

The scheme in Figure \ref{conversionhyp} shows that the denser the nuclear matter, the more Fermi energy exists to exceed the limit of the hyperon's mass, so that their creation reduces the system's energy and pressure, being a preferential state. The higher the system's pressure, the greater the probability that two nucleons will interact, which increases the probability of producing hyperons in new Fermi seas to decrease the system's energy.

\begin{figure}[ht!]
\centering
\includegraphics[width=0.50\textwidth]{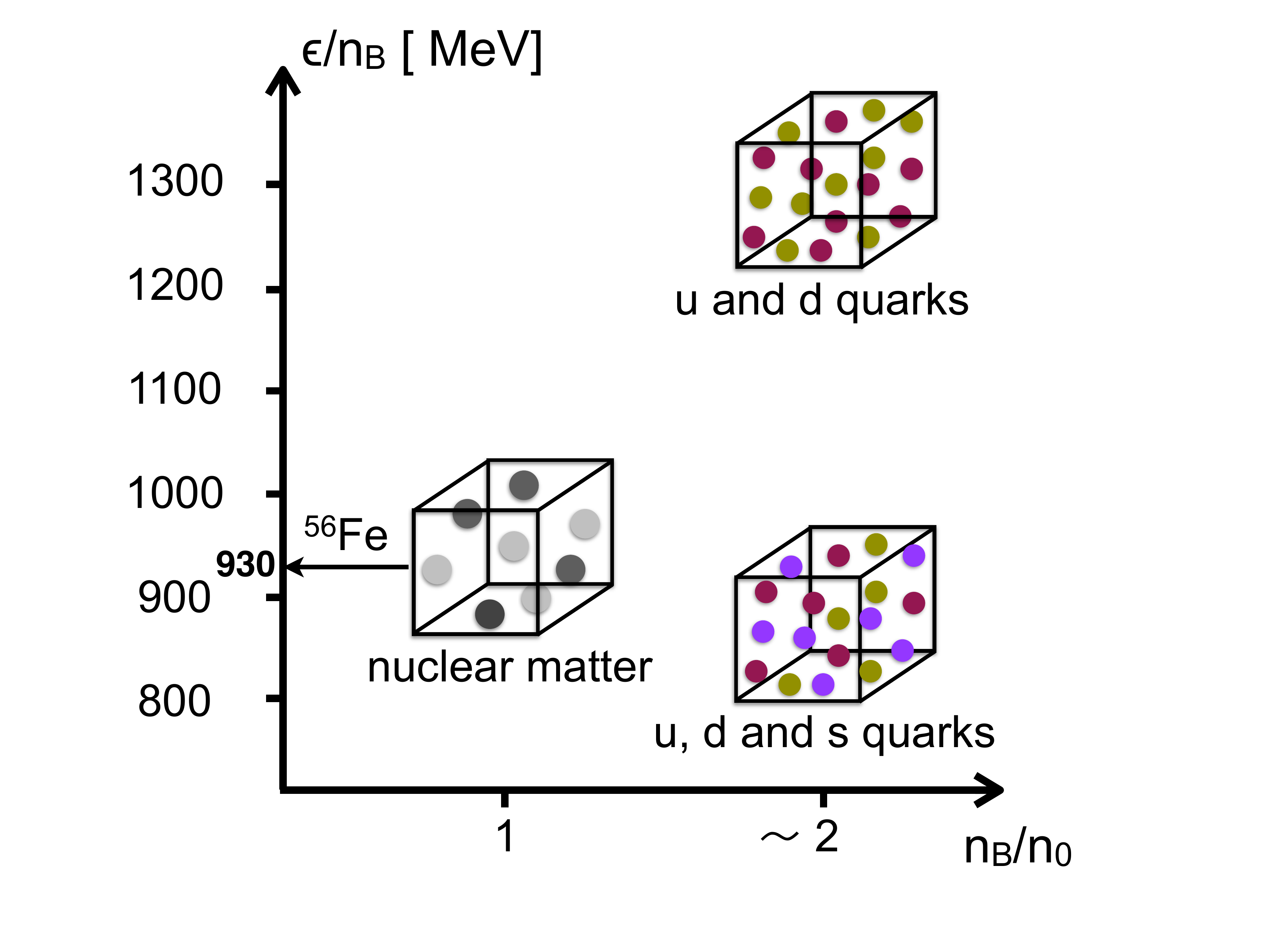}
\caption[Mass per nucleon of $\mathrm {^{56}Fe}$  and quark matter in the strange matter hypothesis.] {Comparison scheme of the mass per nucleon of $ \mathrm{^{56}Fe} $ and quark matter for the case of two flavors ($u$ and $d$ quarks) and three flavors ($u,d,s$ quarks). In theory, the mass per nucleon of strange matter could be below 930 MeV, which would make it more stable than ordinary nuclear matter.}
\label{smhyp}
\end{figure}

Although hyperon presence in NSs seems to be inevitable from the energy point of view, the corresponding EoS results in maximum masses, in the construction of NSS's families, not compatible with the observations. Solving this puzzle is not easy, and it is currently a very active research topic, especially after the detection of the PSR J1614-2230 and PSR J0348 + 0432 pulsars, which rule out many of the currently proposed equations of state with hyperons ( both microscopic and phenomenological). To solve this problem, a mechanism is necessary that can eventually provide the additional repulsion necessary to make the EoS stiffer, that is, lower energy density at the same pressure. This makes the matter less compressible and, therefore, increases the maximum theoretical attainable mass. Several attempts were made to modify the hadronic models, so that they include hyperons, but in which some repulsive interaction channel between them is also considered. However, beyond the success of hadronic models, after the discovery that quarks were the constituent particles of hadrons, a new question arose: What would happen if, above certain density, the conversion of hadrons to free quarks were energetically more favorable, instead of continuing to create hyperons? This questioning, on the one hand, started a search that continues today, that of models that describe both hadronic matter and quark matter. On the other hand, a new question arose from the Bodmer and Witten hypothesis \cite{Bodmer:1971we, Witten:1984rs}: what if strange quark matter was more stable than nuclear matter? This would mean that at zero pressure, the energy per baryon for quark matter should be less than the mass per nucleon for the most stable nucleus, which is $\mathrm{^{56}Fe}$, as shown schematically in Figure \ref{smhyp}-

This hypothesis is valid given that the model that was used (and is used today in many analyzes) is the well-known MIT bag model \cite{Chodos:1974je, Chodos:1974pn, Farhi:1984qu}. This model consists in thinking that quarks and leptons behave as free particles within hadrons, but that they must exceed a bag pressure $ B $ to exit the hadronic structure, so that the pressure is written as
\begin{equation}
 P  = \sum_f P^f - B,
\end{equation}
where $ P $ is the total pressure, and $ P^f $ corresponds to the pressure of free particles of the species $ f $. The detailed analysis of this model is well known and is far from being one of the objectives of this work, however, a quick and brief analysis will be useful. If the particle's chemical potential is small enough, the pressure will be negative and the particles will be confined to hadrons. For a sufficiently large chemical potential, the particles will behave as free fermions, and for a particular value of the chemical potentials, it will happen that $ \sum_f P^f = B $ so the pressure will be zero and the system will be in the interface between hadrons and free quarks. It is worth saying that there are different variants of this model, which include gluonic effects or superconductivity \cite {Alford:2005wj}, however, the way to satisfy the strange matter hypothesis is equivalent in all of them. Starting from the model with two quark flavors $ u $ and $ d $, for ordinary nuclear matter to be more stable than free quarks, it is obtained that the energy per baryon of the latter has to be greater than that of the former, reaching the level $ B^{1/4}> 145.9 $ MeV. When the model of three flavors of quarks is considered, since the pressure must be the same in the transition to the hadronic matter, it can be asked that at zero pressure the strange matter is more stable than the hadronic matter, obtaining a different bound. This bound together with the previous one leads to the relation $ 145.9 \, \mathrm {MeV} <B^{1/4} <162.8 \, \mathrm {MeV} $ \cite{Weber:2004kj} for simplest MIT bag model. It is worth mentioning that this hypothesis does not conflict with the existence of ordinary nuclei, since to convert an atom of mass number $ A $ to strange matter, a number $ A $  of $ u $ and $ d $ quarks needs to transition to $s$ quarks, which implies average lifetimes on the order of $ 10 ^ 6 $ years. However, if strange matter existed as the most stable state of ordinary matter, quark stars would exist, and they could have arbitrarily small radii, as shown in Figure \ref{mradio}.

It is worth highlighting something again: for this hypothesis to be true, it is necessary that at finite densities (non-zero chemical potentials), there exists free quark matter at zero pressure. This is automatically fulfilled in the MIT bag model, since by construction, when the pressure of the free part of the model equals the bag constant, there is zero pressure. However, as we will see later, in more sophisticated models such as that of Nambu Jona-Lasinio, the interface between chiral condensates (the NJL model does not automatically confine, but forms condensates), and free quark matter, is given at finite densities at non-zero pressure. Furthermore, free quark matter never reaches zero pressure, which is why the strange matter hypothesis is not satisfied in these models. However, what is possible is to build hybrid EoS that mix hyperonic matter with strange quark matter, which, as we said, is one of the objectives of this work. In the next section we will see how to build this type of EoS.

\section{Hybrid equation of state: phase transition according to Maxwell and Gibbs formalisms}
\label{construcciones}

So far we have defined the components we need for the construction of the EoS. However, we did not say anything about how the hadronic and quark EoS combine for the construction of hybrid stars. Taking into account the possibility of quark deconfinement inside the NSs, a phase transition from hadronic matter to quark matter could occur when the pressure of the quark phase is equal to that of the hadronic phase. As we anticipated at the beginning of this chapter, the nature of the phase transition depends on the surface tension at the hadron-quark interface, $ \sigma_ {HQ} $, which is still quite uncertain. Recent work suggests values $ \sigma_{HQ} \sim 30 \mathrm {MeV/fm^2}$, although this value could also be higher than 100 $\mathrm {MeV/fm^2} $ (see \cite{JPGreview} and its references). For this type of construction we can identify two types of phase transitions: abrupt (or Maxwell-style) or mixed (or Gibbs-style). If $\sigma_{HQ} \gtrsim 70 \mathrm{MeV/fm^2}$ the hadron-quark phase transition will be abrupt and at constant pressure, with an EoS presenting a discontinuity in the energy density. If $ \sigma_{HQ} \lesssim 70 \mathrm{MeV/fm^2}$, the phase transition will result in the formation of a mixed phase in which hadrons and quarks coexist, gradually converting hadronic matter into deconfined quark matter, as system pressure increases.

\begin{figure}[ht!]
\begin{center}
\subfigure{\includegraphics[scale = 0.26]{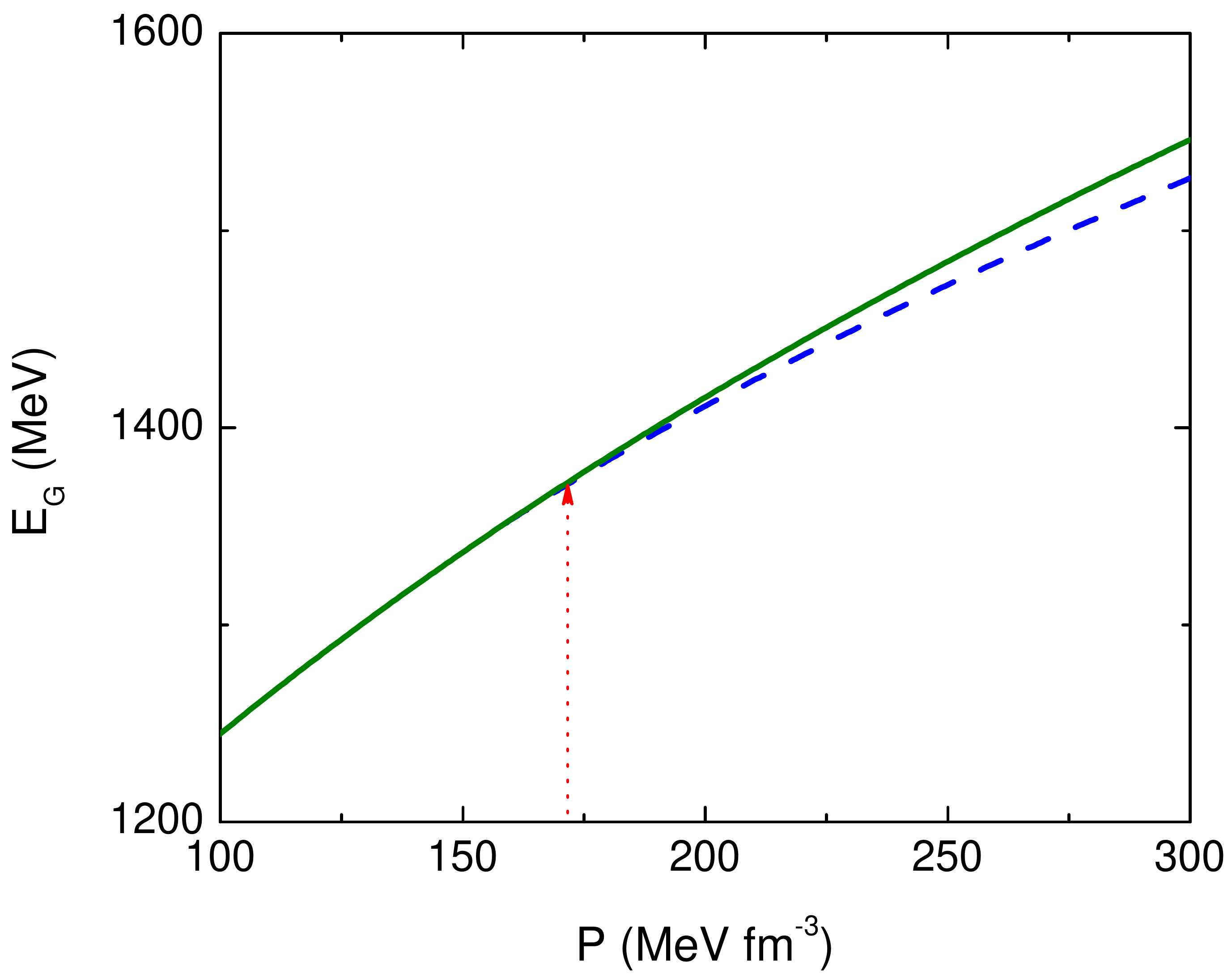}}
\subfigure{\includegraphics[scale = 0.26]{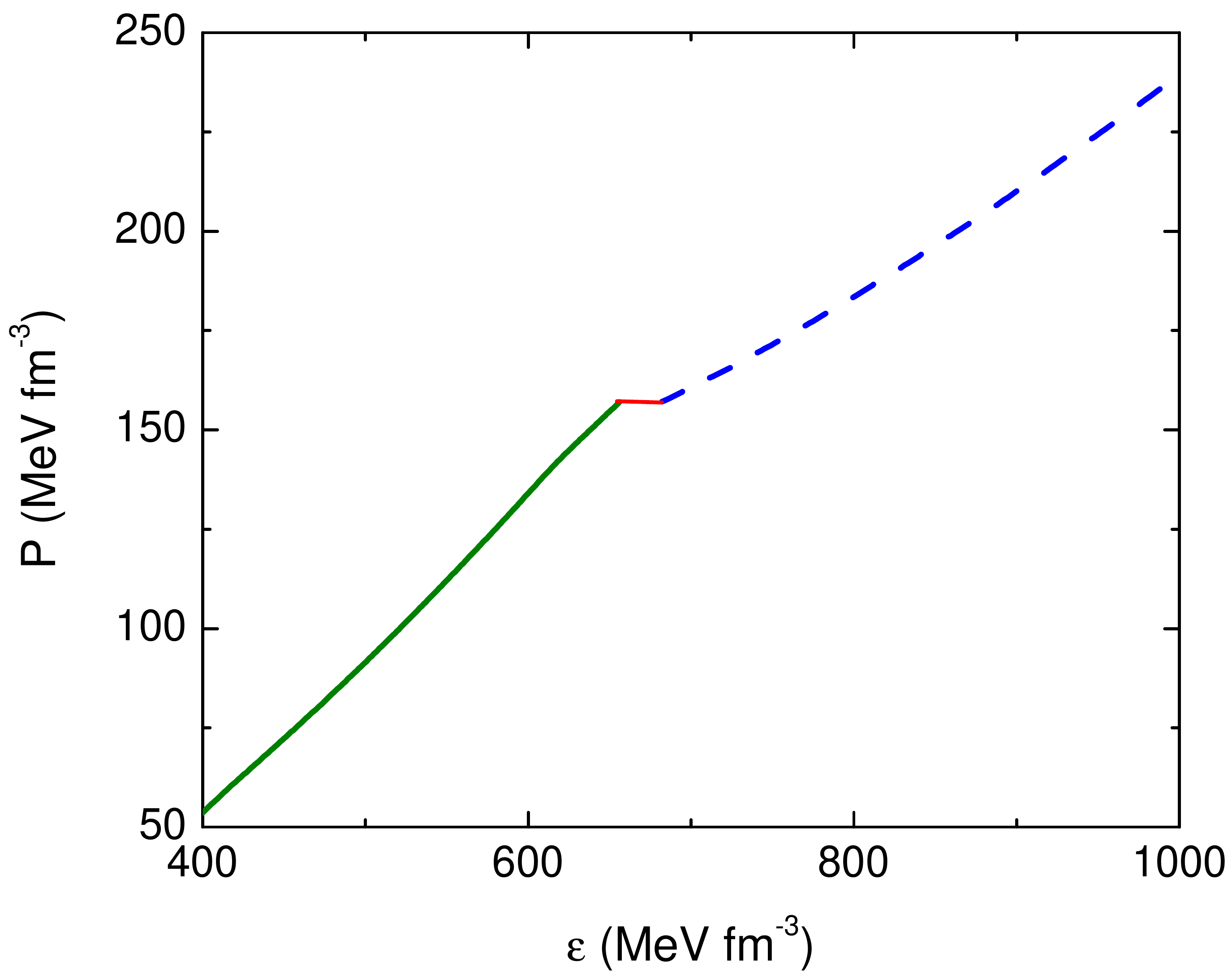}}
\caption[Construction of hybrid Maxwell's equations of state] {Left: Maxwell's construction for the phase transition between quarks and hadrons. The curve with the lowest Gibbs energy is the favored phase, the solid line corresponds to a hadronic matter EoS and the dotted one to a quark matter EoS. Right: Maxwell's construction for pressure as a function of energy density for phase transition from hadrons to quarks, the solid line corresponds to the energy jump between the phases. }
\label{constmaxwell}
\end{center}
\end{figure}

The condition for the transition from a hadron phase to a more stable phase is fulfilled if the Gibbs free energy of the quark phase is less than the Gibbs free energy of the hadron phase, for pressures higher than the transition pressure. In the case of a Maxwell phase transition, calling the hadronic $ (H) $ and Quarks $ (Q) $ phases, the conditions to be fulfilled in the interface for the case of three flavors of quarks are the following:
\begin{eqnarray}
 P^H (\mu_B) &=& P^Q(\mu_Q/3) , \\
 T^H (\mu_B) &=& T^Q(\mu_Q/3) , \\
 \epsilon^H (\mu_B) &<& \epsilon^Q(\mu_Q/3) , \\
 E_G^H(\mu_B) &=& E_G^Q(\mu_Q/3) , \\
 \mu_B &=& \mu_Q/3 = (\mu_u+ \mu_d + \mu_s)/3,
\end{eqnarray}
where $ \{P, T, \epsilon, E_G \} $ correspond to the pressure, temperature, energy density and Gibbs energy per particle. It should be noted that these types of constructions (Maxwell and Gibbs) are not exclusive of phase transitions from hadronic matter to quark matter, but rather they serve for general constructions of different types of phase transitions in thermodynamic equilibrium. In the literature, the series of conditions mentioned is usually found, without mentioning the Gibbs energy and it is built only by finding the contact points between the equations of state in the plane $ (P, \ mu) $ for equal temperatures. This is valid at zero temperature, when the Gibbs energy is equal to the chemical potential, or when the particles are the same in the two phases, since the definition of the Gibbs energy is given by
\begin{equation}
 E_G = \frac{G}{n_B} = \frac{\epsilon - TS  + P}{n_B} = \frac{\sum_i \mu_i n_i}{n_B},
\end{equation}
where $ G $ is the Gibbs energy per unit volume, $ \epsilon $ is the energy density per unit volume, $ T $ is the temperature, $ S $ is the entropy per unit volume, $ n_B $ is the baryon density and $ n_i $ is the particle density  of the species $ i $. When the system is homogeneous and we have the same class of particles in the two phases, asking that the Gibbs energy per particle be the same in both phases is equivalent to asking that the chemical potentials be equal. In the case of having different types (and quantities) of particles, such as going from a phase with hyperons to a phase with quarks, what we must ask is that the Gibbs energies per particle are equal. This brings us to situations as illustrated in Figure \ref{constmaxwell}.
Once the point is found where the Gibbs energies are equal, the EoS is constructed from $ P = 0 $ to the transition pressure with the lower Gibbs energy curve, and we continue from the transition pressure with the other curve, as long as it meets the condition of being the favored phase. In this way, when the EoS is plotted in the ($\epsilon, P) $ plane, there is a jump in energy density as seen in Figure \ref{constmaxwell} (right).

In the case of considering finite temperature, there are two options: either isothermal curves are created, or isentropic curves are created, depending on the required problem to be addressed. For the isothermal curves, the construction is the same as explained above, making sure that the EoS of both phases are at the same temperature. For isentropic curves, however, a problem appears. If, for example, one calculates the quark and hadronic EoS for entropy per baryon (for example) $ s = 2 $, even if the curves intersect, it is highly unlikely that at the crossing point the temperatures are equal. It is then necessary to make a modification to the isentropic construction, as follows: you start by looking for isothermal curves of any desired temperature, and finding the crossing point. Once the transition point $ (P_T, E_{G, T}) $ has been found, the entropy of each phase is calculated. By having the entropy of each phase, the isentropic equations of state are calculated and the crossing is made again, thus ensuring that the transition occurs at the same temperature in both phases. This generates an equation of state that is not strictly isentropic, but that maintains constant entropy in each phase, and fulfills all the thermodynamic conditions required for the interface.
For the Gibbs construction, the procedure is different, and more complex. What is assumed here is that the matter will be purely hadronic up to a certain density, then it will be a mixture of free quarks and hadrons, finally having only free quarks. This construction, which is naturally known as the mixed phase, is carried out as follows. With the same definitions of pressure, temperature, entropy and baryon density as before, plus a parameter $\chi$ that takes values between 0 and 1, we write the conditions as in the work of the reference \cite{Chen:2013tfa} :
\begin{eqnarray}
 P^H (\mu_B) &=& P^Q(\mu_Q/3) \label{Gibbs1}, \\
 \chi n_B^H(\mu_B) + (1 - \chi)n_B^Q(\mu_Q/3) &=& n_B, \\
 \chi n_c^H(\mu_B) + (1 - \chi)n_c^Q(\mu_Q/3) &=& 0, \label{neutcargagibbs} \\
 \chi n_L^H(\mu_B) + (1 - \chi)n_L^Q(\mu_Q/3) &=& n_B Y_L, \\
 \chi S^H(\mu_B) + (1 - \chi)S^Q(\mu_Q/3) &=& n_B S \label{entropiafijagibbs},
\end{eqnarray}

\begin{figure}[ht!]
\begin{center}
\subfigure{\includegraphics[scale = 0.26]{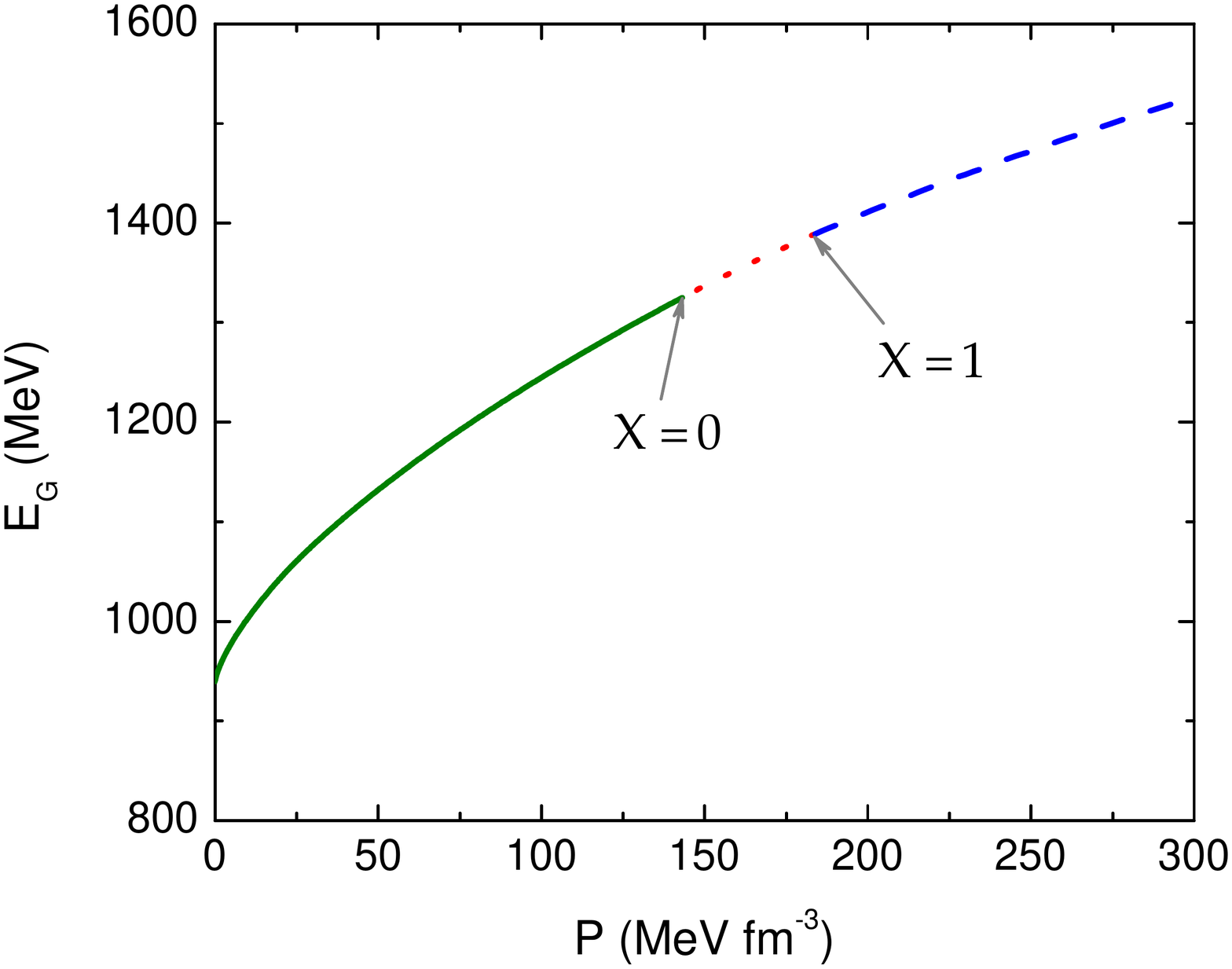}}
%\hspace{0.02\textwidth}
\subfigure{\includegraphics[scale = 0.26]{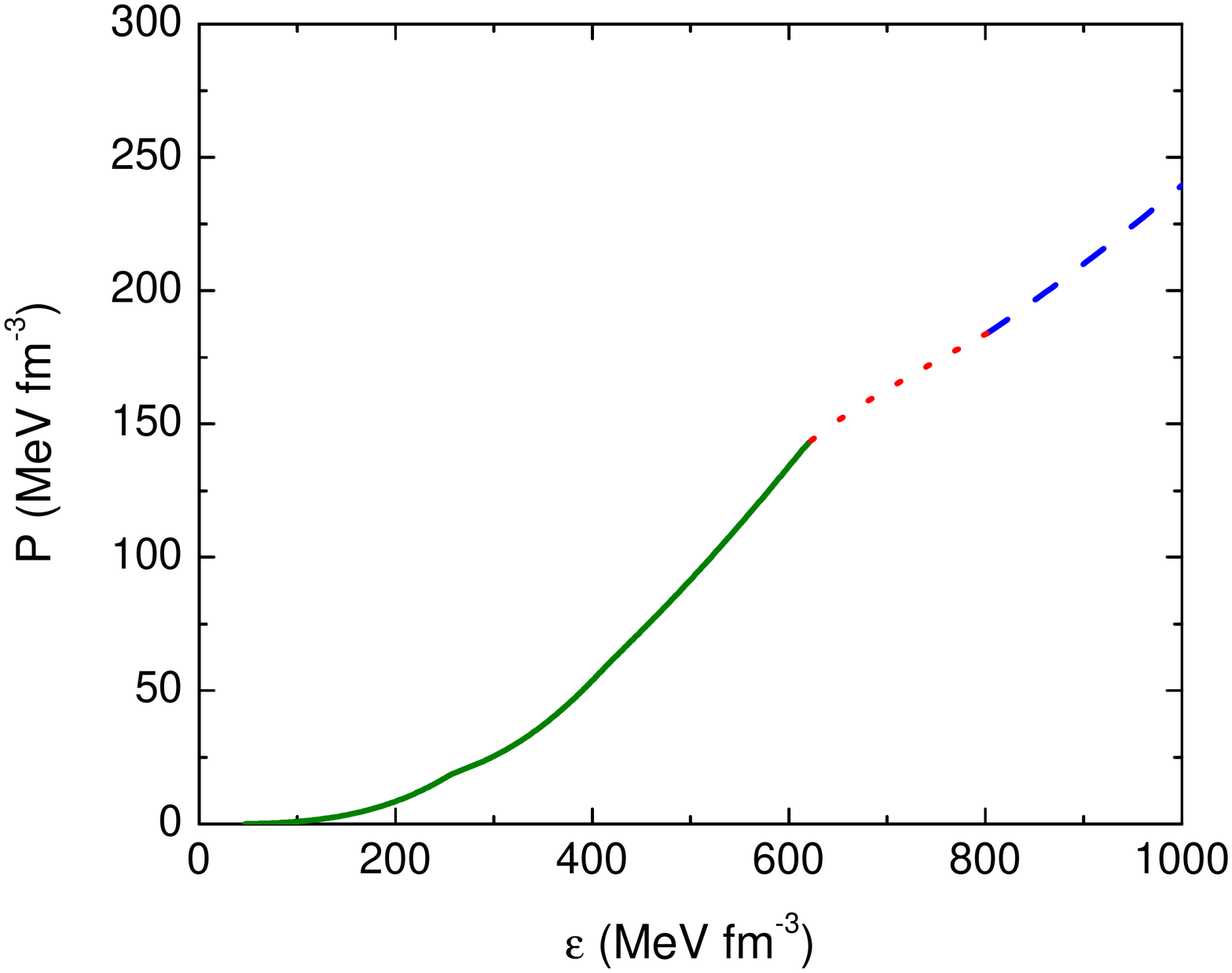}}
\caption[Construction of hybrid Gibbs equations of state.] {Left: Gibbs construction for the mixed phase between quarks and hadrons. Right: result of Gibbs construction for pressure as a function of energy density for phase transition from hadrons to quarks. In both cases the solid line corresponds to a hadronic matter EoS, the dashed line to a quark matter EoS, and the dotted line in the middle to the mixed phase EoS.}
\label{constgibbs}
\end{center}
\end{figure}

where now the thermodynamic quantities are defined as a heavy weight by the variable $ \chi $ between the two phases, and the last equation is only used in the case of requiring fixed entropy in the mixed phase. It is worth highlighting in this case the equation \eqref{neutcargagibbs}, where $ n_c $ is the charge density of each phase. In the case of the Maxwell construction, charge neutrality was a local requirement in each phase, so this equation did not appear. In the mixed phase construction, charge neutrality, lepton number and entropy are required to be conserved globally between the two phases. As we will see in the results that we will show later, this leads to very different particle distributions than in the previous construction. As for the equation \eqref{entropiafijagibbs}, in this case the temperature becomes a variable to be adjusted numerically within the system of equations. Unlike Maxwell's construction, solving the mixed phase requires including both the quark and hadron numerical codes in a code that contains them, to establish the system of equations that solves the thermodynamic quantities consistently, as opposed to just needing to cross curves and just erase the not needed part of each EoS. Once the mixed phase is resolved, the hadronic curve is used up to the point $ \chi = 1 $ where the mixed phase begins, and then the quark curve from the point $ \chi = 0 $ where it ends, as illustrated in the Figure \ref{constgibbs}. Another difference with the previous case is that now for the limits $ \chi = 1 $ and $ \chi = 0 $ the mixed phase construction must coincide in all of its variables with the pure hadronic and quark constructions in each case.

So far we have explained all the components we need at the thermodynamic and hydrostatic level to build the EoS and structure that allow us to model hybrid stars. However, we said nothing about the particular models to be used to describe each phase (hadrons or quarks). In this work we will use, for the two types of constructions, the non-linear Walecka model with density-dependent coupling constants to model the hadronic phase, and the non-local  extension of the Nambu-Jona-Lasinio model with Polyakov's loop to model the quark phase, which we will develop in the following chapters.

\graphicspath{{Nambu/}}
\chapter{\label{ch:Nambu}Non-local quark model}

In this chapter we will describe in detail the non-local extension of the Nambu-Jona-Lasinio model \cite{Nambu:1961fr}. This model was originally conceived to describe the interactions between nucleons, so an interpretation of it must be redone if one wants to use quarks as degrees of freedom instead of nucleons. However, unlike the MIT bag model, this model mantains several of QCD's global symmetries, and interactions can be incorporated that explicitly break the symmetries that both experimental and theoretical evidence show should be broken in certain regimes. Particularly, one of its advantages lies in having the same spontaneous breaking chiral symmetry breaking mechanism as QCD, which allows its parameters to be adjusted to the meson's masses, which makes it especially more robust than other models. Likewise, the NJL model non-local extension implies replacing the interactions in the Lagrangian by non-local interaction terms, modulating the interaction of the fields by an appropriate form factor. Among other advantages (see\cite {Orsaria:2013hna} and its references), non-local interactions generate quark's dynamic masses that depend on the momentum, as suggested by \textit{Lattice QCD} \cite{PhysRevD.73.054504}.

\section{QCD overview}
To model the quark phase, we need to understand that QCD is a theory that cannot be solved both in the regime of neutron stars \cite{Itzykson:1980rh} and in the regime of low temperatures and densities. It is necessary, therefore, to appeal to effective models such as those of the MIT bag model or the NJL model, as well as to understand basic aspects of this theory, to base the choice of each effective model used. To see some general aspects then, let's start with the QCD Lagrangian, which is given by
\begin{equation}
\mathcal{L}_{QCD}= \overline{\psi}(x)(i\gamma_{\mu}D^{\mu}-\hat{m})\psi(x)- \\
\frac{1}{4}G_{\mu\nu}^{a}G^{\mu\nu}_{a}, \label{lagQCD}
\end{equation}
where the fields $ \psi (x) = (\psi_{uc} (x), \psi_{dc} (x), \psi_{cc} (x), \psi_{sc} (x), \psi_{ tc} (x), \psi_{bc} (x)) $ have 18 degrees of freedom, six for each flavor of quarks and three for each color $ c = \{r, g, b \} $, the bare masses matrix of quarks $ \hat m $ have only diagonal components in flavor. The covariant derivative $ D ^ \mu $ relates to gluonic fields in the way
\begin{eqnarray}
 D_{\mu} &=& \partial_{\mu}-ig\lambda^{a}A_{\mu}^{a} \\
 G_{\mu\nu}^{a} &=& \partial_{\mu}A_{\nu}^{a}-\partial_{\nu}A_{\mu}^{a}+gf^{abc}A_{\mu}^{b}A_{\nu}^{c},
\end{eqnarray}
with $ \lambda_a $ the Gell-Man matrices, which are the generators of the SU (3) group of color symmetries of the theory and $ f^{abc} $ the antisymmetric structure constant of the algebra of that group. The quantity $ g $ is the coupling constant between quarks and gluons.

Due to the non-abelian structure of the SU (3) symmetry group of the theory, it is worth mentioning certain peculiarities that do not appear in abelian theories (such as QED for example):

\begin{itemize}
 \item $ \mathcal {L}_{QCD} $ has self-coupling between gluons (three and four gluons per vertex), therefore gluons have color charge.
 \item QCD is asymptotically free. Coupling becomes weak at short distances (or high transferred Euclidean moments $ Q $). At one loop order, for the strong coupling constant, we have to
     \begin{equation}
	\alpha_s(Q^2)\equiv \frac{g^2(Q^2)}{4\pi} = \frac{4\pi}{(11-\frac{2}{3}N_f)\mathrm{log}(Q^2/\Lambda^2_{QCD})},
	\label{alphaQCD}
     \end{equation}
  where $ N_f $ is the number of flavors and $ \Lambda_{QCD} $ is the theory's scale parameter, that can be determined by fitting large experimental data $ Q^2 $ and depends strongly on the number of flavors and the renormalization scheme used. The fact that the theory is asymptotically free allows it to be treated perturbatively at high energy scales ($ \alpha_s \to $ 0 when Q $ \to \infty $).
  
\item The theory becomes non-perturbative at low energy scales. The equation \eqref{alphaQCD} shows that the coupling is intensified for small Q. For transferred moments less than $ \Lambda_{QCD} \sim 200 $, $ \alpha_s> 1 $ which makes it not a suitable parameter for a series expansion in the treatment of theory. For this reason, it cannot be used to describe hadrons of masses less than about 1-2 GeV. This could (or not) be related to confinement, that is, the fact that non-neutral colored objects are not observed as degrees of freedom in a vacuum, whether they are quarks or gluons.
\end{itemize}

Another characteristic that will be of great importance in this work is that QCD has an approximate chiral symmetry. This means that it is symmetric under global transformations of $ SU (N_f)_L \times SU (N_f)_R $ with $ L $ and $ R $ corresponding to the left and right. This is the same as saying that it is invariant to vector and axial transformations in the isospin space
\begin{eqnarray}
 SU(N_f)_V: \psi \to e^{(i\theta_a^V \tau_a)} \psi, \\
 SU(N_f)_A: \psi \to e^{(i\theta_a^A \gamma_5 \tau_a)}\psi ,
\end{eqnarray}
with $ \gamma_5 $ the Dirac matrix, $ \tau_a $ the $ SU (N_f) $ group generators, and $ \theta $ the transformations parameters. These symmetries are not strictly exact, but they are in the case of $ N_f = 2 $ for up and down quarks if the same mass is taken for the two quarks, and it can even be used to analyze the case $ N_f = 3 $ when the strange quark is included, which is the case that we will deal with in this work.

The QCD vacuum also roughly respects the $ SU (N_f)_V $ symmetry, the hadron spectrum has almost a degenerate multiplet in $ SU (N_f) $. If this were exactly true for axial symmetry, each hadron should have a degenerate chiral partner of opposite parity.Since this does not happen, it turns out that the chiral symmetry is spontaneously broken in a vacuum. On the other hand, if the chiral symmetry is spontaneously broken this gives $ (N_f^2 -1) $ Goldstone bosons. If we interpret, for example, the pions as Goldstone bosons, the mixture of the spontaneous symmetry breaking plus the explicit one (corresponding to the nonzero masses of the quarks $ u $ and $ d $), can perfectly explain the pion's low mass. This work was originally done by {\bf{Nambu}} \cite{Nambu:1960tm}, which led to the actual QCD model that we will use throughout this work, albeit in a modified version including non-local interactions and color superconductivity.

On the other hand, at the classical level and at the zero mass limit for quarks, the Lagrangian is invariant under global transformations 
\begin{equation}
 UA(1): \psi \to e^{(i\alpha \gamma_5)} \psi.
\end{equation}
This symmetry is not a true symmetry of the QCD, since it is broken at the quantum level. \cite {PhysRevD.14.3432}, that is, the current $ J_{\mu 5} (x) = \bar {\psi} (x) \gamma_\mu \psi (x) $ has non-zero divergence when $ \bar {\psi} $ and $ \psi $ are considered as fields. When this is taken into account it results
\begin{equation}
\label{divaxial}
 \partial_\mu J^\mu_5(x) = \frac{g^2 N_f}{8 \pi^2}E_a B_a,
 \end{equation} 
 where $ E_a $ and $ B_a $ are the color electric and magnetic fields. The Yang-Mills equations admit instanton-type classical solutions, for which the Euclidean integral $ \int d^4x E, B $ is finite. Using the equation \eqref {divaxial}, these instantons are coupled to the quarks. In the work of 't Hooft \cite{tHooft:1976rip} it is shown that this effective interaction induced by instantaneous, has the form  
 \begin{equation}
 \label{determinantethooft}
  \mathrm{det}[\psi_i (1 + \gamma_5)\psi_i + \mathrm{det}[\psi_j (1 - \gamma_5)\psi_j,
 \end{equation}
 where the indices $ i $ and $ j $ run through the three flavors of quarks, thus ending up being an effective interaction of six fermions.

Finally, among the possible interactions between gluons and quarks, there are channels in the one gluon exchange regime, which are attractive between quarks. As we will see in the chapter \ref{ch:Superconductividad}, this should necessarily generate superconducting color phases, and diquarks condensates. These condensates are crucially dependent on the number of flavors and colors considered, and developing the general theory of the existing number of these objects for QCD is beyond the scope of this work. For the moment, it is worth noting only that since there are attractive channels, this phenomenon should in principle exist.

So far we have made a summary of the most important symmetries and characteristics of QCD. As we said in the introduction, our purpose in this work is to study the presence of quark matter in the interior of NSs, modeling them with an EoS that accounts for the hadronic part and the part of quarks that compose it. As the regime in which we will work the QCD is non-perturbative, it is not solvable. We therefore need an effective model that accounts for the aforementioned aspects, which in short, are:

\begin{itemize}
 \item Chiral symmetry $ SU (N_f)_L \times SU (N_f)_R $ spontaneous breaking mechanism.
 \item Axial symmetry $UA(1)$ breaking mechanism.
 \item Confinement.
 \item Color superconductivity.
\end{itemize}

Next we will see how to introduce the first two elements in one of the simpler versions of the non-local model of Nambu-Jona-Lasinio, and in the following chapters we will deal with confinement and color superconductivity.

\section{Theoretical formalism of the non-local NJL model}

We will now develop one of the simplest variants of the non-local NJL model. After understanding the most fundamental aspects, it will be quite straightforward to add different types of interactions. We then start from the Euclidean action \footnote {The integrals in this chapter are made over the entire d-dimensional Euclidean space of the differential of each integral, or the complete functional space in the case of functional integrals.}:

\begin{eqnarray}
\label{accionNJL}
 S_E &=& \int d^4 x \Big\{\overline{\psi}\left(x\right)\left( i \slashed{\partial} + m \right)\psi\left(x\right) -\frac{G}{2}\left[ j_a^s\left(x\right)j_a^s\left(x\right) + j_a^p\left(x\right)j_a^p\left(x\right)\right]      \nonumber \\ 
 &-& \frac{H}{4}A_{abc}\left[j_a^s\left(x\right)j_b^s\left(x\right)j_c^s\left(x\right) - 3 j_a^s\left(x\right)j_b^p\left(x\right)j_c^p\left(x\right)\right]\Big\},
\end{eqnarray}
where the Dirac matrices correspond to the Euclidean representation, and the scalar and pseudoscalar currents are defined as

\begin{eqnarray}
 j_a^s\left(x\right) &=& \int d^4 z g(z)\overline{\psi}\left(x + \frac{z}{2}\right)\lambda_a \psi\left(x-\frac{z}{2}\right) \\
 j_a^p\left(x\right) &=& \int d^4 z g(z)\overline{\psi}\left(x + \frac{z}{2}\right)i \gamma_5\lambda_a \psi\left(x-\frac{z}{2}\right)  ,
\end{eqnarray}
where $ \lambda_a $ are the eight Gell Mann matrices plus $ \lambda_0 = \sqrt {2/3} \; \bm {1}_{3 \times 3} $. The function $ g (z) $ corresponds to a non-local regulator, for which we choose a Gaussian type, being

\begin{eqnarray}
 g(z) &=& \int \frac{d^4 p}{(2\pi)^4}e^{-izp} g(p) \label {regulador} \\
 g(p) &=& e^{-\frac{p^2}{\Lambda^2}}.
 \label{reguladortransformado}
\end{eqnarray}

As we can see, we have two types of currents, scalars $ j_a^s \left (x \right) $ and pseudoscalars $ j_a^p \left (x \right) $. The coupling that mixes the same types of currents (which we will call scalar coupling for simplicity) has the coupling constant $ G $ and the one that mixes different types corresponds to the term of t 'Hooft \cite {tHooft:1977nqb}, which is responsible for the axial symmetry breaking that we described in the previous section. The regulator, in addition to fulfilling the function of generating non-locality, has an important advantage over the local model: in that model, it is necessary to introduce a cutoff $ \Lambda $ in the integrals which adjusts to reproduce the pion's mass in the expectation values of the condensates $ \langle \overline {\phi} \phi \rangle $. The problem with this is that this parameter works in the low-density regime, where condensates exist, but the use of a cutoff when working at high densities is at least contradictory, since when passing to finite density (as we will see) the chemical potential of each quark is coupled to the zero component of the momentum. In the non-local case, the regulator prevents the cut for being abrupt as the momentum increases, therefore it is more reasonable to work with these models at high densities. The problem with the non-local model is that the explanation for the spontaneous chiral symmetry breaking becomes more extensive and less obvious to the naked eye. Due to this, we will give a simplified explanation in the following sections, inviting the reader to review the references \cite{Buballa:2003qv} for a more detailed explanation.

\section{Bosonization}
\label{bosonizacion}
From the relationship between statistical mechanics and field theory we know that the partition function can be written as $ {\cal {Z}} = \int {\cal {D}} \overline {\psi} {\cal {D} } \psi e^ {- S_E} $. Our goal now is to bosonize the action so that it remains in terms of bosonic fields that are easier to deal with, for that we introduce the identity

\begin{equation}
 f\left(j_a^s, j_a^p\right) = \int {\cal{D}}S_a{\cal{D}}P_a \delta\left(S_a - j_a^s\right)\delta\left(P_a - j_a^p\right)  f\left(S_a,P_a\right),
\end{equation}
where $ S_a (x) $ and $ P_a (x) $ are the auxiliary fields and we write the deltas as follows:

\begin{eqnarray} \delta\left(S_a - j_a^s\right) &=& \int {\cal{D}}\sigma_a \mathrm{exp}\{\int d^4 x\, \sigma_a \left(x\right)\left[S_a\left(x\right) - j_a^s\left(x\right)\right]\} \\
 \delta\left(P_a - j_a^p\right) &=&\int {\cal{D}}\pi_a \mathrm{exp}\{\int d^4 x \, \pi_a \left(x\right)\left[S_a\left(x\right) - j_a^p\left(x\right)\right]\}.
\end{eqnarray}
where $ \sigma_a (x) $ and $ \pi_a (x) $ are the mesonic, scalar and pseudoscalar fields respectively. Then the current's functional is
\begin{eqnarray}
 f\left(j_a^s,j_a^p\right) &=&  \mathrm{exp}\int d^4 x \Big\{ \frac{G}{2}\left[ j_a^s\left(x\right)j_a^s\left(x\right) + j_a^p\left(x\right)j_a^p\left(x\right)\right]     \nonumber \\
 &-& \frac{H}{4}A_{abc}\left[j_a^s\left(x\right)j_b^s\left(x\right)j_c^s\left(x\right) - 3 j_a^s\left(x\right)j_b^p\left(x\right)j_c^p\left(x\right)\right]\Big\}.
\end{eqnarray}

Using the delta's identities and the current function, the idea is to write the partition function in terms of the new fields. The terms that are explicitly dependent on the $ \psi $ fields and the currents (which depend on $ \psi $ within the integrals), cannot be taken out of the functional integral. The rest can be taken out of the fermionic fields integral, thus obtaining
\begin{eqnarray}
 {\cal{Z}} &=& \int {\cal{D}}\sigma_a {\cal{D}} \pi_a {\cal{D}} S_a {\cal{D}} P_a \,\mathrm{exp} \Big\{ \int d^4x \Big[\frac{G}{2}\left(S_a S_a + P_a P_a \right) + \frac{H}{4}\left(S_a S_b S_c - 3S_a P_b P_c \right)\Big] \nonumber\\
 &+& \sigma_a S_a + \pi_a P_a \Big\}\int {\cal{D}}\overline{\psi} {\cal{D}} \psi \,\mathrm{exp}\left\{ \int d^4x \left[\overline{\psi}\left( i \slashed{\partial} + m \right)\psi - \sigma_a j_a^s - \pi_a j_a^p\right]\right\} \label{particionpsi}.
\end{eqnarray}

After performing the fermionic fields integrals (explained in Appendix \ref{ApendiceA}), and writing $\mathrm{det}(A) = \mathrm{exp}\left\{\mathrm{log}\left[\mathrm{det}(A)\right]\right\}$, we can write the bosonized partition function as
\begin{eqnarray}
 {\cal{Z}}^{\mathrm{Bos}} &=& \int {\cal{D}}\sigma_a {\cal{D}} \pi_a {\cal{D}} S_a {\cal{D}} P_a \, \mathrm{exp}\Biggl\{ \mathrm{log}\left[\mathrm{det}(A)\right] \nonumber \\
 &+& \int d^4x\Biggl[\sigma_a S_a + \pi_a P_a + \frac{G}{2}\left(S_a S_a + P_a P_a\right) \nonumber \\
 &+& \frac{H}{4}A_{abc}\left(S_a S_b S_c - 3 S_a P_b P_c\right)\Biggr] \Biggr\}.
\end{eqnarray}
With which, knowing that the partition function is the functional integral of the negative of the exponential of the action, we obtain that the bosonized action is

\begin{eqnarray}
\label{Sbos}
 S_E^{\mathrm{Bos}} = - \mathrm{log\left[\mathrm{det}\left(A\right)\right]} &-& \int d^4x\Biggl[\sigma_a S_a + \pi_a P_a + \frac{G}{2}\left(S_a S_a + P_a P_a\right) \nonumber \\ 
 &+& \frac{H}{4}A_{abc}\left(S_a S_b S_c - 3 S_a P_b P_c\right)\Biggr],
\end{eqnarray}
where for notation simplicity we have avoided the dependencies, but it is worth remembering that the operator $ A $ depends on the momentums and the auxiliary fields, and the bosonic fields as well as the auxiliary $ (\sigma_a, \pi_a, S_a, P_a) $ depend of the coordinates $ x $ in Euclidean space.
               
\section{Mean Field approximation}

To continue with the objective of finding the great thermodynamic potential and the gap equations, we will perform the mean field approximation (MFA). For this, first it is worth noting that the $ \sigma_a (x) $ fields have non-zero mean values for $ a = \{0,3,8 \} $, and the $ \pi_a (x) $ fields have null mean values, since the vacuum of QCD has to be invariant to charge conjugation and parity. That said, we expand the fields around their mean values:

\begin{eqnarray}
 \sigma_a (x) &=& \overline{\sigma}_a + \delta_{\sigma_a}(x) \\
 \pi_a(x) &=& \delta_{\pi_a}(x). 
\end{eqnarray}

Once this is done, the bosonized action of the equation \ eqref {Sbos} can be expanded at different orders of the fields fluctuations. The zero order in fluctuations corresponds to the mean field approximation. This approximation assumes that bosonic fields maintain a value equal to their mean value throughout space. The first order development in the fluctuations is canceled, since the bosonized fields minimize the action. The next leading order would be the quadratic one, so that
\begin{equation}
 S_E^{\mathrm{Bos}}  = S_E^{\mathrm{Bos}, \mathrm{MFA}} + S_E^{\mathrm{Bos}, \mathrm{Quad}} + ...
\end{equation}

It should be noted that the quadratic order is the one used to adjust the coupling constants of the model to the physical observables such as the mesons masses, the decay constants, etc. In this work, the development at that order is an excess, since to obtain the thermodynamic potential and the necessary quantities for astrophysical calculations, the mean field approximation is sufficient. Using this approximation, we have that the operator $ A (p, p') $ of equation \eqref{operadorA} is written as
\begin{eqnarray}
 A(p,p') &=& \left[-\slashed{p} + M(p)\right](2\pi)^4 \delta^4 (p - p')\,\, \mathrm{con} \\
 \label{Masas1}
  M(p) &=& m + g(p) \lambda_a \overline{\sigma}_a,
\end{eqnarray}
where we rewrite the regulator's argument $ g (p) $ so that the quadrimomemtum delta remains as a common factor. Then we can decompose the operator's determinant so that we have
\begin{equation}
 \mathrm{det}\left[A(p,p')\right] = \mathrm{det}\left[-\slashed{p} + M(p)\right]  \mathrm{det} \left[ (2\pi)^4 \delta^4 (p - p') \right].
\end{equation}

It is easy to see (using Fourier transform for example) that the delta's determinant in momentum space together with the factor $ (2 \pi)^4 $ is equal to unity. Then, we are left to calculate only the first determinant, which being within a logarithm, we can replace everything with the trace of the logarithm, thus remaining
\begin{equation}
 \mathrm{log}\left[\mathrm{det}(A)\right] =  \mathrm{Tr}\left\{\mathrm{log}[A(p,p')]\right\} = \int \frac{d^4p}{(2\pi)^4} \frac{d^4p'}{(2\pi)^4}\mathrm{Tr}\left\{ \mathrm{log}\left[-\slashed{p} + M(p)\right]\right\}.
\end{equation}
In this case, the trace is carried out on all the spaces considered, that is, flavor, color, Dirac and momentums. It should be noted that the trace has to continue to be carried out on the operator's original number of degrees of freedom, so as not to undercount. In this case, we have to continue doing the integral over $ p '$, which, having no argument, ends up giving us the space's volume, therefore

\begin{equation}
 \mathrm{log}\left\{\mathrm{det}\left[A\left(p,p'\right)  \right]\right\} =  V^{(4)}\int \frac{d^4p}{(2\pi)^4} \mathrm{Tr}\left\{\mathrm{log}\left[-\slashed{p} + M(p)\right]\right\}.
\end{equation}

We can continue working the term of the trace, using matrices properties we have to
\begin{eqnarray}
\label{trazaslogs}
 &\,& \mathrm{Tr}\left[\mathrm{log}\left(-\slashed{p} + M \right)\right] \nonumber \\
 &=& \frac{1}{2} \left\{\mathrm{Tr}\left[\mathrm{log}\left(-\slashed{p} + M \right)\right] + \mathrm{Tr}\left[\mathrm{log}\left(-\slashed{p} + M \right)\right] + \mathrm{Tr}\left[\mathrm{log}\left(-\slashed{p} + M \right)^{\dagger}\right]  - \mathrm{Tr}\left[\mathrm{log}\left(-\slashed{p} + M \right)^{\dagger}\right] \right\} \nonumber \\
 &=& \frac{1}{2} \mathrm{Tr}\left[\mathrm{log}\left(-\slashed{p} + M \right)\left(-\slashed{p} + M \right)^{\dagger}\right] + \frac{1}{2}\mathrm{Tr}\left[\mathrm{log}\left(-\slashed{p} + M \right)^{\dagger}\left(-\slashed{p} + M \right)^{-1}\right].
\end{eqnarray}
the last term in the equation \eqref{trazaslogs} vanishes. This can be seen by exchanging the logarithm's trace for the logarithm of the determinant and checking that it does not equal the unit, therefore the logarithm vanishes. The first term, after making the matrix product and the trace over Dirac components, results
\begin{equation}
 \label{arglog} 
  \mathrm{log}\left\{\mathrm{det}\left[A\left(p,p'\right)  \right]\right\} = 2\,\mathrm{Tr} \left\{\mathrm{log} \left[ p^2 + M^2(p) \right]\right\},
\end{equation}
where the trace to be made now is about flavor and color only.

For the rest of the bosonized action's terms, it is only necessary to replace $ \sigma_a (x) $ with $ \overline {\sigma}_a $ and cancel $ \pi_a (x) $ since its mean value is null . Doing this and replacing in the equation \eqref {Sbos} we have that
\begin{eqnarray}
\label{SMFA}
 S_E^{\mathrm{MFA}} &=& - V^{(4)}\int \frac{d^4p}{(2\pi)^4} 2\, \mathrm{Tr}\left\{\mathrm{log}\left[p^2 + M^2(p)\right]\right\} \nonumber \\
 &-&\int d^4x\left[\overline{\sigma}_a S_a + \frac{G}{2}\left(S_a S_a + P_a P_a\right) + \frac{H}{4}A_{abc}\left(S_a S_b S_c - 3 S_a P_b P_c\right)\right],
\end{eqnarray}
where we already assume that it is the bosonized action, and we have replaced the name so that it is understood that it is in the mean field approximation.

\section{Stationary Phase Approximation}

So far we have done everything possible to obtain a solvable partition function that allows us to calculate the thermodynamic quantities that we are going to need for the development of our study. However, the equation \eqref {SMFA} leads to an equation for the partition function that is not quadratic in the fields, and it is difficult to solve. It is necessary to point out that if the 't Hooft mixing term were not present ($ H = 0 $), the action would be quadratic in the fields and perfectly solvable. As we said, this term is of interestbecause it is the one that breaks the axial anomaly, and also because, as we will see later, it influences the disappearance of the $ s $ quark condensates and that generates changes both in the phase diagram and in the EoS. For these reasons it is interesting to keep this term, and to perform what is known as the stationary phase approximation (SFA). In this approximation we assume that the integral is performed on the path that minimizes the action. We can then define the auxiliary fields $ \tilde {S_a} $ and $ \tilde {P_a} $ in such a way that they minimize the integrand, such that
\begin{eqnarray}
 \frac{\delta}{\delta_{S_a}} S_E^{\mathrm{MFA}} \bigg\rvert_{S_a = \tilde{S_a}; P_a = \tilde{P_a}} = 0, \\
 \frac{\delta}{\delta_{P_a}} S_E^{\mathrm{MFA}} \bigg\rvert_{S_a = \tilde{S_a}; P_a = \tilde{P_a}} = 0,
\end{eqnarray}
with which we arrive at two equations for the new auxiliary fields, given by
\begin{eqnarray}
 \overline{\sigma}_a + G \tilde{S_a} + \frac{3}{4} H A_{abc}\left(\tilde{S_b} \tilde{S_c} - \tilde{P_b}\tilde{P_c}\right)  &=& 0, \\
 \label{ecuacionindices}
 G\tilde{P_a} - \frac{3}{2}H A_{abc}\left(\tilde{S_b}\tilde{P_c}\right)  &=& 0. 
\end{eqnarray}

Now, by replacing the equation \eqref{ecuacionindices} in the \eqref{SMFA}, and using the tensor's properties $ A_ {abc} = A_ {cab} $, one can, after properly accommodating the expressions, arrive at an action that no longer depends on the fields $ \tilde {P_a} $, obtaining
\begin{eqnarray}
\label{SMFA2}
 S_E^{\mathrm{MFA}} &=& - V^{(4)}\int \frac{d^4p}{(2\pi)^4} 2\, \mathrm{Tr}\left\{\mathrm{log}\left[p^2 + M^2(p)\right]\right\} \nonumber \\
 &-&\int d^4x\left(\overline{\sigma}_a \tilde{S_a} + \frac{G}{2}\tilde{S_a}\tilde{ S_a}  + \frac{H}{4}A_{abc}\tilde{S_a}\tilde{S_b}\tilde{S_c}\right).
\end{eqnarray}
But since we said that the auxiliary fields $ \tilde {S_a} = \tilde {S_a} (\overline {\sigma}_a) $ and $ \overline {\sigma}_a $ no longer depend on the spatial coordinate, we can write the action per unit volume, as follows
\begin{eqnarray}
\label{SMFA3}
 \frac{S_E^{\mathrm{MFA}}}{V^{(4)}} = &-& \int \frac{d^4p}{(2\pi)^4} 2\, \mathrm{Tr}\left\{\mathrm{log}\left[p^2 + M^2(p)\right]\right\} \nonumber \\
 &-& \overline{\sigma}_a \tilde{S_a} -  \frac{G}{2}\tilde{S_a}\tilde{ S_a}  - \frac{H}{4}A_{abc}\tilde{S_a}\tilde{S_b}\tilde{S_c},
\end{eqnarray}
which leads us to a partition function that integrates over constant fields, except for the integral at momentums. Now, using that the $ \tilde {S_a} $ fields minimize the action, the so-called "gap" equations are obtained for the SFA of the nl-NJL model. (depending on what values are given to $ a $)
\begin{equation}
\label{gapsindiagonalizar}
 \overline{\sigma}_a + G \tilde{S_a} + \frac{3}{4} H A_{abc}\tilde{S_b} \tilde{S_c} = 0,
\end{equation}
where the auxiliary fields are defined minimizing the action with respect to $ \overline {\sigma}_a $

\begin{equation}
 -2 \frac{\delta}{\delta_{\overline{\sigma}_a}} \int \frac{d^4p}{(2\pi)^4} \, \mathrm{Tr}\left\{\mathrm{log}\left[p^2 + M^2(p)\right]\right\}  = \tilde{S_a}.
\end{equation}
\section{Diagonalization, gap equations and grand potential at T = 0}
As we said earlier, only $ \overline {\sigma} _0 $, $ \overline {\sigma} _3 $ and $ \overline {\sigma} _8 $ are nonzero. Then we define a diagonal matrix
\begin{equation}
 \mathrm{diag}\left(\overline{\sigma}_u,\overline{\sigma}_d,\overline{\sigma}_s\right) = \overline{\sigma}_0 \lambda_0 + \overline{\sigma}_3 \lambda_3 + \overline{\sigma}_8 \lambda_8.
\end{equation}

Let us note, to begin with, that the dynamic masses defined in the equation \eqref{Masas1}, are automatically diagonals in flavor by the sum that runs in the index $ a $. For this reason, from now on we will say that $ M_f (p) = m_f + \sigma_f g (p) $ with $ f = \{u, d, s \} $. On the other hand, rewriting the equations of the gap \eqref{gapsindiagonalizar} for the indices $ a = \{0,3,8 \} $, multiplying and adding each one by the corresponding factor to write everything in terms of $ \overline {\sigma}_u $, $ \overline {\sigma}_d $ and $ \overline {\sigma}_s $, one can identify the new auxiliary fields, which result
\begin{eqnarray}
 \tilde{S_u} &=& \sqrt{\frac{2}{3}}\tilde{S_0} + \tilde{S_3} + \sqrt{\frac{1}{3}}\tilde{S_8}, \\
 \tilde{S_d} &=& \sqrt{\frac{2}{3}}\tilde{S_0} - \tilde{S_3} + \sqrt{\frac{1}{3}}\tilde{S_8}, \\
 \tilde{S_s} &=& \sqrt{\frac{2}{3}}\tilde{S_0} - \sqrt{\frac{2}{3}}\tilde{S_8}. \\
\end{eqnarray}
Reversing these relations, writing all the non-null components of the mixing term $ A_{abc} \tilde {S_a} \tilde {S_b} \tilde {S_c} $ and rearranging the corresponding expressions, we arrive at the action written in terms of the new fields
\begin{equation}
 \label{SMFAfinal}
 \frac{S_E^{\mathrm{MFA}}}{V^{(4)}} = -2 \int \frac{d^4p}{(2\pi)^4}\,\mathrm{Tr} \left\{\mathrm{log}\left[ p^2 + M^2(p)\right] \right\} - \frac{1}{2}\sum_{i = u,d,s} \left(\sigma_i S_i + \frac{G}{2}S_i S_i\right) - \frac{H}{4}S_u S_d S_s,
\end{equation}
where we eliminate the hat and tilde notation because we will no longer make approximations and these will be the fields that we will use from now on.
If we now minimize with respect to $ \sigma_i $, we obtain the equations for the auxiliary fields, which are
\begin{eqnarray}
 -\frac{1}{2}S_i &-& 2 \frac{\delta}{\delta_{\sigma_i}} \int \frac{d^4p}{(2\pi)^4}\,\mathrm{Tr_c} \left\{\mathrm{log}\left[ p^2 + M_i^2(p)\right] \right\} = 0 \, \implies \nonumber \\
 \label{S_i_t0}
 &S_i& = -8 \int \frac{d^4p}{(2\pi)^4}\,\mathrm{Tr_c}\left[\frac{M_i(p) g(p)}{ p^2 + M_i^2(p)}\right],
\end{eqnarray}
where now the trace is done on color only. Finally, the gap equations for the three flavors of quarks are obtained by minimizing $ S_E^{MFA} $ with respect to each auxiliary field $ S_i $ and they result

\begin{eqnarray}
 \label{gapU} \sigma_u + G S_u + \frac{H}{2}S_d S_s &=& 0, \\
 \sigma_d + G S_d + \frac{H}{2}S_u S_s &=& 0, \\
 \sigma_s + G S_s + \frac{H}{2}S_u S_d &=& 0.
\end{eqnarray}

Let's see here in a simplified way, how these equations, in addition to solving the grand potential, give us an idea of what happens with the quarks masses. If we remember that $ M_f (p) = m_f + \sigma_f g (p) $, from the previous equations, for example for the quark $ u $, equation \eqref {gapU}, we have that

\begin{equation}
 M_u(p) = m_u - \left[G S_u + \frac{H}{2}S_d S_s\right]g(p).
\end{equation}

Let us assume for a moment the simple case where $ H = $ 0. Since the $ S_u $ field is also dependent on the dynamic mass $ M_u $, what we will have is a self-consistent equation for that mass, even in the case that $ m_u = 0 $. That is, depending on the values that $ G $ takes, we will (or not) have a dynamic mass generation for the quark $ u $, with its corresponding chiral symmetry breaking associated. This happens in the same way if the interaction with non-null values of $ H $ is added. The analysis of the $ G $ values that meet (or not) these conditions, as well as the adjustment of these constants to reproduce masses and mesonic decay constants, are outside the scope of this work. It is worth saying, however, that for each flavor of $ i $ of quarks, its corresponding field $ \sigma_i $ serves as an order parameter of breaking of the chiral symmetry, since these fields are those that influence the values of the dressed mass. That is, for null values of the $ \sigma_i $ fields, the quarks only have their bare masses $ m_i $, and in the opposite case, they behave like quarks with dressed masses $ M_i $.

\graphicspath{{PNJL/}}
\chapter{\label{ch:PNJL}Non-local PNJL model}
%\chapter{\label{ch:asdasd}asdad}

So far we have explained how to include chiral and axial symmetry breaking mechanisms in our quark model. Now, although this model produces a transition from a phase with spontaneously broken chiral symmetry to another in which that symmetry is restored and vice versa, it does not account for the confinement mechanism itself. That is to say, the density at which chiral symmetry is restored could perfectly well not be the same at which the quarks are deconfined. To have a confinement mechanism in our model, we have to include what is known as the Polyakov loop. There are \textit {Lattice QCD} calculations for the value of the Polyakov loop trace at finite temperature, so that it is possible to construct an effective potential associated with this loop according to the estimates of \textit {Lattice QCD} \cite {PhysRevD.75.034007}. Therefore let's see before how to extend the developed model to finite chemical potentials and temperature.

\section{Extension to finite temperature and chemical potentials}

For this procedure, we will use the imaginary time formalism and Matsubara frequencies. We already use the first one when we write the action in Euclidean space in the equation \eqref{accionNJL}, let's see how to implement the second. Let's start by considering the case of free fermions, which is the most interesting for the procedure since it is where the Matsubara frequencies are both explicitly and easily seen as included, and then let's see how it is modified with the interactions already proposed. 

For a grand canonical ensemble with Hamiltonian $ H $ and conserved load $ Q $, the partition function can be written as \cite{Pathria:1996hda}
\begin{equation}
 {\cal{Z}} = \mathrm{Tr}\left[e^{-\beta \left(H - \mu Q\right)}\right],
\end{equation}
where $ \beta = 1 / T $ is the inverse of the temperature in Boltzmann's constant units. In the case of our model, the conserved charge corresponds to
\begin{equation}
 Q = \int d^3 x \,\psi^{\dagger}(x) \psi(x).
\end{equation}

Since the Hamiltonian is multiplied by $ \beta $, that gives us the integration limit for the temporary variable $ \tau $ of the Euclidean time formalism, then, passing to the language of path integrals we have that the partition function is written as
\begin{equation}
\label{ZfiniteT}
 {\cal{Z}} = \int i{\cal{D}}\psi^\dagger{\cal{D}}\psi \,\mathrm{exp}\left\{\int_o^\beta d\tau\int d^3 x\, \overline{\psi}(\tau,\bm{x})\left[-\gamma^0 \frac{\partial}{\partial_\tau} + i \bm{\gamma} . \bm{\nabla} - m + \mu \gamma^0\right]\psi(\tau,\bm{x})\right\},
\end{equation}
where the condition now is that the fields are antiperiodic at intervals of $ \beta $, that is, $ \psi (0, \bm {x}) = - \psi (\beta, \bm {x}) $. Now it is convenient to develop this formalism in momentum space, so we write the fields as follows
\begin{eqnarray}
\psi(\tau,\bm{x}) &=& \frac{1}{\beta}\sum_{n = -\infty}^{\infty} \int \frac{d^3 p}{(2\pi)^3} e^{i(\bm{p}.\bm{x} + \omega_n \tau)}\psi_n(\bm{p}),  \\
\psi^\dagger(\tau,\bm{x}) &=& \frac{1}{\beta} \sum_{n = -\infty}^{\infty} \int \frac{d^3 p}{(2\pi)^3} e^{-i(\bm{p}.\bm{x} + \omega_n \tau)}\psi_n^\dagger(\bm{p}),  \\
\end{eqnarray}
with $ w_n = (2n + 1) \pi T $. Rewriting the exponential argument of equation \eqref {ZfiniteT} in terms of $ \psi $ and $ \psi^\dagger $, and using that

\begin{eqnarray}
 \int d^3 x \, e^{i\bm{x}(\bm{p} - \bm{p'})} &=& (2\pi)^3\delta^3(\bm{p} - \bm{p'}), \\
 \int _0^\beta d\tau \,e^{i\tau(w_n - w_{n'})} &=& \beta \delta_{w_n , w_{n'}},
\end{eqnarray}
e can rewrite the partition function as
\begin{equation}
\label{ZfiniteT2}
 {\cal{Z}} =\prod_{n,n',\alpha} \int i{\cal{D}}\psi^\dagger_{n,\alpha}{\cal{D}}\psi_{n,\alpha}\,\mathrm{exp}\left\{\frac{1}{\beta^2}\int \frac{d^3p\, d^3 p'}{(2\pi^6)} i\psi_{\alpha,n}^\dagger(\bm{p})  D_{\alpha \rho} \psi_{\rho,n'}(\bm{p'})\right\},
\end{equation}
with the $ D $ operator defined as
\begin{equation}
 D = -i\beta\left(-i\omega_n + \mu  - \gamma^0 \bm{\gamma}.\bm{p} - m\gamma_0\right)\delta^3(\bm{p} - \bm{p'})\delta_{w_n,w_{n'}}.
\end{equation}

In the last equation, the Dirac and Kronecker deltas were left explicitly to remember the spaces where $ D $ operates. Again, as we saw in the previous sections, the partition function turns out to be the determinant of the $ D $ operator. However, one difference is worth noting: the factor $ (1 / \beta ^ 2) $ that appears before the integral in equation \eqref{ZfiniteT2} is not accidental. Although it comes from the renormalization used for the Fourier expansions of the fields, this factor cannot be absorbed in the $ D $ operator while maintaining its adimensionality. As later we will have to calculate the determinant and that will lead us to use the logarithm, it is necessary to leave the mentioned factor outside the operator. Then, using the properties we saw of determinants, logarithms, traces, and Dirac matrices, it is easy to obtain the desired result. It should be noted that when calculating the traces, instead of obtaining the quadrivolume as we obtained previously, what is obtained is the factor $ V / T $, where now $ V $ is the three-dimensional volume of our system. Taking this into account, the result of these calculations is
\begin{equation}
\label{accionfreesinreg}
 -\frac{T \mathrm{log}\left[{\cal{Z}}(T,\mu)\right]}{V} = \frac{T \,S_E(T,\mu)}{V} = -2T\sum_{n = -\infty}^\infty \int \frac{d^3 p}{(2\pi)^3}\mathrm{log}\left[\frac{\left(w_n - i \mu\right)^2 + \bm{p}^2 + m^2}{T^2}\right],
\end{equation}
where the reason for multiplying the action by the temperature and dividing it by the volume, is that this will then lead us to the great thermodynamic potential that we will need to calculate the EoS, as we will see in the next section. Before that, let us note that up to now we have extended the case of free fermions to finite temperature. To move on to the interacting  theory that we described earlier, two things should be noted about the equation \eqref{SMFAfinal}. First, that the explicit dependence of the action on temperature after making the midfield and stationary phase approximations, will only come from the term within the integral. This is because the fields are defined based on the argument of the integral, and what should be discretized for the Matsubara frequencies is what is inside the logarithm. Second, the logarithm's argument in \eqref{accionfreesinreg} is exactly the same as in the free case, if we replace $ m $ by $ M (p) $. As the moment-dependent regulator appears in the definition of that mass, what must be done is to include the dependence with the new discretized frequencies. The rest remains the same as in the free case, so without losing generality we can write that

\begin{eqnarray}
 \label{SMFATfinita}
 \frac{T\, S_E^{\mathrm{MFA}}(T,\mu)}{V} &=& -2 T \sum_{n = -\infty}^{\infty}\int \frac{d^3p}{(2\pi)^3}\,\mathrm{Tr} \left\{\mathrm{log}\left[\frac{\left(\omega_n - i \mu\right)^2 + \bm{p}^2 + M^2(\omega_n,\bm{p})}{T^2}\right] \right\} \nonumber \\
 &-& \frac{1}{2}\sum_{i = u,d,s} \left(\sigma_i S_i + \frac{G}{2}S_i S_i\right) - \frac{H}{4}S_u S_d S_s.
\end{eqnarray}

From here on, we'll just make a change in notation to make writing the equations easier. We define the Euclidean vector $ \hat {\omega}_n = (\omega_n -i \mu, \bm {p}) $, and we write the action in terms of this quantity

\begin{eqnarray}
 \label{SMFATfinita2}
 \frac{T\, S_E^{\mathrm{MFA}}(T,\mu)}{V} &=& -2 T \sum_{n = -\infty}^{\infty}\int \frac{d^3p}{(2\pi)^3}\,\mathrm{Tr} \left\{\mathrm{log}\left[\frac{\hat{\omega}_n^2 + M^2(\hat{\omega}_n^2)}{T^2}\right] \right\} \nonumber \\
 &-& \frac{1}{2}\sum_{i = u,d,s} \left(\sigma_i S_i + \frac{G}{2}S_i S_i\right) - \frac{H}{4}S_u S_d S_s.
\end{eqnarray}

\section{Regularization}
\label{Regularizacion}
So far we regularize the nl-NJL model with the simplest interactions, at finite temperature and chemical potential. As we said in the previous section, the quantity of interest to us is the grand thermodynamic potential. This quantity is from which we will not only derive the relevant quantities such as susceptibilities, condensates, entropy, but the behavior of the grand potential itself is what will give us the the phase transitions points that we study. It happens that at certain points in the plane $ (T, \mu) $ the grand potential has more than one solution that minimizes it, and then the solution with the lowest Gibbs energy is the most favored one. Therefore, as we have been doing implicitly, we write its form

\begin{equation}
\label{omegagenerico}
 \Omega^{MFA}(T,\mu) = -\frac{T}{V}\mathrm{log} {\cal{Z}}^{MFA}(T,\mu) = \frac{T}{V}S_E^{MFA}(T,\mu) = -p(T,\mu),
\end{equation}
where $ p (T, \mu) $ is the pressure. For completeness, we rewrite it with its name, so that
\begin{eqnarray}
 \label{Omegasinreg}
 \Omega^{MFA}(T,\mu) &=& -2 T \sum_{n = -\infty}^{\infty}\int \frac{d^3p}{(2\pi)^3}\,\mathrm{Tr} \left\{\mathrm{log}\left[\frac{\hat{\omega}_n^2 + M^2(\hat{\omega}_n^2)}{T^2}\right] \right\} \nonumber \\
 &-& \frac{1}{2}\sum_{i = u,d,s} \left(\sigma_i S_i + \frac{G}{2}S_i S_i\right) - \frac{H}{4}S_u S_d S_s.
\end{eqnarray}

The grand potential written like this has a problem, the integral is clearly divergent. To solve this problem, it is convenient to identify where the divergences come from, as is usual in regularization processes in quantum field theories. Note that the integral of the extension at finite temperature of the equation \eqref{accionfreesinreg} corresponding to the free part is also divergent. So, to solve this problem we propose that the divergences of the equation \eqref{Omegasinreg} come only from the free part and we say that the large regularized potential will be

\begin{equation}
\label{defomega}
 \Omega^{MFA}_{Reg}(T,\mu) = \Omega^{MFA}(T,\mu) - \Omega^{free}(T,\mu) + \Omega^{free,Reg}(T,\mu) - \Omega_0,
\end{equation}
where $ \Omega^{Free} (T, \mu) $ is the theory's grand potential without interactions, which as we said, is divergent. The quantity $ \Omega^{Free, Reg} (T, \mu) $ is the non-divergent part of the previous one, and $ \Omega_0 $ is a constant that is fixed so that at zero temperature and chemical potential the result is $ \Omega^{MFA}_{Reg} (T = 0, \mu = 0) = 0 $. Let's start with $ \Omega^{Free} (T, \mu) $ to regularize it, which was written as

\begin{eqnarray}
 \label{Omegasinreg2}
 \Omega^{free}(T,\mu) &=& -2 T \sum_{n = -\infty}^{\infty}\int \frac{d^3p}{(2\pi)^3}\,\mathrm{Tr} \left\{\mathrm{log}\left[\frac{\left(\omega_n - i\mu \right)^2 + E_p^2}{T^2}\right] \right\}, 
\end{eqnarray}
where we define $E_p^2 = \bm{p}^2 + m^2$. To regularize it, let us note that the following identity applies
\begin{eqnarray}
\label{regid1}
 &2&\sum_{n=-\infty}^{\infty}\mathrm{log}\left[\frac{\left(\omega_n - i\mu \right)^2 + E_p^2}{T^2}\right]  \nonumber \\
 &=& \sum_{n=-\infty}^{\infty} \left\{\mathrm{log}\left[\frac{\omega_n^2   + \left(E_p + \mu\right)^2}{T^2}\right] + \mathrm{log}\left[\frac{\omega_n^2   + \left(E_p - \mu\right)^2}{T^2}\right] \right\},
\end{eqnarray}
and we write each term in the way
\begin{equation}
\label{regid2}
 \frac{w_n^2  + \left(E_q \pm \mu\right)^2}{T^2} = \left(2n + 1 \right)^2\pi^2 + \left(\frac{E_q \pm \mu}{T}\right)^2.
\end{equation}

Then, taking into account that the expression \eqref{regid2} is a logarithm's argument in \eqref{regid1}, we can write it as an integral of the form

\begin{eqnarray}
\label{regid3}
 &\,& \mathrm{log}\left[\left(2n + 1 \right)^2\pi^2 + \left(\frac{E_q \pm \mu}{T}\right)^2\right] \nonumber \\
 &=& \int_1^{\left(\frac{E_q \pm \mu}{T}\right)^2} \frac{d\left(\theta\right)^2}{\theta^2 + \left(2n+1\right)^2\pi^2} + \mathrm{log}\left[1 + \left(2n+1\right)^2\pi^2\right],
\end{eqnarray}
and using a last identity for the sum, which is
\begin{equation}
\label{regid4}
 \sum_{n=-\infty}^{\infty} \frac{1}{\theta^2 + \left(2n+1\right)^2\pi^2} = \frac{1}{\theta}\left(\frac{1}{2} - \frac{1}{1+e^\theta}\right),
\end{equation}
we can use \eqref{regid1}, \eqref{regid2}, \eqref{regid3} y \eqref{regid4} to write
\begin{equation}
\label{reg5}
 \sum_{n=-\infty}^{\infty}\mathrm{log}\left[\frac{\left(\omega_n - i\mu \right)^2 + E_p^2}{T^2}\right] = \frac{E_q}{T} + \mathrm{log}\left[1 + e^{-\frac{E_q + \mu}{T}}\right] + \mathrm{log}\left[1 + e^{-\frac{E_q - \mu}{T}}\right] + \mathrm{const},
\end{equation}
where `` const '' refers to terms that do not depend on $ T $ or $ \mu $. It is clear that the first term of \eqref{reg5} is divergent, since $ E_q $ is linear in the momentums for great momentums values, and that term goes inside a quadrimomentum integral. Since we said that we want to keep only the non-divergent part, we can eliminate this term together with the constants, which do not contribute to the subtraction. Discarding these terms and putting everything into equation \eqref{Omegasinreg2}, we can define the regularized free grand potential:

\begin{eqnarray}
 \label{OmegaReg}
 \Omega^{free, Reg}(T,\mu) &=& -2 T \int \frac{d^3p}{(2\pi)^3}\,\mathrm{Tr} \left\{\mathrm{log}\left[1 + e^{-\frac{E_q + \mu}{T}}\right] + \mathrm{log}\left[1 + e^{-\frac{E_q - \mu}{T}}\right] \right\}, 
\end{eqnarray}
and with that, plus the definition \eqref{defomega}, we finally have the regularized grand potential at mean field
\begin{eqnarray}
 \label{Omegareg2}
 \Omega^{MFA}_{Reg}(T,\mu) &=& -2 T \int \frac{d^3p}{(2\pi)^3}\,\mathrm{Tr} \Biggl\{\sum_{n = -\infty}^{\infty}\mathrm{log}\left[\frac{\hat{\omega}_n^2 + M^2(\hat{\omega}_n^2)}{\hat{\omega}_n^2 + m^2}\right]  \nonumber \\
  &+&\mathrm{log}\left[1 + e^{-\frac{E_q + \mu}{T}}\right] + \mathrm{log}\left[1 + e^{-\frac{E_q - \mu}{T}}\right] \Biggr\} \nonumber \\
 &-& \frac{1}{2}\sum_{i = u,d,s} \left(\sigma_i S_i + \frac{G}{2}S_i S_i\right) - \frac{H}{4}S_u S_d S_s.
\end{eqnarray}

Now, with this we have the grand potential at regularized mean field, what remains to be seen is how the auxiliary fields extend at finite temperature and chemical potential. For these fields, which are obtained by minimizing the grand potential with respect to $ \sigma_i $, it is valid to use the following substitution, for an integrand that is a generic $ f (p_0, \bm {p})$ function, it valid to say that:

\begin{equation}
\label{recetamala}
 \int \frac{dp_0}{(2\pi)}\int \frac{d^3p}{(2\pi)^3}f(p_0,\bm{p}) \rightarrow T \sum_{n = -\infty}^{\infty} \int \frac{d^3p}{(2\pi)^3}f(w_n - i\mu, \bm{p}).
\end{equation}

Note that either using \eqref{recetamala} or calculating the $ S_i $ as we did before, minimizing the grand potential with respect to $ \sigma_i $ leads us to the same result

\begin{eqnarray}
 -\frac{1}{2}S_i &-& 2 T\frac{\delta}{\delta_{\sigma_i}} \sum_{n = -\infty}^{\infty}\int \frac{d^3p}{(2\pi)^3}\,\mathrm{Tr_c} \left\{\mathrm{log}\left[\frac{\hat{\omega}_n^2 + M^2(\hat{\omega}_n^2)}{\hat{\omega}_n^2 + m^2}\right] \right\} = 0 \, \implies \nonumber \\
 \label{S_i_TF}
 &S_i& = -8 T \sum_{n = -\infty}^{\infty}\int \frac{d^3p}{(2\pi)^3}\,\mathrm{Tr_c}\left[\frac{M_i(\hat{\omega}_n^2) g(\hat{\omega}_n^2)}{\hat{\omega}_n^2 + M_i^2(\hat{\omega}_n^2)}\right].
\end{eqnarray}

It is worth clarifying that the recipe \eqref{recetamala} serves only for quantities and auxiliary fields after the grand potential was regularized. If we had used this method to extend to finite temperature and chemical potential in the action without interactions (for example), we would have missed the $ T^2 $ dividing term in the logarithm of \eqref{accionfreesinreg}, and the regularization procedure would have been wrong. That is, you can't help but develop the Matsubara method in detail for the regularization of the integral, but then the auxiliary fields can be extended to finite $ T $ and $ \mu $ easily with the mentioned recipe. We will use this form of extension again in subsequent sections, when we add interactions (and auxiliary fields) to the model we have been studying.

Finally, it is useful to have the expression of the grand potential at zero temperature and finite $ \mu $, since it will allow us to determine the thermodynamic properties in that regime. In the first logarithmic term of the integral of \eqref{OmegaReg}, by continuing explicitly with the summation over n, one can change $ {\hat {w}}_n^2 $ to $ p^2 $ ( the four-momenta), and change the integral to one in $ d^4p $. The second term (the exponentials  one) is calculated by taking the $ T \to 0 $ limit within the integrand, and then performing the $ d^3p $ integral, obtaining

\begin{eqnarray}
 \label{Omegareg3}
 \Omega^{MFA}_{Reg}(0,\mu) &=& -2  \int \frac{d^4p}{(2\pi)^4}\,\mathrm{Tr} \left\{\mathrm{log}\left[\frac{p^2 + M^2(p^2)}{p^2 + m^2}\right]\right\}  \nonumber \\
  &-&\mathrm{Tr}\left\{\frac{\Theta \left(\mu - m\right)}{24\pi^2}\left\{ \left[\left(2\mu^2 - 5m^2\right)\mu \sqrt{\mu - m}\right] + \left[    3\mu^4 \mathrm{log}\left(\frac{\mu + \sqrt{\mu - m}}{m}\right) \right] \right\} \right\} \nonumber \\
 &-& \frac{1}{2}\sum_{i = u,d,s} \left(\sigma_i S_i + \frac{G}{2}S_i S_i\right) - \frac{H}{4}S_u S_d S_s.
\end{eqnarray}

So far, we have developed one of the variants with simpler interactions of the nl-NJL  model at finite temperature and chemical potential. Let's see how, now that we have incorporated the temperature ingredient, we can extend the model to more realistic cases.

\section{PNJL-nl model extension}

So far we have developed the NJL nonlocal model at finite chemical potentials and temperatures. As we said earlier, one of our goals is to include a confinement mechanism, known as the Polyakov loop, in the model. Let's see then a brief introduction to its formalism, and then include it in the model.
\subsection{The Polyakov loop}
\label{Polyakov}

The Polyakov loop was proposed by A.M. Polyakov \citep{Polyakov:1976fu} as an application of the Wilson loop \cite {Wilson:1974sk} to the thermal properties problem of gauge fields, in particular as a mechanism that explains the deconfining of quarks at a certain temperature, known as deconfinement temperature. To better explain it, we must first briefly comment on the symmetries that are taken into account.

Following the work of 't Hooft \cite{tHooft:1977nqb}, a global symmetry $ Z (N) $ appears in a local gauge $ SU (N) $ theory . To see this, we start from a Lagrangian density, which includes the interaction of quarks with gluonic fields
\begin{equation}
\mathcal{L}=\frac{1}{2}\mathrm{Tr} G_{\mu\nu}^2 + \bar{\psi}i \gamma^\mu D_\mu \psi, %\not \!\! D
\label{eq:densi_L_PL}
\end{equation}
where
\begin{equation}
D_\mu=\partial_\mu-i g A_\mu~~,~~G_{\mu\nu}=\frac{1}{-i g} [D_\mu,D_\nu];
\end{equation}
$A_\mu=A_\mu^a t^a$, with the normalized  $SU(N)$ generators written as Tr$(t^a,t^b)=\delta^{ab}/2$.
This Lagrangian is invariant under $ \Omega $ gauge transformations of $ SU (N) $,  given by
\begin{equation}
D_\mu \rightarrow \Omega^{\dagger}D_\mu \Omega~~,~~\psi\rightarrow \Omega^{\dagger} \psi.
\end{equation}
Being $ \Omega $ an element of $ SU (N) $, it satisfies that
\begin{equation}
\Omega^{\dagger}\Omega=\leavevmode\hbox{\small1\kern-3.8pt\normalsize1}~~,~~ \mathrm{det} \Omega=1.
\end{equation}
Since $ \Omega $ is a local gauge transformation, it is generally a function of spacetime.
Let's consider a gauge transformation given by a constant phase by the unit matrix:
\begin{equation}
\Omega_c=e^{-i \varphi} \leavevmode\hbox{\small1\kern-3.8pt\normalsize1}.
\end{equation}
For this transformation to be an element of $ SU (N) $, the determinant must be equal to one, which requires that
\begin{equation}
\varphi=\frac{2 \pi j}{N}~~,~~j=0,1,...(N-1).
\end{equation}
Since an integer cannot change continuously from one point to another, this defines a global $ Z (N) $ symmetry.

Being a subgroup of the gauge transformations, the rotations of the group $ Z (N) $ are always a Lagrangian symmetry. However, we will see later that in the presence of dynamic quarks, the rotations of $ Z (N) $ are not a symmetry of the theory since they violate the required boundary conditions.

Working in Euclidean spacetime at a temperature $ T $, the imaginary time coordinate $ \tau $ is of finite extent, $\tau$: $0\rightarrow\beta=1/T$. The boundary conditions that the fields must satisfy are given by the statistics of each of them. That is, gluons (bosons) must be periodic in $ \tau $, while quarks (fermions) must be anti-periodic:
\begin{equation}
A_\mu(\vec{x},\beta)= +A_\mu(\vec{x},0)~~,~~\psi(\vec{x},\beta)=- \psi(\vec{x},0).
\end{equation}
Obviously, any gauge transformation that is periodic on $ \tau $ respects these boundary conditions. However, 't Hooft found that more general gauge transformations can be considered, which are periodic at less than $ \Omega_c $:
\begin{equation}
\Omega(\vec{x},\beta)=\Omega_c~~,~~\Omega(\vec{x},0)=1.
\end{equation}
The adjoint color fields are invariant under this transformation, while those in the fundamental representation are not, consequently, the pure $ SU (N) $ gauge theories have a global $ Z (N) $ symmetry, which is destroyed by including dynamic quarks.

In a pure gluonic theory, an order parameter for the symmetry $ Z (N) $ is constructed using the thermal Wilson line:
\begin{equation}
L(\vec{x})= \mathcal{P}\exp \left(ig\int_0^\beta A_0(\vec{x},\tau)d\tau\right),
\end{equation}
where $ g $ is the gauge coupling constant, and $ A_0 $ is the potential vector in the temporal direction. The symbol $ \mathcal {P} $ denotes path ordering, so that the thermal Wilson line transforms as an adjoint field under local $ SU (N) $ gauge transformations:
\begin{equation}
L(\vec{x})\rightarrow \Omega^{\dagger}(\vec{x},\beta)L(\vec{x}) \Omega(\vec{x},0).
\label{eq:wilson_line}
\end{equation}
The Polyakov loop \cite{Polyakov:1976fu} is defined as the trace of the thermal Wilson line, and is therefore gauge invariant:
\begin{equation}
\Phi(\vec{x})=\frac{1}{N}\mathrm{Tr}L= N^{-1}\mathrm{Tr}\mathcal{P}\exp \left(ig\int_0^\beta A_0(\vec{x},\tau)d\tau\right)
\label{eq:def_PL}
\end{equation}
Under global transformations $ Z (N) $, the Polyakov loop $ \Phi $ transforms as a field with unitary charge:
\begin{equation}
\Phi\rightarrow e^{i \varphi}\Phi.
\end{equation}

At very high temperature, the theory is almost ideal ($ g \approx0 $), so one would expect $ \langle \Phi \rangle \sim 1 $. However, the allowed vacuum exhibits a degeneration of $ N $ sheets. This is,
\begin{equation}
\langle\Phi\rangle=\exp\left(\frac{i 2\pi j}{N}\right) \Phi_0~~,~~j=0,1,...,(N-1),
\end{equation}
where $ \Phi_0 $ is a real function, which also fulfills that $ \Phi_0 \rightarrow1 $ when $ T \rightarrow \infty $. Any value of $ j $ is equivalent, so the usual choice results in the spontaneous breaking of the global symmetry $ Z (N) $.

At zero temperature, confinement implies that $ \Phi_0 $ is canceled \cite{tHooft:1977nqb}. Therefore, there must be a certain temperature value $ T_c $, after which $ \Phi_0 $ is no longer null and deconfinement occurs. This is
\begin{equation}
\left\{
\begin{aligned}
&\Phi_0=0~~si~~T<T_c,& \\
&\Phi_0>0~~si~~T>T_c.&
\end{aligned}
\right.
\end{equation}

As usual, if $ \ Phi_0 $ goes nonzero continuously around $ T_c $, the transition is second order; while if it jumps suddenly in $ T_c $, it is of the first order. What is unusual is that the symmetry $ Z (N) $ is spontaneously broken at high temperatures instead of at low values of $ T $. We will not enter into the discussion of this matter here, but if a heuristic explanation can be found in \citep{Pisarski:2002ji}.

On the other hand, in the presence of dynamic quarks the symmetry $ Z (N) $ is explicitly broken. In this case then, the Polyakov loop is no longer a rigorous order parameter, but it still serves as an indicator of a fast \textit{crossover} towards deconfinement. In the next section we will see how to add the $ \Phi $ field to the effective quark model with which we will work.

\subsection{$\Phi$ field inclusion in nl-NJL model}
\label{polynambu}

In order to include the $ \Phi $ field in our model, we start by replacing the normal derivative with the covariant derivative in equation \eqref{accionNJL}, so that
\begin{equation}
 \overline{\psi}\left(x\right)\left( i \slashed{\partial} + m \right)\psi \rightarrow \overline{\psi}\left(x\right)\left(  i\gamma^\mu D_\mu + m \right)\psi,
\end{equation}
and adding the Polyakov potential $ {\cal {U}} (\Phi) $ to the Lagrangian, with $ D_\mu = \partial_\mu - iA_\mu $, and $ A_\mu $ the color gauge fields.

Regarding gluonic fields, we assume that they provide a constant color field $ A_4 = i A_0 = ig \, \delta _ {\mu 0} \, G^\mu_a \lambda^a/2 $, where $ G^\mu_a $ is the color gauge fields tensor. So, the trace of the Polyakov loop is given by $ \Phi = \frac {1} {3} {\rm Tr} \, \exp (i \phi / T) $, where $ \phi = i A_0 $ and $ A_0 $ is the same gluon field developed in section \ref{Polyakov}. We will work on the Polyakov gauge, in which the matrix $ \phi $ is given by a diagonal representation $\phi = \phi_3 \lambda_3 + \phi_8 \lambda_8$. This leaves only two independent variables, $\phi_3$ y $\phi_8$. At zero chemical potential, due to the charge conjugation properties of the QCD Lagrangian, the mean field of the Polyakov loop trace is expected to be a real quantity. Since $ \phi_3 $ and $ \phi_8 $ have to be real \citep{Rossner:2007ik}, this implies that $ \phi_8 = 0 $. In general, this need not be true at finite $ \mu $ \citep{Dumitru:2005ng,Elze:1986gz, Fukushima:2006uv}. As in the references \citep{Roessner:2006xn,Rossner:2007ik,Elze:1986gz,Fukushima:2006uv,Abuki:2008ht,GomezDumm:2008sk}, we will assume that the potential $ \cal U $ is such that the condition $ \phi_8 = 0 $ is satisfied, at least for the $ \mu $ and $ T $ value range of interest. The mean field of the traced Polyakov loop is therefore given by $\Phi = \bar{\Phi} = [ 1 + 2\,\cos (\phi_3/T)]/3$. The way of coupling with the quarks is evident since $ A_4 = iA_0 = \phi $ is coupled to the zero component of the derivative, generating a shift of the chemical potential in such a way that if before we had to $\hat{\omega}_n = (\omega_n -i\mu, \bm{p})$, now we have to perform the replacement

\begin{equation}
\hat{\omega}_n = (\omega_n -i\mu, \bm{p}) \rightarrow \omega_{nc} = (\omega_n - i\mu + \phi_c, \bm{p}),
\end{equation}
with $\phi_c = c\phi_3$, y $ c= {1,-1,0}.$ for $r,g,b$ respectively. The Polyakov potential that we will use is given by \cite{Roessner:2006xn}
\begin{equation}
 \mathcal{U}(\Phi,T) = \left\{ -\dfrac{1}{2} a(T)\,\Phi^2 + b(T) \mathrm{log} \left[ 1 - 6\, (\Phi^2 + 8\Phi^3 - 3\, \Phi^4 \right]\right\} T^4 \ ,
\end{equation}
with the coefficients
\begin{equation}
 a(T) = a_0 +a_1 \left(\dfrac{T_0}{T}\right) + a_2\left(\dfrac{T_0}{T}\right)^2 \qquad ; \qquad
b(T) = b_3\left(\dfrac{T_0}{T}\right)^3.
\end{equation}

The values $T_0$, $a_i$ y $b_3$ are fixed to \textit{Lattice QCD} results, being then $a_0 = 3.51$, $a_1 = -2.47$, $a_2 = 15.2$, $,b_3 = -1.75$ y $T_0 = 195$  MeV, in the same way as in the reference \cite{Carlomagno:2013}. With these ingredients, we can write the grand potential with the Polyakov loop included, this being

\begin{eqnarray}
 \label{OmegaregPoly}
 \Omega^{MFA}_{Reg}(T,\mu) &=& -2 T \int \frac{d^3p}{(2\pi)^3}\,\sum_{f,c}\Biggl\{\sum_{n = -\infty}^{\infty}\mathrm{log}\left[\frac{\omega_{nfc}^2 + M_f^2(\omega_{nfc}^2)}{\omega_{nfc}^2 + m_f^2}\right]  \nonumber \\
  &+&\mathrm{log}\left[1 + e^{-\frac{E_q + \mu_f + i\phi_c}{T}}\right] + \mathrm{log}\left[1 + e^{-\frac{E_q - \mu_f - i\phi_c}{T}}\right] \Biggr\} \nonumber \\
 &-& \frac{1}{2}\sum_{i = u,d,s} \left(\sigma_i S_i + \frac{G}{2}S_i S_i\right) - \frac{H}{4}S_u S_d S_s + {\cal{U}}(\Phi).
\end{eqnarray}

In the last equation, the sum over color is done on $ c = {1, -1,0} $ and the subscript $ f $ added to both frequencies and chemical potentials is set for the purpose of writing the most generic case where the chemical potentials of each flavor are different, so that $\omega_{nfc} = (\omega_n - i\mu_f + \phi_c, \bm{p})$. It is worth noting that the $ \phi_c $ field is also included in the exponentials in the integral's second term. Although that term came from regularizing the non-interacting grand potential, we assume that the color fields are still present at a fundamental level in the theory. That is, the derivative $ D_\mu $ is still present in the `` free '' theory, therefore the color field must be included in the regularization term.

Finally, for the system to be solvable, we must ask $ \Phi $ to minimize the grand potential, taking a system of equations to a higher degree than the one we had previously, this being

\begin{eqnarray}
 \sigma_u + G\,S_u + \frac{H}{2}S_d S_s = 0, \\
 \sigma_d + G\,S_d + \frac{H}{2}S_s S_u = 0, \\
 \sigma_s + G\,S_s + \frac{H}{2}S_u S_d = 0, \\
 \frac {\partial \Omega}{\partial \phi_3} = 0,
\end{eqnarray}

where the new contributions of the Polyakov loop are given by

\begin{eqnarray}
 \frac {\partial \Omega}{\partial \Phi} &=&  \sum_{f,c} \int \frac{d^3p}{(2\pi)^3}\Bigg\{ 4 cT \sum_{n=-\infty}^{\infty} \omega_{0nfc} \left[ \frac{1}{\omega_{nfc}^2 + M_{fnc}^2}  - \frac{1} {\omega_{nfc}^2 + m_{f}^2} - \frac{2 M_{fnc} \sigma_f g(\omega_{nfc}^2)}{(\omega_{nfc}^2 + M_{fnc}^2)\Lambda^2}\right]  \nonumber \\
  &+& \frac{i c}{1 + e^{\frac{E - \mu_f - ic\phi_3}{T}}} - \frac{i c}{1 + e^{\frac{E + \mu_f + ic\phi_3}{T}}} \Bigg\} \sqrt{\frac{3T^2}{1 + (2 - 3\Phi)\Phi}} + \frac{\partial {\cal{U}}( \Phi , T)}{\partial \Phi} ,
\end{eqnarray}
donde $\omega_{0nfc} = \omega_n - i\mu_f +c\phi_3$, abreviating  $M_f$ instead of $M_{fnc}(\omega_{nfc}^2)$ and the terms coming from the potential
\begin{eqnarray}
 \frac{\partial {{\cal{U}} }( \Phi , T)}{\partial \Phi} &=&  T^4 \Phi \left[- a -\frac{12b}{1 + 2 \Phi - 3 \Phi^2}   \right], \, \, \mathrm{con} \\
 \phi_3 &=& T \, \mathrm{arccos}\left(\frac{3\Phi - 1 }{2}\right).
\end{eqnarray}

\section{Vector interactions inclusion}
\label{intvec}
So far we have developed the non-local NJL model taking into account only scalar and pseudoscalar interactions between quarks. It is necessary, however, to note that two ingredients are missing: color superconductivity, and vector interactions. As we will see later, vector interactions play an important role in the competition between the appearance of diquark condensates and the restoration of chiral symmetry. Although by increasing this interaction, both the appearance of diquarks and the restoration of chiral symmetry appear at higher densities, they do not do so in the same way. It is then possible to find values of vector couplings that lead to these two phenomena happening at the same density, which could lead to unexpected phase diagrams, such as a structure of two critical points in the $(T,\mu)$ plane  \cite{Asakawa:1989bq,Klimt:1990ws,Buballa:1996tm}. On the other hand, it is worth noting that from a renormalization group analysis point of view \cite{Evans:1998ek,Schafer:1998na}, the instanton anti-instanton chiral model \cite{Schafer:1996wv}, and the truncated Dyson-Schwinger QCD model \cite{Tandy:1997qf}, all support the idea of the existence of vector interactions of four-point quarks. These interactions could also be responsible for the vacuum properties of vector mesons in effective low-energy theories of QCD. \cite{Ebert:1985kz,Klimt:1989pm}. Let's see then how to include them in our model. The idea is to add to the original Lagrangian, an interaction of the type

\begin{eqnarray}
{\cal{L}}_V &=& \frac{G_V}{2}j_{a}^{\mu}(x)j_{a}^{\mu}(x);\,\,\,\,\, \mathrm{con} \\
j_{a}^{\mu}(x) &=& \int d^4z g(z)\overline{
\psi} \left(x + \frac{z}{2} \right) \lambda_a \gamma^{\mu}\psi \left(x
- \frac{z}{2} \right),
\end{eqnarray}
where again the $ \lambda_a $ are the Gell-Man matrices (with the one proportional to the identity for $ a = 0 $ added) that can adopt the values $ a = \{0,3,8 \} $ like for scalar interactions. In the mean field approximation, the only current that gives rise to a non-zero expectation value is $j_{a}^{0}(x)$, for which we will have to introduce three new fields that we will call $ \theta_a $ and their respective auxiliary fields $ V_a $, as in section \ref{bosonizacion}. It is easy to see that by having this component the matrix $ \gamma ^ 0 $ in between, what will happen is that it will generate a shift in the zero component of the moment, which results in $p_0 \rightarrow p_0 + i\theta_a \lambda_a g(p)$. Analogously to the chapter \ref{ch:Nambu}, remember that the scalar currents that were not multiplied by matrices, generated the shift in the mass so that $m\rightarrow  m + \sigma_a  \lambda_a g(p)$ in \eqref{Masas1}. In the same way that we did before, the mean field approximation together with the stationary phase, and the diagonalization, are straightforward and are even simpler than the development carried out in section \ref{bosonizacion}. Because of this, we can directly write the grand potential including the vector interaction, which results:

\begin{eqnarray}
 \label{OmegaregPolyVec}
 \Omega^{MFA}_{Reg}(T,\mu) &=& -2 T \int \frac{d^3p}{(2\pi)^3}\,\sum_{f,c}\Biggl\{\sum_{n = -\infty}^{\infty}\mathrm{log}\left[\frac{q_{nfc}^2 + M_{fnc}^2}{\omega_{nfc}^2 + m_f^2}\right]  \nonumber \\
  &+&\mathrm{log}\left[1 + e^{-\frac{E_q + \mu_f + i\phi_c}{T}}\right] + \mathrm{log}\left[1 + e^{-\frac{E_q - \mu_f - i\phi_c}{T}}\right] \Biggr\} \nonumber \\
 &-& \frac{1}{2}\sum_{i = u,d,s} \left(\sigma_i S_i + \frac{G_S}{2}S_i S_i + \theta_f V_f - \frac{G_V}{2}V_f^2\right) \nonumber \\
 &-& \frac{H}{4}S_u S_d S_s + {\cal{U}}(\Phi),
\end{eqnarray}
where $q_{nfc} = (\omega_{nfc} + i\theta_{f}\,g(\omega_{nfc}),\bm{p})$, and we have made the change $G \rightarrow G_S$, so that the scalar interaction is explicitly differentiated from the vector interaction. It should be noted that the new frequencies $q_{nfc}$ do not appear as an argument of the $g(\omega_{nfc})$ regulators, since the regulators are introduced at Lagrangian level before performing the mean field approximations, so they keep their original arguments.In the same way, the chemical potential of the exponentials of the second term of the integral is not corrected either, because they come from the regularization of the free part, in which the vector interaction has no role. The $ V_f $ auxiliary fields are obtained by minimizing the grand potential relative to $ \theta_f $, so that

\begin{eqnarray}
 -\frac{1}{2}V_f &-& 2 T\frac{\delta}{\delta_{\sigma_i}} \sum_{n = -\infty}^{\infty}\int \frac{d^3p}{(2\pi)^3}\,\mathrm{Tr_c} \left\{\mathrm{log}\left[\frac{q_{nfc}^2 + M_{fnc}^2}{\omega_{nfc}^2 + m_f^2}\right] \right\} = 0 \, \implies \nonumber \\
 \label{V_f_TF}
 &V_f& = -8 T \sum_{n = -\infty}^{\infty}\int \frac{d^3p}{(2\pi)^3}\,\mathrm{Tr_c}\left[\frac{i q_{0fnc} g(\omega_{nfc}^2)}{\omega_{nfc}^2 + M_{fnc}^2}\right],
\end{eqnarray}
where $q_{0nfc} = \omega_{nfc} + i\theta_{f}\, g(\omega_{nfc})$ is the zero component of $q_{nfc}$. Finally, new equations are added to the gap equations, which arise from minimizing the grand potential with respect to the $ V_f $ fields, thus having the system:
\begin{eqnarray}
\label{gapequationsvec}
 \sigma_u + G_S S_u + \frac{H}{2}S_d S_s = 0, \\
 \sigma_d + G_S S_d + \frac{H}{2}S_s S_u = 0, \\
 \sigma_s + G_S S_s + \frac{H}{2}S_u S_d = 0, \\
 \frac {\partial \Omega}{\partial \Phi} = 0, \\
 \theta_u - G_V V_u = 0, \\
  \theta_d - G_V V_d = 0, \\
  \label{gapequationsvec2}
  \theta_s - G_V V_s = 0, 
\end{eqnarray}
where have we made the replacement $G\rightarrow G_S$ to make it explicit that it is the scalar coupling constant, and the derivative with respect to $ \Phi $ slightly modified from the previous case:

\begin{eqnarray}
 \frac {\partial \Omega}{\partial \Phi} &=&  \sum_{f,c} \int \frac{d^3p}{(2\pi)^3}\Bigg\{ 4 cT \sum_{n=-\infty}^{\infty} \left[ \frac{q_{0nfc}}{q_{nfc}^2 + M_{nfc}^2}  - \frac{\omega_{0nfc}} {\omega_{nfc}^2 + m_{f}^2} - \frac{2 \omega_{0nfc}\,g(\omega_{nfc}^2)\left(M_f \sigma_f + i q_{0nfc}\right) }{(q_{nfc}^2 + M_{nfc}^2)\Lambda^2}\right]  \nonumber \\
  &+& \frac{i c}{1 + e^{\frac{E - \mu_f - ic\phi_3}{T}}} - \frac{i c}{1 + e^{\frac{E + \mu_f + ic\phi_3}{T}}} \Bigg\} \sqrt{\frac{3T^2}{1 + (2 - 3\Phi)\Phi}} + \frac{\partial {\cal{U}}( \Phi , T)}{\partial \Phi},
\end{eqnarray}
finally, the $ G_V $ parameter is defined in terms of the $ G_S $ parameter, so that $\zeta_V = G_V/G_S$ y $ 0< \zeta_V < 0.5 $ ,according to the limits calculated by the reference \cite{Buballa:2003qv}. 

The system of equations \eqref{gapequationsvec} is the complete system that needs to be solved to describe quark matter, now with vector interactions, finite temperature and chemical potential. As we said in the previous chapters, our goal is to calculate the EoS in the high-density regime to describe the quark matter inside the NSs. However, two components would be missing for the EoS generated by the aforementioned system to meet the requirements for matter within the NSs: on the one hand color superconductivity, and on the other the inclusion of leptons in the model, in order to have electric charge neutrality. Regarding color superconductivity, it is impossible to include it at the same time as the Polyakov loop, so we will develop it separately in subsequent sections. On the other hand, the electric charge neutrality is easy to add, but first it is necessary to emphasize that this inclusion generates changes in first order line location in the phase diagram, since the quarks chemical potentials are affected by the electrons, muons and neutrinos chemical potentials due to the chemical equilibrium condition. In order to guarantee that the EoS that we compute for the interior of the NSs are of deconfined quark matter, we should in principle ensure that the chemical potentials we work with fall in the corresponding area of the phase diagram. On the other hand, to verify that our model is complete, it is also useful to verify how it behaves at null density and finite temperatures. That is, unlike the MIT bag model, it serves to describe properties of mesons at low densities and temperatures, and at the same time that the phase transition lines are within the regions suggested by the experiments. Therefore, before including leptons or developing the theory with color superconductivity, we will see in the next section how to construct the phase diagram, and what results we obtain with our model for different parameterizations of the vector coupling that was developed in this section.

\section{Phase diagram construction}
\label{const_diag_fases}

In this section we will explain how to construct the phase diagram of the nl-NJL model. As we will show some results as an example, we will give the values of the parameterizations that we use to obtain them. The quark masses $ u $ and $ d $, the coupling constants $ G_S $ and $ H $, and the model parameter $ \Lambda $ were adjusted to reproduce the phenomenological values for the pion decay constant $f_\pi=92.4$ MeV, and the mesons$S$ quark mass was chosen according to the updated phenomenological value of $ m_s = 95.00 $ MeV, and $ m_s / m_u \ simeq 26 $ in accordance with the data published by the Particle Data Group \cite{Tanabashi:2018oca}.

\begin{figure}[ht!]
\centering
\includegraphics[scale = 0.4]{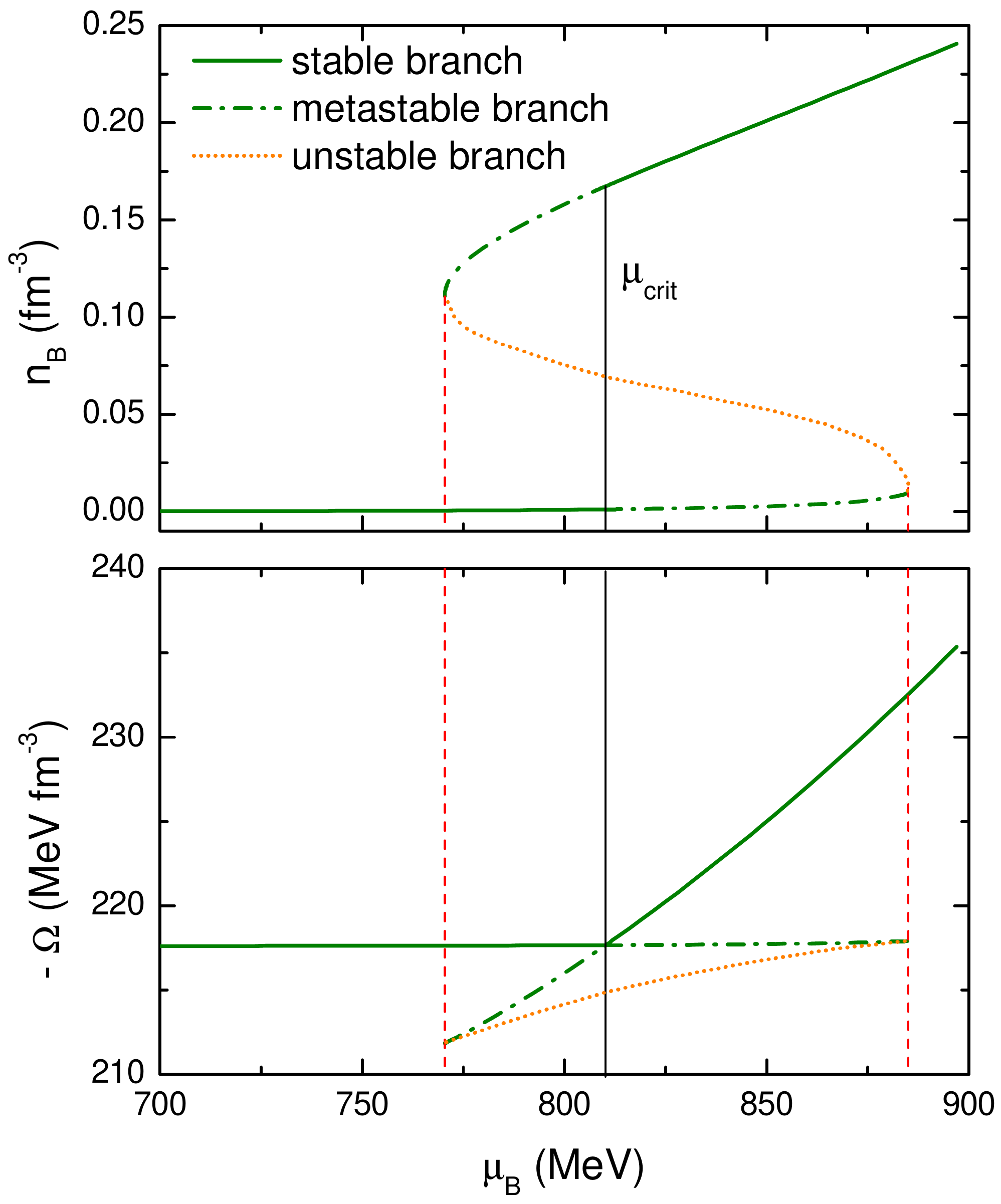}
\caption[Grand potential and density as a function of the baryonic chemical potential for the construction of the first order transition for the quark model.]{Baryon density as a function of the chemical potential for two branches of the baryon density (top) and the grand thermodynamic potential (bottom) run at fixed temperature: from low chemical potentials and non-zero condensates to high chemical potentials, and vice versa. In the upper figure, the two values of densities corresponding to the line of $ \mu_ {crit} $ correspond to the two values of the phase diagram in the $(n_b, T)$ plane. In the lowe figure the crossing of the two branches determines the transition point of the phase diagram for each value of $(T,\mu)$.}
\label{omegamu}
\end{figure}

To obtain the first order phase transition, we proceed to perform the Maxwell construction as explained in the \ref{construcciones} section. However, it is important to note that we will not be working with hadronic and quark EoS as explained in that section, but rather that we will use the same EoS that we obtain from the nl-NJL model, calculating two different branches. The first, going from low chemical potentials ($ \mu \simeq 10 $ MeV) to high ones ($\mu \simeq 500$ MeV), and the second going from high to low chemical potentials.In this case, since we have the same type of particles for both the confined and deconfined phases, we can simply plot the grand potential as a function of $ \mu $, as we see in Figure \ref{omegamu}. 

\begin{figure}[ht!]
\centering
\includegraphics[scale = 0.3]{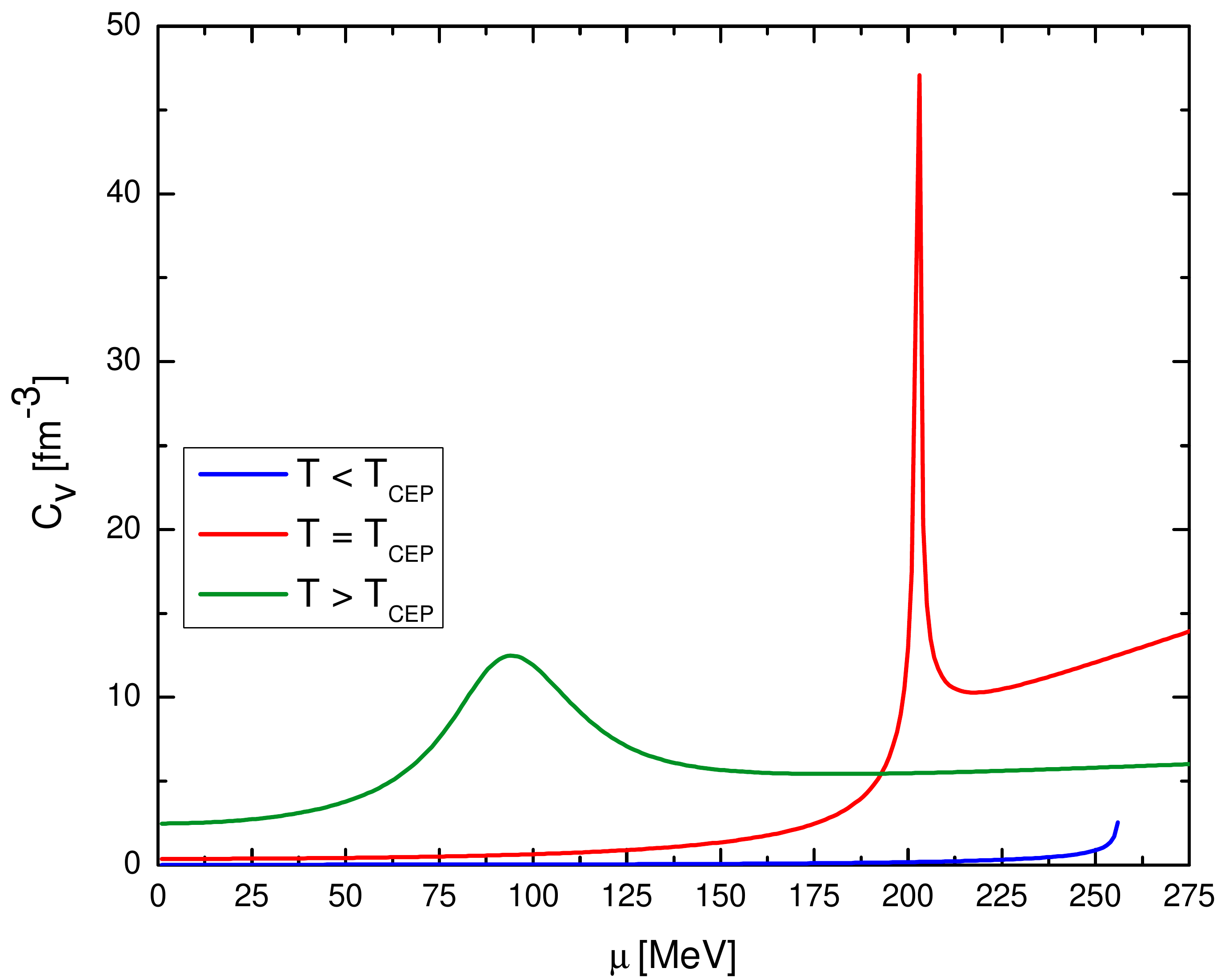}
\caption[Specific heat of quarks for the construction of the crossover transition.]{Specific heat for different temperatures. The maximum determines the crossover point for each value of $ (T, \mu) $ for temperatures greater than the critical temperature. For lower temperatures, the specific heat diverges and ceases to have a defined maximum, as the transition becomes first order. }
\label{susceptu}
\end{figure}

For each given temperature, when traversing the phase diagram, the grand potential in the two branches is plotted and the chemical potential at which they cross, together with the temperature chosen to plot, generate a point ($ T, \mu $) of the phase diagram. It starts from low temperatures and this procedure is carried out by sweeping a certain temperature range. At a critical point, the transition is no longer a first order one, and there are no longer two different branches for the grand potential. At the point where this happens is the critical end point (CEP), resulting in a second order transition. After that point, and towards lower chemical potentials, the phase transition becomes smooth or crossover type. To determine it, we use the peak position of the specific heat, being this $C_V = -T\frac{\partial^2 \Omega}{\partial T^2}$, as shown in Figure \ref{susceptu}.

\begin{figure}[ht!]
\centering
\includegraphics[scale = 0.45]{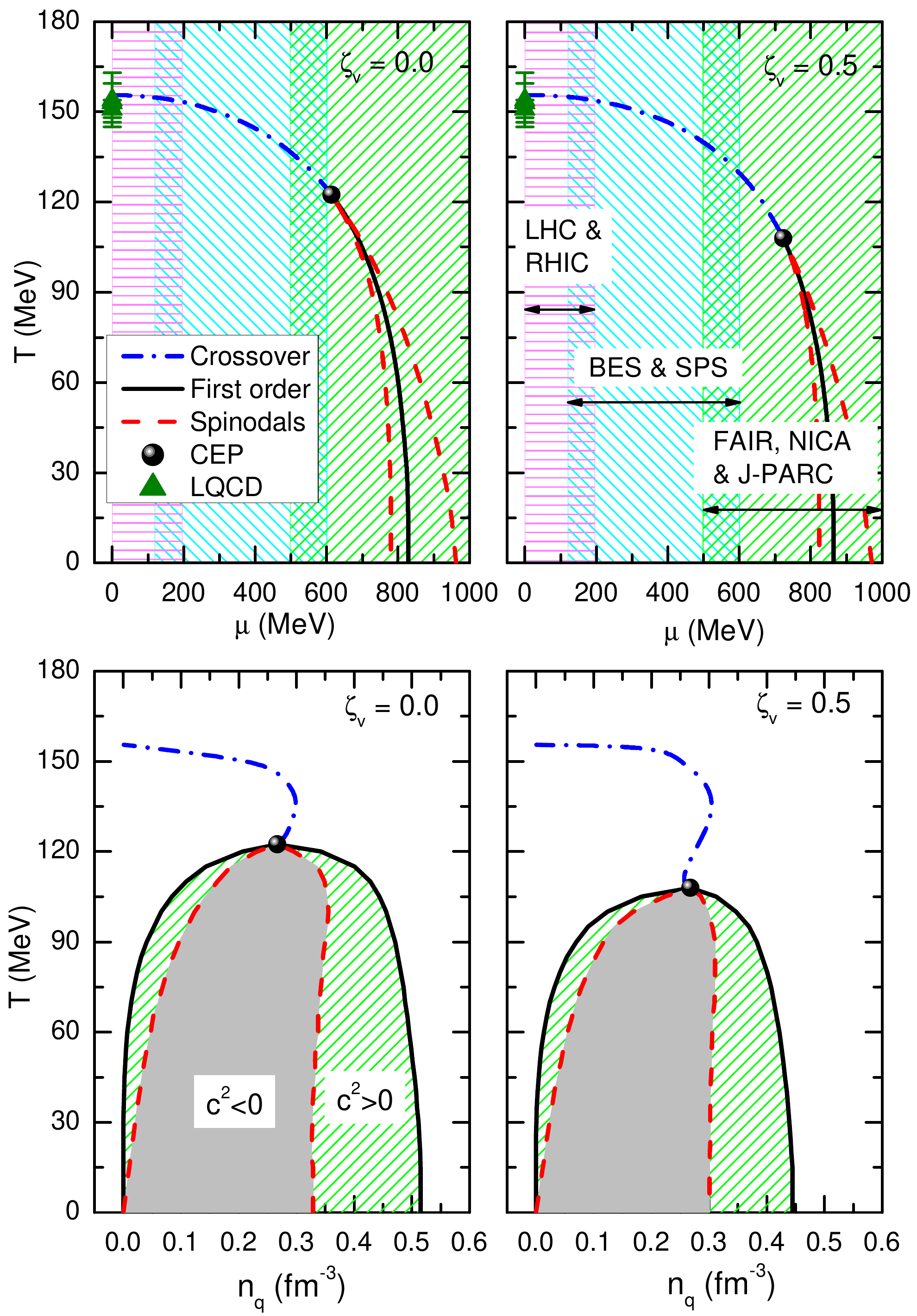}
\caption[Phase diagrams of the quark model in the ($T,\mu$) and ($T,n_b$) planes .]{Top: phase diagrams in the $ (T, \mu) $ plane for different values of $ \zeta_V = G_V/G_S $. Bottom: the same as the left panel but on the  $ (T, n_B) $ plane.}
\label{diagramas_fase}
\end{figure}

Once the points corresponding to the first-order transition and the crossover transition have been obtained, it is possible to construct the two types of phase diagrams, in the $(T,\mu)$ and $(T, n_b)$ planes, as shown in Figure \ref{diagramas_fase}. As is known of extensions of \textit{Lattice QCD} to finite chemical potentials for non-zero masses \cite{Philipsen:2012nu,Borsanyi:2013bia,Bellwied:2015rza}, there should be a crossover phase transition at low chemical potentials. On the other hand, some extrapolations for 2 + 1 flavors at the limit of the continuum indicate a critical temperature for zero chemical potential $T_c(\mu=0) \simeq 155$ MeV \cite{Bazavov:2011nk,Aoki:2009sc,Bazavov:2016uvm}. These two results are in agreement with those obtained with the parameterizations used in this work.

On the other hand, at high chemical potentials and low temperatures, a first order transition is expected as we discussed earlier. The end of the transition line at the CEP is not known, but it is currently being investigated in laboratories such as NICA; FAIR, J-PARK, while the intermediate density regions where the \textit {crossover} type transition is found is among the objectives of the BED, SPS laboratory at RHIC and CERN. The regions corresponding to each experiment are also indicated in Figure \ref{diagramas_fase} (top). The dotted lines of the mentioned figure indicate the limit of the metastable regions, which will be explained later.o.

As demonstrated in the reference work \cite{Fukushima:2008is}, the inclusion of repulsive vector interactions between quarks shrinks the first-order transition line, moving the CEP towards lower temperatures and higher chemical potentials, causing it to eventually disappear for sufficiently strong interactions. However, the values used in this work always allow first order phase transitions at low temperatures. Furthermore, it can be seen that the effect of vector interactions is to move the first order transition line to higher chemical potentials. Finally, it can be seen in the bottom panel of Figure \ref{diagramas_fase} that these interactions cause the metastable regions (marked in gray) to tend to disappear for sufficiently large values of $ \zeta_V $.

In addition to the mentioned construction, spinodal lines can be obtained, corresponding to metastable regions. The density fluctuations associated with these lines can be analyzed in terms of the isothermal speed of sound, given by \cite{randrup:2009,randrup:2010}
\begin{equation}
c_{s}^2 = \frac{n_q}{\epsilon + P} \left( \frac{\partial P}{\partial
  n_q} \right)_T \, .
\end{equation}
The gray marked areas in Figure \ref{diagramas_fase} show unstable regions of the phase diagram where $c^2 < 0$. These regions are surrounded by metastable areas indicated in green, where $ c^2> 0 $ and the red dotted curves show the spinodal lines determined when $ c^2 = 0 $. In regions where $ c^2 <0 $, the compressibility $\kappa \propto n_q(\frac{\partial P}{\partial n_q})_T$ is negative and the system reacts to an increase in density (compression) by increasing low-density fluctuations. As this region is unstable, the density fluctuations that normally occur within the area delimited by the spinodal lines, will separate the system into regions of low and high density. The metastable areas are determined by the region between the spinodals and the first order lines. In these regions, the density fluctuations grow through the addition of quark condensates (left branch), or shrink due to the evaporation of said condensates (right branch). It is worth mentioning that in order to build an EoS of deconfined quark matter, one must work with chemical potentials that are greater than the right branch of the spinodal lines, in order to avoid that pertirbative effects generate the formation of mesons.

\begin{figure}[ht!]
\centering
\includegraphics[scale = 0.5]{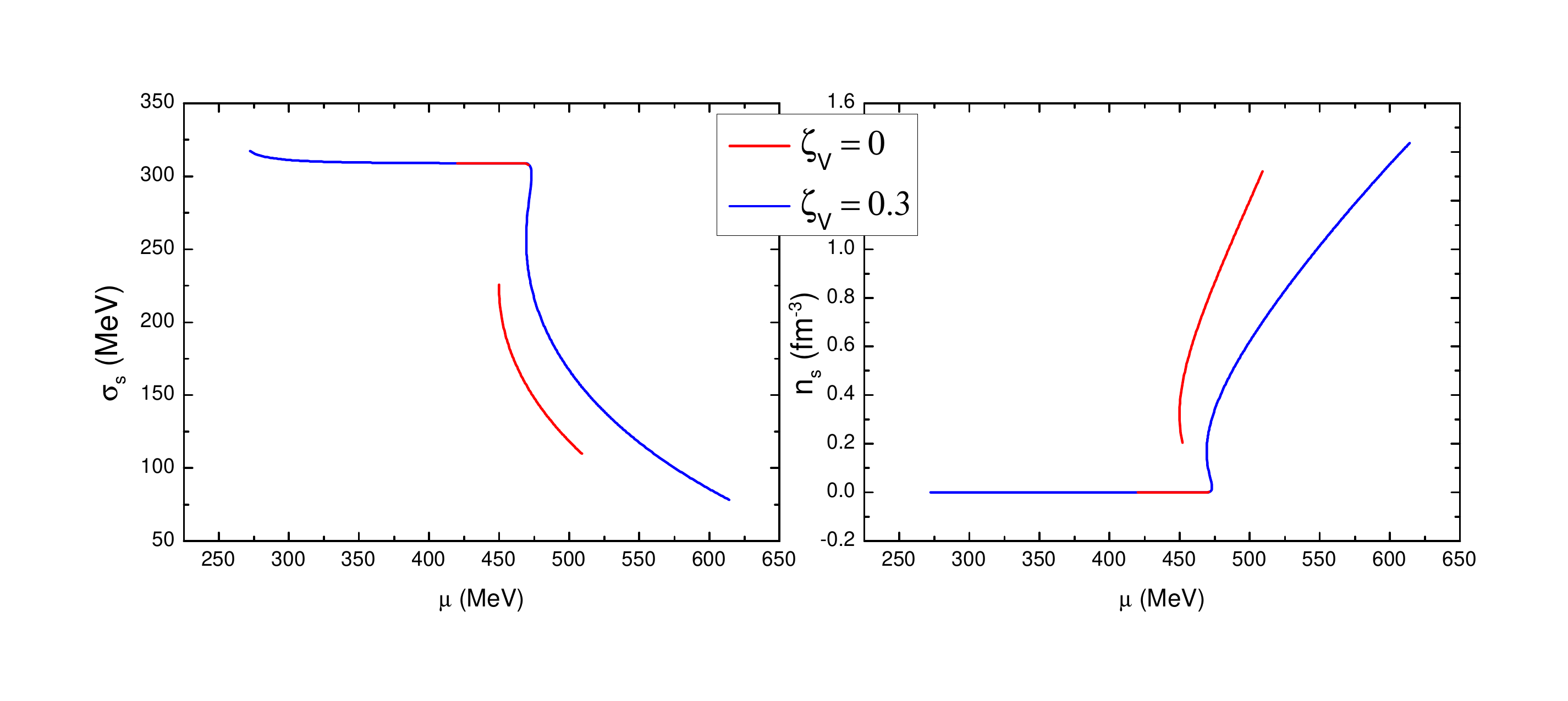}
\caption[$\sigma_s$ and $n_s$ as a function of $ \mu $ for different vector coupling values.]{Left: $ \sigma_s $ a $ T = 0 $ as a function of the chemical potential for $ \zeta_V = 0 $ and $ \zeta_V = 0.3 $. Right: the same as the left panel but for $ n_s $. It can be seen how despite being at greater potentials than the deconfinement of the $u$ and $d$  quarks, the $s$ quark density is zero until another breaking point, from which the field $ \sigma_s $ starts to go down. It is also seen how by increasing the intensity of the vector coupling, the transition goes from being of the first order to being a crossover type. }
\label{sigma_s_dens}
\end{figure}

 It is necessary, however, to indicate a subtlety regarding the type of construction of the phase diagram that was explained: since we are working with three flavors of quarks, it is possible that this construction leads us to an error. To see that the grand potential has discontinuities in the solutions, and constructing the first order line is correct, as long as we do not assume that on the right side of the phase diagram the quarks behave like free quarks. In fact, what happens is that the discontinuity occurs because the order parameters of the chiral transition of the $u$ and $d$ quarks become zero, so we have chiral symmetry restoration for those quarks. However, the same is not the case with the $ s $ quark. This one, being heavier than the others, keeps the $ \sigma_s $ field constant at much higher chemical potentials, and only for potentials on the order $ \mu \simeq 450 $ MeV does it begin to behave as a free quark. This can be clearly seen in Figure \ref{sigma_s_dens}. In the case of $ \zeta_V = G_V / G_S = 0 $ there is a first order phase transition due to the restoration of the chiral symmetry of the heavier quark. On the other hand, by varying the values of the vector coupling, this transition becomes crossover and both the $ \sigma_s $ field and the density vary continuously, as can be seen in the same figure for the case of $ \zeta_V = 0.3 $. It is possible to analyze the construction of phase diagrams including the chiral symmetry restoration line for the $s$ quark, but they strongly depend on the vector interaction used and, as we will see later, it also depends on the inclusion (or lack of) of color superconductivity in the model. It is useful to know, however, that for all vector interactions used in this work $ 0 \leq \zeta_V \leq 0.5 $, the chiral symmetry restoration for the $ s $ quark  occurs for a maximum chemical potential whose value is found between 420 MeV and 450 MeV. We mention this because to try to use this model for the description of free quarks, it is important to know that one should work at chemical potentials greater than those of the restoration of this symmetry, for all quarks. Otherwise, we would be working with free $ u $ and $ d $ quarks, but with the $ s $ quark still confined to its chiral condensate.
 
\section{Lepton inclusion}

As we saw in section  \ref{Estadios}, the construction of the equation of state to model both hadronic and quark matter in NSs has to include leptons, to satisfy the conditions of chemical equilibrium and electric charge conservation, whether global or local. As we saw, there are three types of leptons that are added to the model: electrons ($ e^-) $, muons ($ \mu^- $), and electron neutrinos $ (\nu_e) $. The interaction between leptons and quarks is given by weak interaction processes ($\beta$ decay and inverse $\beta $ decay), which gives a relationship between the chemical potentials of each particle. However, at Lagrangian level (and consequently at the grand potential level), leptons are added to the quark model as free particles. Therefore, no bosonization treatment is necessary, and the regularization used is the same as that developed in section \ref{Regularizacion}. Therefore, we can directly write the regularized generic contribution of a lepton of mass $ m_l $ to the grand potential in the way

\begin{equation}
\label{granpoleptones}
 \Omega^l_{Reg}(T,\mu_l) = 2 T \int \frac{d^3p}{(2\pi)^3}\left\{\mathrm{log}\left[1 + e^{-\frac{E_l + \mu_l }{T}}\right] + \mathrm{log}\left[1 + e^{-\frac{E_l - \mu_l}{T}}\right]\right\},
\end{equation}
where $E_l = \sqrt{\bm{p}^2 + m_l}$ is the lepton $l$ energy, $m_l$ its mass and $\mu_l$ its chemical potential. It should be noted that in the case of neutrinos, which we will consider massless, the result is $ m_ {\nu_e} = 0 $ but the equation \eqref{granpoleptones} is still valid. In this way, the new grand potential will be
\begin{equation}
 \Omega^{MFA}_{Reg}(T,\mu) \rightarrow \Omega^{MFA}_{Reg}(T,\mu) + \sum_l \Omega^l_{Reg}(T,\mu_l), 
\end{equation}
where the sum over $ l $ is carried out on the different leptons as the case corresponds. For simplicity and in order toavoid adding indexes, we will continue to refer to the grand potential with leptons included in the model as $ \Omega^{MFA}_{Reg} (T, \mu) $ and we will distinguish any other case explicitly.

Until now, the chemical potentials of the quarks were distinguished in flavor only in the subscripts, that is, we had that $ \mu_u = \mu_d = \mu_s = \mu $, where $ \mu $ is the generic chemical potential of the quarks.

Following what was explained in the section \ref{Estadios} according to the different stages of a NS, we will need two different ways to add leptons. One that includes electrons and muons, and another that includes electrons and neutrinos, in either case these inclusions will result in the quarks chemical potentials  no longer being equal to each other. For the first case, that of electrons and muons, the relationship that these potentials have to meet to maintain chemical and beta equilibrium is given by

\begin{eqnarray}
 \mu_u &=& \mu - \frac{2}{3}\mu_e  \\
 \mu_d &=& \mu + \frac{1}{3}\mu_e  \\
 \mu_s &=& \mu + \frac{1}{3}\mu_e  \\
 \mu_\mu &=& \mu_e,
\end{eqnarray}

Two things follow from these equations. The first is that until now we had a single free chemical potential that, together with temperature, were the two parameters that were set in the euqations system. Now we still have that potential, since $ (\mu_u + \mu_d + \mu_s) = 3 \mu $, but another free parameter appears which is the electron's chemical potential. This brings us to the second observation, which is that we must find a way to fix the electron's chemical potential. This cannot be done arbitrarily, but is done by adding one more equation to our system, to ensure that the EoS is locally charge neutral. This equation is written as
\begin{equation}
\label{neutralidadconmu}
\frac{2}{3} n_u - \frac{1}{3}\left(n_d + n_s\right) - n_e - n_\mu = 0, 
\end{equation}
which must be added to our equations system to solve in \eqref {gapequationsvec}. In equation \eqref{neutralidadconmu} the $ n_i $ refers to each particle densities that are obtained by deriving the grand potential with respect to the corresponding chemical potential so that

\begin{equation}
 n_i = \frac{\partial\,\Omega^{MFA}_{Reg}(T,\mu_)}{\partial \mu_i}.
\end{equation}

In the case in which the leptons to be included are neutrinos and electrons, the relationships between the chemical potentials change, and they are written as
\begin{eqnarray}
 \mu_u &=& \mu - \frac{2}{3}(\mu_e - \mu_{\nu_e})  \\
 \mu_d &=& \mu + \frac{1}{3}(\mu_e - \mu_{\nu_e})  \\
 \mu_s &=& \mu + \frac{1}{3}(\mu_e - \mu_{\nu_e}), \\ 
\end{eqnarray}
where now muons are not considered. In this case, we see that by including the electron neutrino's chemical potential, we have one more free parameter for our equations system. To fix it, we need to add the equation that defines the lepton fraction of the system, so we add the following relation to our system
\begin{equation}
 \frac{n_e + n_{\nu_e}}{n_B} - Y_L = 0 ,
\end{equation}
where now $n_B = (n_u + n_d + n_s)/3$ is the baryonic number density and $Y_L$ the lepton fraction per baryon that will enter as a fixed parameter of our system. It should be noted that the lepton fraction is fixed on the baryon number density instead of on the total density of quarks, since later the EoS of the quarks will be joined with that of the hadrons and it is necessary to do it consistently. However, when talking purely about quark equations of state, one could work with the quarks density and ask for an arbitrary fraction of leptons over the total quarks density and would not incur errors. Finally, in the case of including neutrinos in the model, the equation \eqref{neutralidadconmu} that ensured charge neutrality must be slightly modified, obtaining

\begin{equation}
 \frac{2}{3} n_u - \frac{1}{3}\left(n_d + n_s\right) - n_e  = 0.
\end{equation}

\graphicspath{{Superconductividad/}}
\chapter{\label{ch:Superconductividad}Color superconductivity in the non-local quarks model}

So far we have developed the nl-PNJL model with attractive scalar and pseudoscalar interactions, and repulsive vector interactions. However, it is known that at zero temperature, for theories that have an arbitrary attractive interaction, fermions are unstable against the formation of bound states \cite{Pathria:1996hda}, hence, it would be natural to include superconductivity. The argument is as follows: at $ T = $ 0, all the states of momentum less than the Fermi momentum are occupied. Since the free energy $ | E_p - \mu | $ vanishes at the Fermi surface, pairs of (anti) particles could be created without energy cost. If there is an attractive type interaction between these particles, the Fermi surface becomes unstable. In the original theory of superconductivity BCS \cite{Bardeen:1957kj} this is fixed with the formation of Cooper pairs, the direct consequence of which is the formation of a superconducting gap (energy needed to break a pair) in the excitation spectrum, causing that there are no more levels with zero free energy. In the case of QCD, where the gluon exchange interaction is attractive in certain channels, it is to be expected that pairs of quark condensates similar to Cooper pairs will be formed, given what has been explained above. Since there are many attractive channels in theory, one can construct a very wide range of interactions that give rise to condensates of diquarks. In this work we will focus on the predominant attractive channel for both gluon exchange interactions and instantons mediated interactions \cite{Rischke:2003mt}.

\section{Diquarks as quarks condensates}

The inclusion of diquarks starts from adding an interaction that mixes different colors and flavors of quarksto the action. As the treatment is rather extensive, in this section we will restrict ourselves only to the Lagrangian  with the interaction between quark-quark currents where each possible color pairing can lead to the formation of condensates. Then we will see that these calculations do not conflict when adding them to the calculations made with the rest of the interactions.In general, in this section we will maintain a certain similarity with what was developed in the work of the reference \cite {Buballa:2003qv}, with the difference that there they work with the local model instead of the non-local one, and they use other matrices definitions from the ones we use here, which are the same as those used in the reference \cite{Blaschke:2007ri}. In this work, the superconducting phase that we will consider is the so-called $ 2SC + s $. Since the mass of the quark $ s $ does not favor the pairing with the other two light quarks, we can consider it uncoupled, so that in the phase $ 2SC + s $ the condensates are formed only by quarks $ u $ and $ d $. So let's start from action
\begin{eqnarray}
\label{acciondiquarks}
 S_E &=& \int d^4 x \left\{\overline{\psi}\left(x\right)\left( i \slashed{\partial} + m \right)\psi\left(x\right) -\frac{G_D}{2} \left[j_D\left(x\right)\right]^\dagger j_D\left(x\right) \right\},
\end{eqnarray}
with the diquarks current
\begin{equation}
j_D\left(x\right)  =  \int d^4 z \,g(z) \overline{\psi}_C \left(x + \frac{z}{2}\right)i\gamma^5 \lambda_A \lambda_{A'}\psi\left( x - \frac{z}{2}\right),
\end{equation}
where $\psi_C (x) = \gamma_2 \gamma_4 \overline{\psi}^T(x)$. The gamma matrices are defined as $(\bm{\gamma},\gamma_4)$ with $\gamma_4 = i\gamma_0$. The $\lambda_A$ y $\lambda_{A'}$ matrices act in space of color and flavor respectively and run only through the indices ${2,5,7}$. This type of interaction results in a condensate matrix that can be written in the way:
\begin{equation}
 s_{AA'}= \langle \overline{\psi}_C \gamma_5\lambda_A \lambda_{A'} \psi\rangle,
\end{equation}
with $C = \gamma_2 \gamma_4$ the charge conjugation operator. Fortunately, this matrix can be reduced by rotating it in color, bringing it into the form
\begin{equation}
s = 
 \begin{pmatrix}
s_{22} & 0 & 0 \\
s_{52} & s_{55} & 0 \\
s_{72} & s_{75} & s_{77} \\
\end{pmatrix}.
\end{equation}
It can be seen (for example in the works of the references \cite{Alford:1998mk, Fritzsch:1973pi}) that in the one gluon exchange regime, the non-diagonal components are negligible with respect to the others, so we will take them as null, and we will be left with only the elements $ s_ {22} $, $ s_ {55} $ and $ s_ {77} $. 
It is worth noting something that we briefly mentioned at the end of the \ref{intvec} section: interference with the Polyakov loop. Although the rotations that were made and the fact of neglecting the non-diagonal components take the condensate matrix to a mathematically treatable form, when we include the Polyakov loop and want to rotate in color, it is practically impossible to bring everything to a diagonal form. This is why in general the literature does not include these two effects at the same time. It could be argued that the order parameter $ \Phi $ of the Polyakov loop is small at low temperatures, and therefore prefer to work with a model that includes a superconducting color phase and not the Polyakov loop in this regime. Although the argument seems valid in principle, what is not yet known is how the inclusion of diquarks in the expectation values of  $ \Phi $ interferes. That is, although the latter at low temperatures is negligible, it could be that its behavior changed drastically when including diquarks precisely because of the non-diagonal terms of the condensate matrix. Therefore, to think that an approximation only including color superconductivity in the model at low temperatures is the same as including both contributions, would be to incur an error. This is why in this work we will analyze the color superconductivity separated from the Polyakov loop theory, except for the case of zero temperature, for which this last contribution is strictly null.
On the other hand, once the flavor rotation is done that allows the diquarks to be identified, the scalar, pseudoscalar and vector interactions that we saw earlier can be written in such a way that they result in the same type of interaction for the fields written in the new base. This is what allows, without losing generality, to add the other interactions after having done the inclusion of diquarks treatment.

Returning to the matrix's treatment, what we have now are three condensates that correspond to different types of pairing between quarks, which are shown in the table \ref{tabladiquarks}.
\begin{table}[htb]
\begin{center}
\begin{tabular}{|c|c|c|c|}
\hline %\noalign{\smallskip}
Condensate  & $s_{22}= \langle \overline{\psi} \gamma_5\lambda_2 \lambda_2' \psi\rangle$& $s_{55}= \langle \overline{\psi}_C \gamma_5\lambda_5 \lambda_5 \psi \rangle$ &  $s_{77}= \langle \overline{\psi}_C \gamma_5\lambda_7 \lambda_7 \psi \rangle$\\ \hline
  Diquark pairs & $(u_r,d_g),(u_g,d_r)$&$(d_g,s_b),(d_b,s_g)$ & $(s_b,u_r),(s_r,u_b)$ \\ \hline
\end{tabular}
 \caption{Color and flavor structure of the various diquark condensates.}
\label{tabladiquarks}
\end{center}
\end{table}
As we have three different quark condensates, the bosonization will lead us to have three new fields, which we will call $ \Delta_i $, with their respective three auxiliary fields, which we will call $ D_i $. Let's see how the structure of the $ A $ operator that we defined above looks like, to which we are going to have to calculate the determinant. Since this operator is the inverse of the $ S (p) $ propagator, we will use the name of the latter to avoid a confuse notation. What will remain in the fermionic path integrals after extending to finite temperature and chemical potential is:
\begin{equation}
S^{-1}(p) = 
 \begin{pmatrix}
-\slashed{p} + \hat{M} + i\hat{\mu}\gamma^4 & i\sum_A \Delta_A^p \gamma_5  \lambda_A \lambda_A \\
i\sum_A (\Delta_A^p)^* \gamma_5  \lambda_A \lambda_A& -\slashed{p} + \hat{M} - i\hat{\mu}\gamma^4  \\
\end{pmatrix},
\end{equation}
where $\Delta_A^p = \Delta_A g(p)$. The matrix $S^{-1}(p)$ previously written is only expanded in the Nambu-Gorkov indices (which appear when writing the fields as a function of the charge conjugates, doubling the operator space). However, now the chemical potentials have flavor and color structure, therefore they are 9x9 matrices in that space, with $ \mu_ {fc} $ indices. The mass matrix $ \hat{M} $ has flavor indices, but it is useful to preserve the flavor and color indices so as not to lose generality. This will help us to include vector and scalar interactions, where the mass is affected by the regulator, which in turn depends on the chemical potential, so it will also have flavor and color indices. If we take into account all the Dirac, flavor, color and Nambu-Gorkov indices, the operator for which we will have to calculate the determinant is a 72x72 matrix, practically impossible to deal with in a straightforward manner.  However, rearranging rows and columns, and writing determinants as trace of logarithms, we can divide it as follows:
\begin{eqnarray}
\label{matricessupercond}
 \mathrm{Tr}\{\mathrm{log}[S^{-1}(p)]\} &=& \mathrm{Tr}[\mathrm{log} (D_{ug,dr})] + \mathrm{Tr}[\mathrm{log} (D_{ur,dg})] + \mathrm{Tr}[\mathrm{log} (D_{ub,sr})] + \mathrm{Tr}[\mathrm{log} (D_{ur,sb})] \nonumber \\
 &+& \mathrm{Tr}[\mathrm{log} (D_{db,sg})]+ \mathrm{Tr}[\mathrm{log} (D_{dg,sb})] +  \mathrm{Tr}[\mathrm{log} (D_{ur,dg,sb})],
\end{eqnarray}
where the matrices $ M $ correspond to different independent blocks of $ S^{- 1} (p) $. The first six mix only two species of flavors and colors (phase 2SC), and the seventh includes the blocking structure of color and flavor (CFL phase). The different condensates that appear correspond to the different combinations of quark flavor, that is,
\begin{eqnarray}
 \Delta_{ud} &=& \Delta_{du} = \Delta_2 = \Delta_2^*\\
 \Delta_{us} &=& \Delta_{su} = \Delta_5 = \Delta_5^*\\
 \Delta_{ds} &=& \Delta_{sd} = \Delta_7 = \Delta_7^*. 
\end{eqnarray}
Due to what was explained at the beginning of the section, we are interested in the phase $ 2SC + s $, assuming that the condensed quarks $ u $ and $ d $ appear before the others, so we will work in the case where $ \Delta_5 = \Delta_7 = 0 $. This is not strictly true, but since the $s$ quark mass is much greater than those of the $ u $ and the $ d $, it is to be expected that if diquarks appear, the condensate $ \Delta_2 $ will predominate over the others, at least at densities at which they are just beginning to form. On the other hand, in the context of the non-local model, the treatment of the determinant corresponding to the CFL phase is quite complex, so that in a first approximation it is very useful to cancel the other condensates. By doing this, the only remaining diquark structure takes the form
\begin{equation}
D_{ud} = 
 \begin{pmatrix}
-\slashed{p} + \hat{M} + i\hat{\mu}\gamma^4 & i\Delta_2^p \gamma^5 \\
i\Delta_2^p \gamma^5 & -\slashed{p} + \hat{M} - i\hat{\mu} \gamma^4  \\
\end{pmatrix}.
\end{equation}
The determinant of this matrix is perfectly solvable, so once we have done that we can calculate the grand potential. Before that, let's see how auxiliary fields that arose due to color superconductivity influence the grand potential. As in section \ref{bosonizacion}  before diagonalizing, the grand potential will now include terms of the form
\begin{equation}
\label{granpodiquarksauxiliares}
 \Omega \supset \Delta D + \frac{G_D}{2}D^2,
\end{equation}
as it happened with the fields $ \sigma_a $ $ S_a $ in \eqref{SMFA}. Now, the diagonalization that is carried out for the fields $\sigma_a$, $S_a$ $\theta_a$ and $V_a$, was generated because the sum of the Gell-Man matrices by the fields  $\lambda_0 \sigma_0 + \lambda_3 \sigma_3 + \lambda_8 \sigma_8$ resulted diagonal in flavor space, so auxiliary fields with $ a $ indices  could be rewritten on a diagonal flavor base. As the field $ \Delta_2 $ and its auxiliary $ D $ are already defined previously without going through these indices and without requiring diagonalization, they will be included in the grand potential without any modification with respect to the equation  \eqref{granpodiquarksauxiliares}. \\
On the other hand when considering $\Delta_5 = \Delta_7 = 0$, the matrices $D_{f,f'}$ of equation \eqref{matricessupercond} which involve the  $s$ quark, have no off-diagonal components, and the only interactions that appear are those coming from the dynamic masses and shifts in chemical potential due to the vector interaction. Therefore, all the terms related to the$s$  quark in the grand potential are decoupled and do not undergo changes with respect to the treatment we did previously. In this way, we can write the grand potential corresponding to the color superconductivity, remembering that we leave the dressed mass $ M $ instead of $ m $ to allow us to later include the shifts generated in the mass by the scalar interactions. Finally, the grand potential is in the form
\begin{eqnarray}
 \Omega^{ACM}_{Diq, ud} =   &-& 2\int\frac{d^4p}{(2\pi)^4}  \sum_c \left\{ \frac{1}{2} \mathrm{log} |A_c|^2 -  \mathrm{Real}\left[ \mathrm{log}\left({p_{uc}^+}^2 + m_u^2 \right)\right] - \mathrm{Real}\left[\mathrm{log}\left({p_{dc}^+}^2 + m_d^2\right)\right]    \right\} \nonumber \\ 
 &-&  \Delta  D + \frac{G_D}{2} D^2 - \sum_{f,c} \frac{\Theta\left(\mu_{fc} - m_f\right)}{24\pi^2}\Biggl[\left(-5m_f^2 + 2\mu_{fc}^2\right)\mu_{fc}\sqrt{\mu_{fc}^2 - m_f^2} \nonumber \\
 &+& 3m_f^4\mathrm{log}\left(\frac{\mu_{fc} + \sqrt{\mu_{fc}^2 - m_f^2}}{m_f}\right) \Biggr],
\end{eqnarray}
donde 
\begin{eqnarray}
\label{Ac}
 A_c &=& \left({p_{uc}^+}^2 + M_{uc}^2\right)\left({p_{dc}^-}^2 + {M_{dc}^*}^2\right) \nonumber \\
 &+& \left(1 - \delta_{bc}\right){\Delta^p}^2\left({\Delta^p}^2 + 2 p_{uc}^+ . p_{dc}^- + 2M_{uc}M_{dc}^* \right), \\
 p_{fc}^\pm &=& \left( p_0 \mp i \mu_{fc} , \bm{p}\right) , \\
\Delta^p &=& \Delta\,g\left(\frac{ \left[{p_{ur}^+} + {p_{dr}^-} \right]^2 }{4}\right).
\end{eqnarray}
Let's now conceptually see how this form of the term $ A_c $ is arrived at. On the one hand, from equation \eqref{matricessupercond}, when eliminating the condensates that include the $ s $ quark, a new matrix $ D_{ur, dg} $ appears, so together with the first term of that equation it allows us to say that the red and green colors are interchangeable for the $ u $ and $ d $ quarks. The $ s $ quark is totally decoupled from the color interactions, as well as the color blue.  Then using this and the complex number property  $|Z|^2 = Z Z^*$, it is easy (although laborious) to arrive at the mentioned form of $ A_c $. Let's see that the last term of equation \eqref{Ac} vanishes both when the color is blue, as well as when $\Delta = 0$. This is because the $ \Delta $ condensate  is only present when the quarks colors are red and green, also if the condensate vanishes one recovers the non-local model described in chapters \ref{ch:Nambu} and \ref{ch:PNJL} as it was expected. For all the remaining effects, the only contribution that the $ A_c $ term will receive when including scalar, pseudoscalar, and vector interactions, will come in a shift of the chemical potential and a contribution to the dressed mass $ M_{fc} $ for each each flavor and color. Introducing these interactions where they correspond, and adding the auxiliary fields that we saw earlier to the grand potential, we complete the model for three flavors and colors of quarks, with all the couplings studied. Thus, the superconducting color phase ($ 2SC + s $) grand potential is written as
\begin{eqnarray}
 \Omega =  &-& 2\int\frac{d^4p}{(2\pi)^4}  \sum_c \mathrm{Real}\left\{ \mathrm{log}\left[\frac{{q_{sc}^+}^2 + M_{sc}^2}{p_{sc}^2 + m_s^2}\right]\right\}\nonumber \\ 
 &-& 2\int\frac{d^4p}{(2\pi)^4}   \sum_c \left\{ \frac{1}{2} \mathrm{log} |A_c|^2 -  \mathrm{Real}\left[ \mathrm{log}\left({p_{uc}^+}^2 + m_u^2 \right)\right] - \mathrm{Real}\left[\mathrm{log}\left({p_{dc}^+}^2 + m_d^2\right)\right]    \right\} \nonumber \\ 
 &-& \frac{1}{2}\left[ \sum_f \left(\sigma_f  S_f  + \frac{G_S}{2} S_f^2 +  \theta_f   V_f - \frac{G_V}{2} V_f^2\right) + \frac{H}{2} S_u  S_d  S_s + 2\Delta  D +G_D D^2\right] \nonumber \\
 &-& \sum_{f,c} \frac{\Theta\left(\mu_{fc} - m_f\right)}{24\pi^2}\Biggl[\left(-5m_f^2 + 2\mu_{fc}^2\right)\mu_{fc}\sqrt{\mu_{fc}^2 - m_f^2} \nonumber \\
 &+& 3m_f^4\mathrm{log}\left(\frac{\mu_{fc} + \sqrt{\mu_{fc}^2 - m_f^2}}{m_f}\right) \Biggr],
 \label{omegasupercond}
\end{eqnarray}
where now the term $ A_c $ includes the shifted momentums for each chemical potential
\begin{eqnarray}
 A_c &=& \left({q_{uc}^+}^2 + M_{uc}^2\right)\left({q_{dc}^-}^2 + {M_{dc}^*}^2\right) \nonumber \\
 &+& \left(1 - \delta_{bc}\right){\Delta^p}^2\left({\Delta^p}^2 + 2 q_{uc}^+ . q_{dc}^- + 2M_{uc}M_{dc}^* \right), \\
 q_{fc}^\pm &=& \left(  p_0 \mp i \left[\mu_{fc} - \theta_f g\left({p_{fc}^\pm}^2 \right) \right], \bm{p} \right), \\
 p_{fc}^\pm &=& \left( p_0 \mp i \mu_{fc}, \bm{p} \right),  \\
 M_{fc} &=& m_f + \sigma_f \, g\left( {p_{fc}^+}^2  \right) , \\
 \Delta^p &=& {\Delta}\, g\left(\frac{ \left[{p_{ur}^+} + {p_{dr}^-} \right]^2 }{4}\right).
\end{eqnarray}
With this we complete the model with diquarks at zero temperature for the case of the ($ 2SC + s $) phase, plus the scalar, pseudoscalar and vector interactions seen previously. The extension to finite, fortunately, is straightforward. In this case, having already made the regularization of the free part, we can reuse the replacement \eqref{recetamala}, taking into account that the terms multiplied by $\Theta\left(\mu_{fc} - m_f\right)$ of equation \eqref{omegasupercond} must be replaced by the integrals of the regularized free part, as we did when we regularized the case of free fermions, and with that the process of analytical calculation is finished. Since the equations are extensive, we invite the reader to see Appendix \ref{ApendiceB} to review them.

\section{Color and electric charge neutrality for diquarks in Neutron Stars}

So far we have seen how to introduce diquarks in the model that we had developed in the chapters \ref{ch:Nambu} y \ref{ch:PNJL}. However, unless we include color and electric charge neutrality in the model, each quark will have three independent color chemical potentials, which together with the three flavors of quarks gives us a 9x9 chemical potentials matrix. This leads us to two problems: first, our system has too many free parameters, second, we need to impose the physical conditions that are required in the context of a NS, which is what we want to model. For this, we know that the matter in these objects is in chemical equilibrium, which, as we saw, leads us to the conditions of charge neutrality including leptons. In order for these conditions to be fulfilled when there are diquarks, however, the matter, in addition to being electrically neutral, must be color neutral. We will then include leptons as in the previous cases, but we will stick to the simplest case of including only electrons and muons. Remembering the charge neutrality conditions of the previous chapter, in the case of not including diquarks we had
\begin{equation}
\label{neutralidadcargaconcolor}
n_Q = \sum_c \left[\frac{2}{3} n_{u,c} - \frac{1}{3}\left(n_{d,c} + n_{s,c}\right)\right] - n_e - n_\mu = 0,
\end{equation}
where we have that this equation is identical to the equation \eqref{neutralidadconmu}, but making a sum over color in each quark density. If we remember that each density is the derivative of the grand potential with respect to each chemical potential, this last condition was implicitly fulfilled in the previous cases, since the chemical potentials did not have color indices (or they were the same for each color, formally). Now, since we have these indices, we can define the color densities in the way
\begin{equation}
 n_c = \frac{\partial \Omega^{ACM}}{\partial \mu_{fc}}.
\end{equation}
Now instead of working with the total color densities, it is useful to define the linear combinations:
\begin{eqnarray}
 n &=& 3 n_B = n_u + n_d + n_s \\
 n_3 &=& n_r - n_g \\
 n_8 &=& \frac{1}{\sqrt{3}}\left(n_r + n_g -2 n_b \right),
\end{eqnarray}
where $ n_B $ is the baryon number density and the pair $ (n_3, n_8) $ describes color asymmetries.  As in our case we have a red and green symmetry, the result is $ n_3 = 0 $ for which $ \mu_ {fr} = \mu_{fg} $. We then have three conserved charges $ \{n, n_8, n_Q \} $ related to three independent chemical potentials $ \{\mu, \mu_8, \mu_e \} $. The relationship between these chemical potentials and those of the quarks is given by the generators of transformations in flavor and color respectively, being
\begin{eqnarray}
 \mu_{ur} &=& \mu_{ug} = \mu - \frac{2}{3}\mu_e + \frac{1}{2\sqrt{3}}\mu_8, \\
 \mu_{ub} &=&  \mu - \frac{2}{3}\mu_e - \frac{1}{\sqrt{3}}\mu_8, \\
 \mu_{dr} &=& \mu_{dg} = \mu + \frac{1}{3}\mu_e + \frac{1}{2\sqrt{3}}\mu_8, \\
 \mu_{db} &=&  \mu + \frac{1}{3}\mu_e - \frac{1}{\sqrt{3}}\mu_8, \\
 \mu_{sc} &=& \mu_{dc}.
\end{eqnarray}

\begin{figure}[ht!]
\begin{center}
 \includegraphics[scale = 0.45]{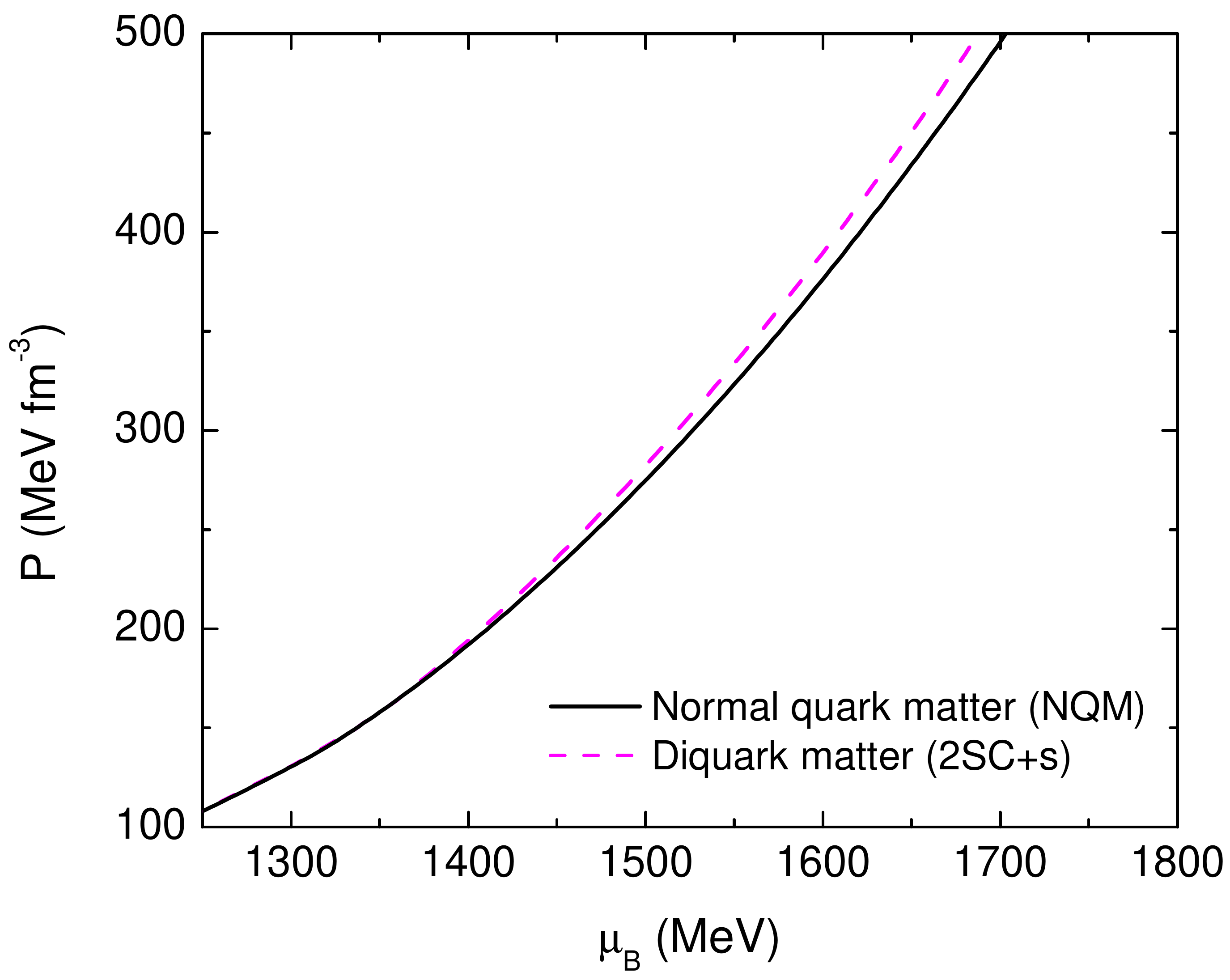}
\caption{Equations of state for the ($ 2SC + s $) phase that of normal quark matter (NQM).}
\label{figEoS2SC}
\end{center}
\end{figure}

Given these relationships, imposing the neutralities $n_8 = n_Q = 0$, together with the equations system \eqref{gapequationsvec}-\eqref{gapequationsvec2}, allow us to keep a single free parameter, which we choose will be the quarks chemical potential $\mu$. Due to its length, the particle densities per color and other thermodynamic quantities for the color superconductivity case are presented in the appendix. \ref{ApendiceB}. The aforementioned conditions ensure a solution for the grand potential, but this solution is not always unique. or example, at high chemical potentials, the superconducting phase $ 2SC + s $ overlaps with the quark gluons plasma phase. In the $(P,\mu_B$) plane, where $\mu_B = 3\mu$, it is easy to see that the diquark solution has more pressure for the same chemical potential than the solution without diquarks. Since in the case of zero temperature the chemical potential is equal to the Gibbs energy per particle, the phase with the lowest Gibbs energy will be the most favorable, thus resulting in the phase $ 2SC + s $ the preferred one, as seen in Figure \ref{figEoS2SC}.

\begin{figure}[ht!]
\begin{center}
 \includegraphics[scale = 0.55]{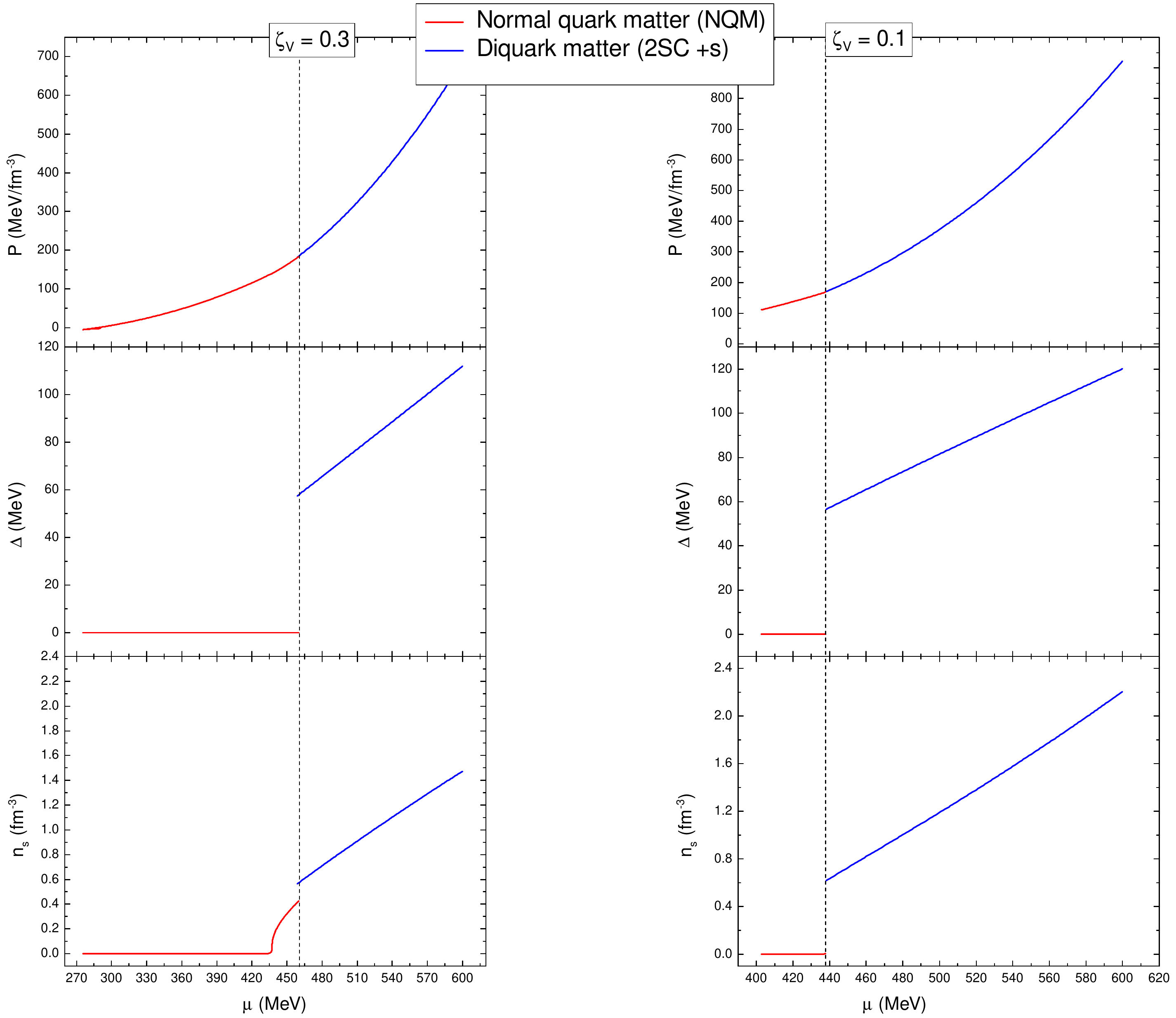}
\caption[Equation of state, $ \Delta $ field  and density $ n_s $ for different values of the vector coupling.]{Equation of state, $ \Delta $ field  and density $ n_s $ for $ \zeta_V = 0.3 $.  In the right panel the same is plotted for $ \zeta_V = 0.1 $. We can see how reducing the vector interaction causes superconducting matter to occur at a lower chemical potential, deconfining more $s$ quarks.}
\label{fig_eos_delta_ns}
\end{center}
\end{figure}

It is worth mentioning that now we have not only two phases in play, but we must continue to consider the phase that has the condensate $ \langle s \bar {s} \rangle $ other than zero, where the $ s $  quark is not yet decoupled and holds that $ n_s = 0 $. For different values of vector interaction and diquark coupling $ (G_V, G_D) $, it is possible that the $s$ quark is deconfined before diquarks are generated, or conversely, that diquarks are created before the $s$ quark is deconfined. If we name ($ \mathrm {NQM_{ud}} $) to the phase that has free $ u $ and $ d $ quarks \footnote{To ease reading in the explanations of this chapter, we understand 'free' as 'deconfined', that is, with the condensate $ \langle q \bar {q} \rangle $ null, but it should not be confused with strictly free particles.}, but null $s$ quark density ($n_s = 0$), (NQM) phase  to which it has the three deconfined quarks and ($ 2SC + s $) phase  to which it has free $ s $ quarks  and non-zero superconducting gap, we can distinguish two transitions shown in Figure \ref{fig_eos_delta_ns}: the first one from  (NQM) to ($2SC+s$). That is, quarks behave as free quarks before transitioning to the color superconducting phase. This happens when the vector interaction is sufficiently high ($\zeta_V \gtrsim 0.2$).  The other transition is from ($ \mathrm {NQM_{ud}} $) phase  to ($ 2SC + s $) phase, that is, the $ s $ quark  is still part of the chiral condensates $\langle \overline{s} s \rangle$, and the transition causes a jump in the density $ n_s $, that is to say that the condensates break. This provokes a larger jump in $ n_s $ from the previous transition, and moves it towards lower chemical potentials. In summary, decreasing the vector interaction causes the transition to superconducting matter to occur at lower chemical potentials, but at the cost of deconfining more $ s $ quarks.

The fact that the color superconducting phase occurs before or after the decoupling of the $s$ quark  from its chiral condensate is currently a matter of debate \cite{Shahrbaf:2019vtf}, and as we have just seen it strongly depends on the model parameters used. Furthermore, changing the diquarks coupling $ \zeta_D = G_D / G_S $ also competes by moving the phase transition to higher (or lower) chemical potentials. Analysis of the structure of the complete phase diagram for quark matter at finite temperature becomes even more complex. The points in the $ (T, \ mu) $ plane at which the ($ 2SC + s $) phase  appears, plus the competition with the $ s $ deconfinement are not only computationally expensive to achieve, but also, depending on the vector coupling used, the transitions can be first order or \textit {crossover}, which generates a variety of phase diagrams that are outside the scope of this work. It is important to note, however, that in order to construct the hybrid EoS with a hadronic model that includes hyperons, the strong condition that must be asked for the quark matter EoS is that it has deconfined $s$ quarks, that is, $ n_s \neq 0 $ density.In this work we will restrict ourselves to the most recommended value in the literature for the diquark coupling constant $ \zeta_D = (3/4) $ \cite {Buballa:1996tm}, and we will take vector coupling as a free parameter. The choices that we will make will depend on the type of hadron-quark interface used for the hybrid EoS, as well as on the observational bounds regarding the determination of the values for the maximum masses measured in NSs. These results and their corresponding analysis will be presented in Chapter \ref{ch:Resultados}.

\graphicspath{{Hadrones/}}
\chapter{\label{ch:Hadrones}Hadronic matter and DDRMF model}

To complete the description of the matter within the NSs we need, in addition to the mentioned quark models, a model that describes the hadronic matter. Although in the past it was thought that nuclear matter was composed solely of protons and neutrons, with the advancement of experimental physics and the discovery of the baryons and mesons spectrum, it was necessary to improve the theoretical models to reproduce the experimental observations. In this context, John Dirk Walecka and his collaborators built a model in 1974 that made it possible to describe nucleon-nucleon interactions. This model is based on a phenomenological treatment of hadronic degrees of freedom, within a quantum field theory. The first versions of the model considered scalar ($ \sigma $) and vector ($ \omega $) mesons, coupled to baryon fields, while new versions also include a meson responsible for baryon-baryon interactions ($ \rho $).  These mesonic fields are generally approximated by their mean values, and are called relativistic mean field theories (RMF).
The first model included only linear interactions between the baryon fields and the $ \sigma $ and $ \omega $ fields, but this led to very high values of the nuclear compressibility modulus $ K_0 = 545 $ MeV, while different analyzes show that it should be in the range $ K_0 \simeq $ 200 MeV - 300 MeV \cite{Blaizot:1995zz,Vretenar:2000bq} . For this reason, it was first thought of including $ \sigma $ meson cubic interactions, thus obtaining better values for the compressibility modulus \cite{Serot:1992ti}. Then, Bodmer \cite{Bodmer:1989hdx} proposed a quartic self-interaction for the $ \omega $ meson to obtain a positive coefficient of the scalar quartic self-interaction, since a negative coefficient for that interaction resulted in an energy spectrum that it was not bounded below. The difference in the predictions of these models for the masses of the NSs are really significant. Models without cubic and quartic interactions predict extremely large masses, on the order of 2.8 $ M_{\odot} $, in contrast to those that predict 1.8 $ M_{\odot} $ and that include such interactions. Currently, these models have been improved, and also include coupling constants dependent on baryon density, whose parameterizations are abbreviated as DDRMF (for \textit{Density Dependent} RMF). This inclusion of the density dependence results in a redundancy of the effect of the cubic and quartic interactions for the $ \sigma $ and $ \omega $ mesons, for which the DDRMF models, in general, do not include the corresponding explicit effective potentials to such interactions. Given that the treatment of DDRMF models largely includes what is necessary to explain the RMF models, in the following sections we will explain the former in detail, briefly commenting on the modifications necessary for treatment without density dependence.

\section{Density Dependant Relativistic Mean Field Theory}

To model hadronic matter in neutron stars, we use the DDRMF approximation that describes baryon-baryon interactions in terms of mesonic fields. 
As we mentioned, DDRMF models are an extension of RMF models whose coupling constants do not depend on density. These types of extensions take into account the effects of the environment, making the meson-baryon coupling constants depend on the density of baryons \cite{PhysRevC.52.3043}. This dependency is extracted from matter properties of finite nuclei, making these models better fit the empirical data than the RMF models \cite{Typel:2009sy}.

Given the impossibility of modeling the interactions between all the particles in an NS, the values of the mesonic fields are equal to their mean values in the RMF approximation. These mesons are: a ($ \sigma $) scalar meson  that describes the attraction between baryons, an ($ \omega $) vector meson  that describes the repulsion and an isovectorial ($ \rho $) meson that describes the baryon-baryon interaction in asymmetric isospin systems. The pion ($ \pi $), which has a primary role in the description of these interactions, has odd parity, so its field vanishes in the RMF approximation \cite{Glendenning:1997wn}. There are other mesons with  $ s $ quarks that describe the interactions between baryons, but due to the lack of existing empirical data to limit the values of the coupling constants, they are usually excluded from these models..

Among the baryons \textit {B} that can exist in NS, there are the octet baryons, made up of the nucleons $N$ = $\{n$, $p\}$, the hyperons $Y$ = $\{\Lambda$, $\Sigma^+$, $\Sigma^0$, $\Sigma^-$, $\Xi^0$, $\Xi^-\}$ and the delta resonances quartet $\Delta(1232)$ = $\{\Delta^{++}$, $\Delta^+$, $\Delta^0$, $\Delta^-\}$, being then $B$ = $\{N$, $Y$, $\Delta\}$. The Lagrangian for baryon interactions is \cite{Weber:2006ep, Glendenning:1997wn}
\begin{equation}
 {\cal{L}}_{Bariones} = \sum_B \overline{\psi}_B \left\{  \gamma_\mu\Big[i\partial^\mu  - g_{\omega B}(n)\omega^\mu - \frac{1}{2}g_{\rho B}(n)\pmb{\tau} \cdot \pmb{\rho}^\mu \Big] - \Big[m_b - g_{\sigma B}(n) \sigma\Big] \right\} \psi_B,
\end{equation}
where $g_{\sigma B}(n)$, $g_{\omega B}(n)$ and $g_{\rho B}(n)$ are the coupling constants between mesons and baryons, $n$ is the numerical baryon density and $\pmb{\tau} = (\tau_1,\tau_2, \tau_3)$ are the Pauli matrices and $\gamma^\mu$ the Dirac matrices. The meson-baryon coupling constants are taken to be density dependent, such that, given the nuclear saturation density $ n_0 $, they result
\begin{equation}
 g_{iB}(n) = g_{iB}(n_0) f_i (x),
\end{equation}
where $i$ scroll through the list of mesons ${\sigma, \omega, \rho}$, $x = n/n_0$ and the  $f_i$ give the functional form of the density dependence. Usually the prescription for these functions is given by \cite{Typel:1999yq}
\begin{equation}\label{f_i_sigmaomega}
 f_i(x) = a_i\frac{1 + b_i(x + d_i)^2}{1 + c_i(x + d_i)^2},
\end{equation}
para $i = {\sigma, \omega}$ y 
\begin{equation}\label{f_rho}
 f_\rho(x) = e^{\left[-a_\rho \left(x-1\right)\right]}.
\end{equation}

The functions parameters for the equations \eqref{f_i_sigmaomega} and \eqref{f_rho} \{$a_\sigma$, $b_\sigma$, $c_\sigma$, $d_\sigma$, $a_\omega$, $b_\omega$, $c_\omega$, $d_\omega$, $a_\rho$\}, the values of the meson-nucleon coupling constants at $n_0$ \{$g_{\sigma N}(n_0)$, $g_{\omega N}(n_0)$, $g_{\rho N}(n_0)$\} and the scalar mesons's mass $m_\sigma$ are all fitted to the saturation properties of nuclear matter at $n_0$, nd the properties of finite nuclei,including binding energies, charge and diffraction radius, spin-orbit spliting, etc.\cite{PhysRevC.71.024312, Typel:2005ba}. The lepton's $\lambda = \{e^-, \mu^-$\}  Lagrangian is
\begin{equation}
 {\cal{L}}_{Leptones} = \sum_\lambda \overline{\psi}_\lambda\left(i\gamma_\mu \partial^\mu - m_\lambda \right) \psi_\lambda,
 \end{equation}
 and the mesons's Lagrangian is given by
\begin{equation}
{\cal{L}}_{Mesones} = \frac{1}{2}\left(\partial_\mu \sigma\partial^\mu \sigma  - m_{\sigma}^2 \sigma^2 \right) - \frac{1}{4}\omega_{\mu \nu}\omega^{\mu \nu} + \frac{1}{2} m_{\omega}^2 \omega_\mu \omega^\mu + \frac{1}{2} m_{\rho}^2 \pmb{\rho}_\mu \cdot \pmb{\rho}^\mu - \frac{1}{4}  \pmb{\rho}_{\mu \nu}\cdot \pmb{\rho}^{\mu \nu},  
\end{equation}
where $\omega_{\mu \nu} = ( \partial_\mu \omega_\nu - \partial_\nu \omega_\mu ) $ y $\pmb{\rho}_{\mu \nu} = (\partial\mu \pmb{\rho}_{\nu} - \partial\nu \pmb{\rho}_{\mu} )$. The final Lagrangian of the system is given by

\begin{equation}
 {\cal{L}}_{DDRMF} = {\cal{L}}_{Bariones} +  {\cal{L}}_{Leptones} + {\cal{L}}_{Mesones}.
\end{equation}

The equations for the baryonic and mesonic fields are obtained by evaluating the Euler-Lagrange equations on the total Lagrangian of the system. After that, the relativistic mean field approximation is taken assuming that the matter is uniformly distributed and static in its state of minimum energy \cite{Glendenning:1997wn}. In this way, the \{$\sigma$, $\omega$, $\rho$\} fields, are replaced by their mean values \{$\overline{\sigma}$, $\overline{\omega}$, $\overline{\rho}$\}, and the baryon currents too. Taking into account this approximation, and the Euler-Lagrange equations, the following system of equations is obtained
\begin{eqnarray} \label{sist_hadrones1}
 \sum_B g_{\sigma_B}(n)\langle\overline{\psi}_B\psi_B\rangle - m_{\sigma}^2 \overline{\sigma} &=& 0 \\
 \sum_B g_{\omega_B}(n)\langle \psi^{\dagger}_B\psi_B\rangle - m_{\omega}^2 \overline{\omega} &=& 0 \\
 \sum_B g_{\rho_B}(n) I_{3B}\langle \psi^{\dagger}_B\psi_B\rangle - m_{\rho}^2 \overline{\rho} &=& 0 \\
 \sum_B \left( \frac{\partial g_{\omega B}}{\partial n} \langle \psi^{\dagger}_B\psi_B\rangle \overline{\omega}   + \frac{\partial g_{\rho B}}{\partial n} I_{3B}\langle \psi^{\dagger}_B\psi_B\rangle \overline{\rho} - \frac{\partial g_{\sigma B}}{\partial n} \langle\overline{\psi}_B\psi_B\rangle \overline{\sigma}         \right) - \Sigma_r &=& 0,
\end{eqnarray}
where $I_{3B}$ is the isospin projection of each baryon in direction 3. The expectation values of the baryon currents in the mean field approximation are the scalar densities, which at zero temperature are given by
\begin{eqnarray}
 n_{B}^S &=& \langle\overline{\psi}_B\psi_B\rangle = \frac{2J_B + 1}{2\pi^2}\int_{0}^{k_b} \frac{m_{B}^* (\overline{\sigma})}{\sqrt{k^2 + m_{B}^* (\overline{\sigma}) }}k^2 \,dk \nonumber \\
 &=& \frac{2J_B + 1}{2\pi^2}\left(\frac{m_{B}^*}{2}\right) \left[k_B E_B^* - {m_{B}^*}^2 \mathrm{log}\left(\frac{E_B^* + k_B}{m_{B}^*}  \right)    \right] ,
\end{eqnarray}
and the baryon number density (vector density)
\begin{equation}
 n_B = \langle \psi^{\dagger}_B\psi_B\rangle = \frac{2J_B + 1}{6 \pi^2}k_{B}^3,
\end{equation}
where $J_B$ is the spin and $k_B$ the Fermi momentum,
\begin{equation}
 m_{B}^* = m_B -  g_{\sigma B} (n) \overline{\sigma}, 
\end{equation}
is the effective mass and $E_B^* = \sqrt{k^2 + m_{B}^* (\overline{\sigma}) }$ is the Fermi energy of baryonic species B.

To determine the energy and pressure density at a fixed baryonic density, the equations of the fields must be solved together with the condition of electric charge neutrality, in the context in which we are interested in working, which is that of NSs, that is,
\begin{equation}
 \sum_B n_B q_B + \sum_\lambda n_\lambda q_\lambda = 0,
\end{equation}
where $q_B$ y $q_\lambda$ are the baryonic and lepton electric charges respectively. The baryon number must be conserved so that
\begin{equation}
 n - \sum_B n_B = 0.
\end{equation}
The Fermi momentum of the baryons is related to that of the neutron so that the matter is in chemical equilibrium, then it has to satisfy, at zero temperature,
\begin{equation}
 \mu_B = \mu_n - q_B \mu_e,
\end{equation}
where $\mu_n$  y $\mu_e$ are the neutron and electron's chemical potentials respectively, and the relationship between the Fermi energy of each baryon and its chemical potential is
\begin{equation}
 E_B^* = \mu_B - g_{\omega B}(n) \overline{\omega} - g_{\rho B}(n)I_{3B}\overline{\rho} - \Sigma_r.
\end{equation}

The fact that the  $\Sigma_r$ term appears in the Fermi energy, is what makes the Fermi moment of each baryon now depend on the local baryon density at each point in the system, which is why the last equation appears in the system \eqref{sist_hadrones1}. Together with charge neutrality and baryon number conservation, the system of six nonlinear equations that needs to be solved to find the mean values of the fields \{$\overline{\sigma}$, $\overline{\omega}$, $\overline{\rho}$\}, the Fermi moments \{$k_n$, $k_e$\} and the rearrangement energy $\Sigma_r$ are written as
\begin{eqnarray} \label{sist_hadrones2}
 \sum_B g_{\sigma_B}(n)n_B^S - m_{\sigma}^2 \overline{\sigma} &=& 0 \\
 \sum_B g_{\omega_B}(n)n_B - m_{\sigma}^2 \overline{\omega} &=& 0 \\
 \sum_B g_{\rho_B}(n) I_{3B}n_B - m_{\rho}^2 \overline{\rho} &=& 0 \\
 \sum_B \left( \frac{\partial g_{\omega B}}{\partial n} n_B \overline{\omega}   + \frac{\partial g_{\rho B}}{\partial n} I_{3B}n_B \overline{\rho} - \frac{\partial g_{\sigma B}}{\partial n} n_B^S \overline{\sigma}         \right) - \Sigma_r &=& 0 \\
 \sum_B n_B - n &=& 0 \\
 \sum_B n_B q_B + \sum_\lambda n_\lambda q_\lambda &=& 0.
\end{eqnarray}

Once this system of equations is solved, we are interested in calculating thermodynamic quantities. To calculate the pressure (or energy density) of the system, first the energy-moment tensor must be found, for which we start from the metric $ g_{\mu \nu} $ and the Lagrangian
\begin{equation}
 {\cal{T}}^{\mu \nu} = - g^{\mu \nu}{\cal{L}} + \sum_\phi \left( \frac{\partial{{\cal{L}}}}{\partial_\mu(\partial \phi)} \right) \partial^\nu \phi,
\end{equation}
with $\phi = \{\sigma, \omega, \rho\}$. As space-time on the scale of interactions in NS is practically flat ($\sim 1 fm$), we can approximate the metric by that of Minkowski $g_{\mu \nu} \sim \eta_{\mu \nu} = \mathrm{diag}(-1,1,1,1)$. Finally, for pressure, at zero temperature we have
\begin{eqnarray}
 P_{DDRMF} &=& \frac{1}{3}\sum_i \langle T_{ii}\rangle \nonumber \\
 &=& \frac{1}{3}\sum_B \frac{2J_B + 1}{2\pi^2}\int_0^{k_B}\frac{k^4\, dk}{\sqrt{k^2 + {m_B^*}^2 (\overline{\sigma})}} + \frac{1}{3\pi^2}\sum_\lambda \int_0^{k_\lambda}\frac{k^4\,dk}{\sqrt{k^2 + m_\lambda^2}} \nonumber \\
 &-& \frac{1}{2}\left[ m_{\sigma}^2 \overline{\sigma}^2  - m_{\omega}^2 \overline{\omega}^2  -  m_{\rho}^2 \overline{\rho}^2  \right] + n\Sigma_r.
\end{eqnarray}

Teniendo en cuenta que $P = -\Omega/V$, from this quantity, or even from the momentum energy tensor, it is possible to derive all the thermodynamic quantities of interest, however, as with quark models, it is necessary to extend the model at finite temperature to study the stages of proto-NSs, so the most relevant amounts will be explicitly written in section \ref{WaleckaTfinita}. It is necessary to remember that, as in the quark model, the charge neutrality equation serves to fix the electron and muon's chemical potential. The inclusion of neutrinos and the lepton fraction equation are exactly the same as those already mentioned in the section \ref{Estadios}. 

\section{RMF(L) models and used parametrizations}

So far we have seen the theoretical development of the density-dependent DDRMF hadronic model. However, non-density dependent RMF models can be slightly modified to achieve results similar to those in which all coupling constants are density dependent.

In general, RMF models are parameterized in order to reproduce certain properties of hadronic matter at nuclear saturation density $ n_0 $, such as the asymmetry energy  $J = E_{sym} (n_0)$. The problem with this approach is that it does not adjust the slope of that energy $ (L_0) $ to density $ n_0 $. As $ L_0 $ is an amount that was being delineated with greater precision over time \cite{Glendenning:1991es} and that could have important implications in the composition of the matter inside NSs, it is necessary to introduce modifications in this type of models. In particular, this can be achieved by including various self-interaction terms between mesons \cite{Costa:2014tpa}. The problem with this approach is that it softens too much the EoS, which does not allow it to meet the maximum mass requirements of two solar masses. Rather than adding more nonlinear terms to those already in the RMF models, another approach is to use density dependence only on isovectorial interactions, as can be seen from the equation \eqref{f_rho}, leaving the $ f_i $ of the equation \eqref{f_i_sigmaomega} equal to the unity . This is convenient, since the $ a_\rho $ coefficient can be adjusted in order to obtain good results for $ L_0 $, without modifying other saturation properties already existing in the RMF parameterization \cite{Drago:2014oja}. This modification in the parametrization of this type of models is known as RMF (L), and unlike the treatment we did previously, the non-linear interaction terms must be added to the Lagrangian, so that
\begin{eqnarray}
{\cal{L}}_{RMF(L)} &=& {\cal{L}}_{DDRMF} + {\cal{L}}_{NL\sigma}  \\
{\cal{L}}_{NL\sigma} &=&  - \frac{1}{3} \tilde{b}_\sigma m_N [g_{\sigma N}(n)
    \sigma]^3 - \frac{1}{4} \tilde{c}_\sigma [g_{\sigma N}(n) \sigma]^4. 
\end{eqnarray}

In our work, we use two types of models, one of the RMF (L) type known as GM1(L)\cite{Spinella:2018bdq}, and one of the DDRMF type whose parameterization is known as DD2\cite{Typel:2009sy}. In both cases the meson-hyperon coupling constants were determined following the prescription of the Nijmegen extension for soft nuclei (ESC08)\cite{ECSO_SU3}. The isovectorial meson-hyperon relative coupling constants were set taking into account properties concerning the hyperons isospin. For the  $\Delta$ resonances we used  $x_{{\sigma}  {\Delta}} = x_{{\omega} {\Delta}} = 1.1$ y $x_{{\rho} {\Delta}} =1.0$, where $x_{ i H} = g_{i H}/g_{i N}$. 

In Table \ref{table:parametrizations} we list the coupling constants of the models used in this work. In Table \ref{table:properties} se muestran las propiedades de los modelos, las cuales son: the nuclear saturation density $n_0$, the energy per nucleon $E_0$, the nuclear incompressibility constant $K_0$, the nucleon's effective mass $m^*/m_N$, the asymmetry energy $J$, the slope of the asymmetry energy $L_0$ and the nucleon's potential $U_N$.

\begin{table}[ht!]
\begin{center}
\begin{tabular}{|c|c|c|}
\hline %\noalign{\smallskip}
$~~${Parameters}$~~$ & $~~$GM1L$~~$
&DD2$~~$\\ \hline
$m_{\sigma}$  (GeV)    & 0.5500      & 0.5462     \\
$m_{\omega}$  (GeV)          &0.7830    & 0.7830  \\
$m_{\rho}$  (GeV)          & 0.7700        & 0.7630    \\
$g_{\sigma N}$             & 9.5722       & 10.6870    \\
$g_{\omega N}$            & 10.6180         & 13.3420     \\
$g_{\rho N}$            & 8.9830         & 3.6269   \\
$\tilde{b}_{\sigma}$         &0.0029                  & 0         \\
$\tilde{c}_{\sigma}$         &- 0.0011                 &   0       \\
$a_{\sigma}$         &1          &1.3576   \\
$b_{\sigma}$         & 0                  & 0.6344          \\
$c_{\sigma}$         & 0                 &  1.0054        \\
$d_{\sigma}$         & 0                 & 0.5758         \\
$a_{\omega}$         &0           &1.3697   \\
$b_{\omega}$         & 0               &0.4965          \\
$c_{\omega}$         & 0                 &  0.8177        \\
$d_{\omega}$         &0                  & 0.6384         \\
$a_{\rho}$         &0.3898           &0.5189   \\ \hline
\end{tabular}
  \caption[GM1L and DD2 parametrization values]{Parameters for the  RMF(L) models with the GM1L parameterization and DDRMF with the DD2 parameterization used in this work.}
\label{table:parametrizations}
\end{center}
\end{table}

\begin{table}[htb]
\begin{center}
\begin{tabular}{|c|c|c|}
\hline %\noalign{\smallskip}
$~~${Saturation properties}$~~$ & $~~$GM1L$~~$
&DD2$~~$\\ \hline
$n_0$  (fm$^{-3}$)    & 0.153      & 0.149     \\
$E_0$  (MeV)          & -16.30    & -16.02  \\
$K_0$  (MeV)          & 300.0        & 242.7    \\
$m^*/m_N$             & 0.70       & 0.56    \\
$J$    (MeV)          & 32.5         & 32.8     \\
$L_0$  (MeV)          & 55.0         & 55.3    \\
$-U_N$ (MeV)        &65.5           &75.2   \\ \hline
\end{tabular}
 \caption[Properties of nuclear matter for the GM1 (L) and DD2 parameterizations.] {Nuclear matter properties for the GM1(L)\cite{Spinella:2018bdq} and DD2\cite{Typel:2009sy}.}
\label{table:properties}
\end{center}
\end{table}

\section{Finite temperature extension}
\label{WaleckaTfinita}

The extension of the model at finite temperature can be done in a simple way, taking into account that we are working with fermions. The difference between the fermions that we treat in this model and the free fermions that we regularize in the previous chapter, lies only in that those of the hadronic model have a shift in the chemical potential and the effective mass, that is,
\begin{eqnarray}
 \mu_B^* &=& \mu_B - g_{\omega B}(n) \overline{\omega} - g_{\rho B}(n)I_{3B}\overline{\rho} - \Sigma_r \\
 m_{B}^* &=& m_B -  g_{\sigma B} (n) \overline{\sigma}.
\end{eqnarray}

It is worth noting that what it was the Fermi energy $E_B^*$  in the case of zero temperature, now becomes the modified chemical potential $\mu_B^*$.  This is understood directly since in the case of zero temperature the chemical potential is equal to the Fermi energy, and in the extension at finite temperature this does not hold. In this case, the energy of each baryon is still $E_B^* (k) = \sqrt{k^2 + m_{B}^* (\overline{\sigma}) }$ but it does not have to be equal to the chemical potential. After making these distinctions, the treatment is the same as for free fermions: construct the partition function, perform the Matsubara sums, and regularize in order to eliminate the divergent term from the vacuum. For simplicity we will not repeat these calculations, since as we said they are equivalent to those of free fermions. Then the pressure is written in the form

\begin{eqnarray}
 P_{DDRMF}(T) &=&  \frac{1}{3}\sum_B \frac{2J_B + 1}{2\pi^2}\int_0^{\infty}\frac{k^4\, dk}{E_B^* (k)}\left[n_B^+ + n_B^- \right] + \frac{1}{3\pi^2}\sum_\lambda \int_0^{\infty}\frac{k^4\,dk}{E_\lambda(k)}\left[n_\lambda^+ + n_\lambda^- \right] \nonumber \\ 
 &-& \frac{1}{2}\left[ m_{\sigma}^2 \overline{\sigma}^2  - m_{\omega}^2 \overline{\omega}^2  -  m_{\rho}^2 \overline{\rho}^2  \right] + n\Sigma_r,
\end{eqnarray}
with dependence on temperature given by occupancy numbers
\begin{eqnarray}
 n_B^\pm(k,T) &=& \left[ 1 + e^{\frac{E_B^*(k) \pm \mu_B^*}{T}}\right]^{-1} \\
 n_\lambda^\pm(k,T) &=& \left[ 1 + e^{\frac{E_\lambda(k) \pm \mu_\lambda}{T}}\right]^{-1},
\end{eqnarray}
with $E_\lambda$ and $\mu_\lambda$ the energy and chemical potential of each lepton respectively, where now instead of the Fermi moments as unknowns variables we will have the chemical potentials of each particle. The rest of the thermodynamic quantities, again, can be obtained from both the impulse energy tensor and the pressure. For completeness, and because it will be necessary later to work with systems with fixed entropy, we calculate this quantity using the fundamental thermodynamic relationship, obtaining
\begin{eqnarray}
 S(T) &=& \frac{\partial P}{\partial T} = \nonumber \\ 
  &=& \sum_B \gamma_B\int_0^{\infty}\frac{k^4\, dk}{E_B^* }\left[\left(n_B^+ - {n_B^+}^2 \right) \left(\frac{E_B^* + \mu_B^*}{T^2}\right) + \left(n_B^- - {n_B^-}^2 \right) \left(\frac{E_B^* - \mu_B^*}{T^2}\right)\right] \nonumber \\ 
  &+& \sum_\lambda \gamma_\lambda\int_0^{\infty}\frac{k^4\, dk}{E_\lambda }\left[\left(n_\lambda^+ - {n_\lambda^+}^2 \right) \left(\frac{E_\lambda + \mu_\lambda}{T^2}\right) + \left(n_\lambda^- - {n_\lambda^-}^2 \right) \left(\frac{E_\lambda - \mu_\lambda}{T^2}\right)\right], 
\end{eqnarray}
being $\gamma_B = (2J_B + 1)/(6\pi^2)$ y $\gamma_\lambda = 1/(3\pi^2)$ the respective degeneracy factors. The last two thermodynamic quantities necessary to calculate the EoS are obtained from the fundamental thermodynamic relation, being then
\begin{eqnarray}
 \varepsilon &=& -P + TS + \sum_i \mu_i n_i \\
 E_{G/B} &=& \frac{\varepsilon - TS + P}{n},
\end{eqnarray}
where the  $i$ subindex runs over all particle flavors (baryons and leptons). The quantity  $\varepsilon$ is the system's total energy density, and $E_{G/B}$ is the Gibbs energy per baryon.

\begin{figure}[ht!]
\begin{center}
\subfigure{\includegraphics[width=0.45\textwidth]{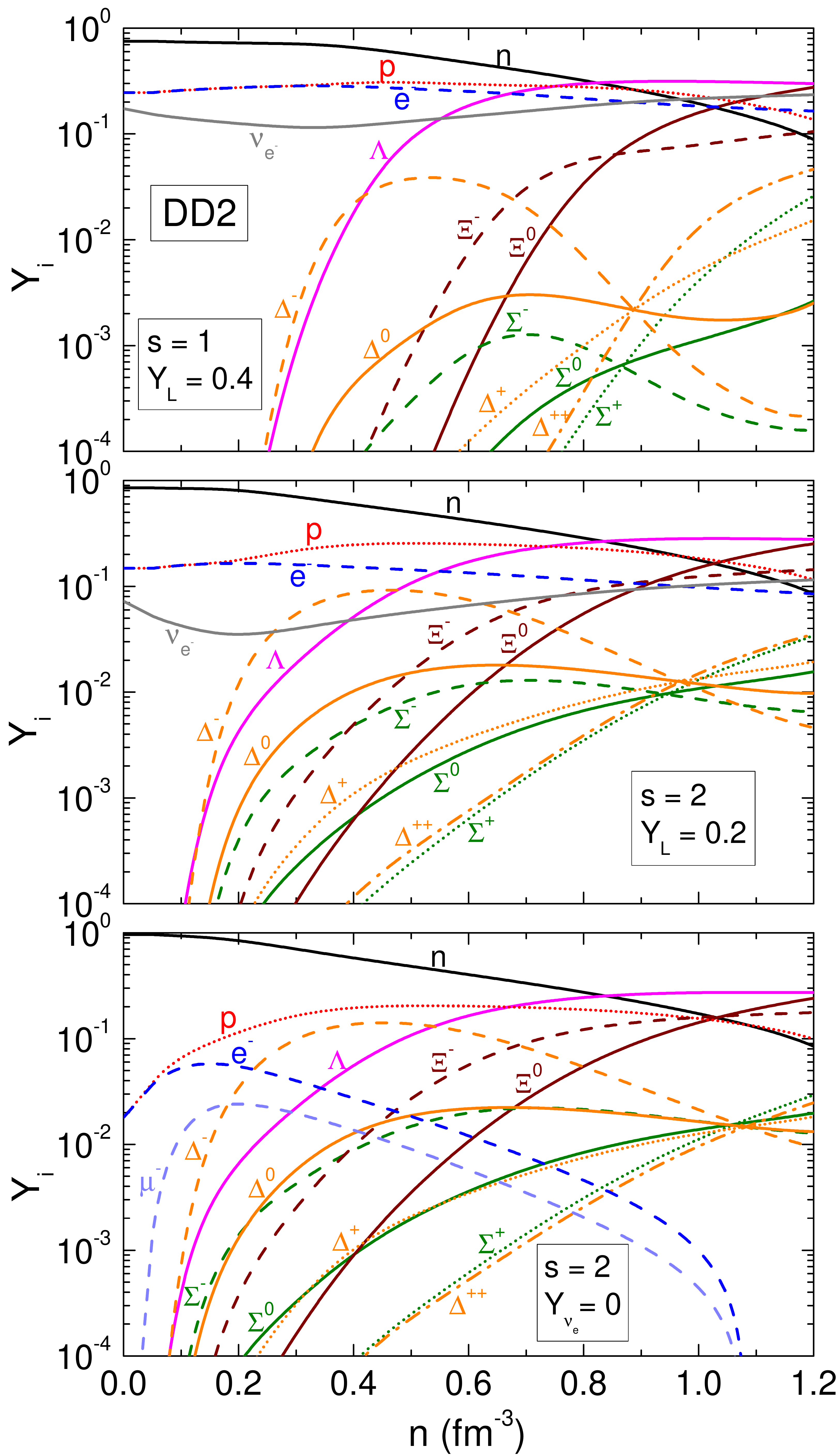}}
\hspace{0.02\textwidth}
\subfigure{\includegraphics[width=0.45\textwidth]{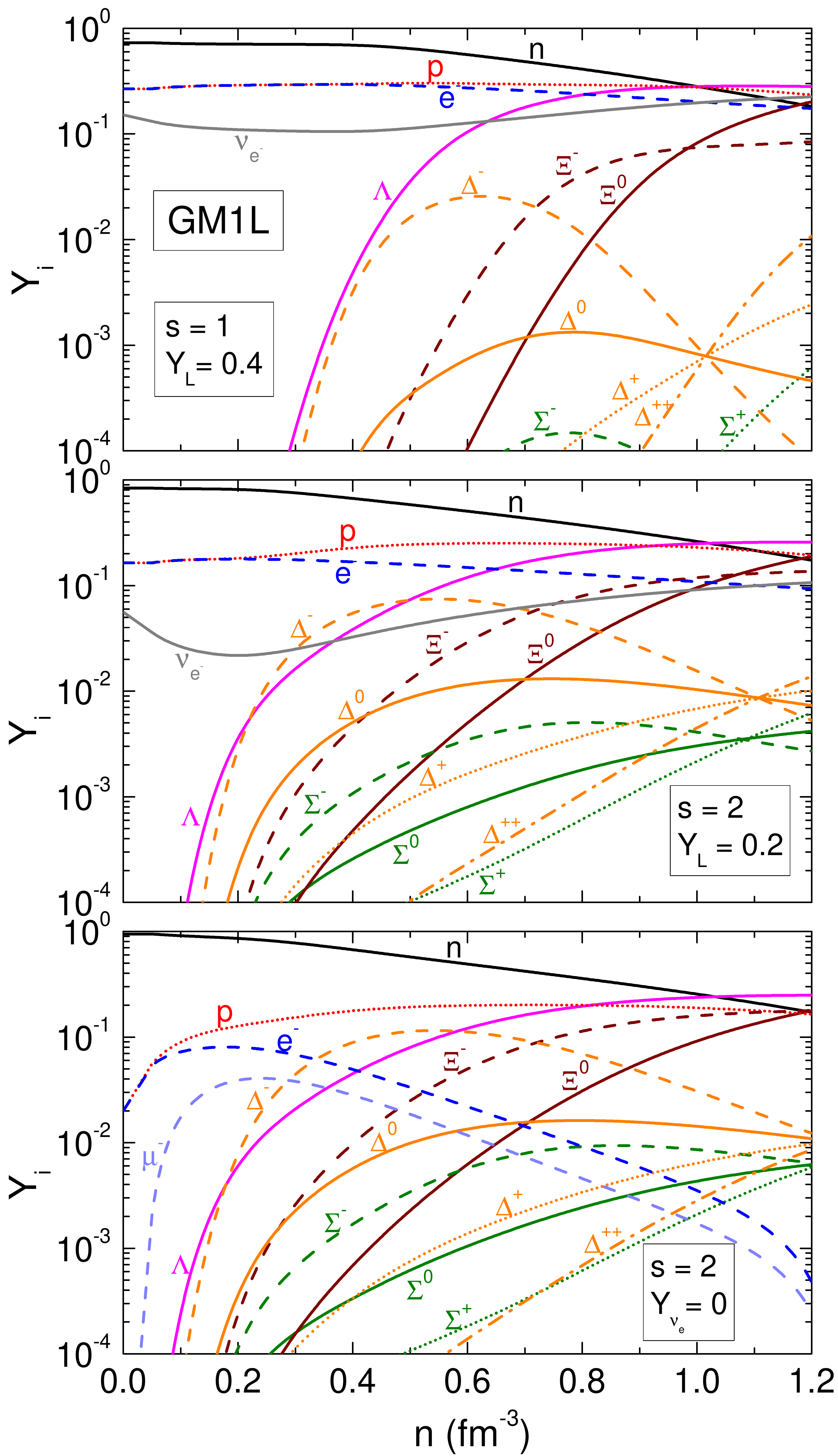}}
\caption[Particle population as a function of density for the DD2 and GM1L parametrizations.]{Left: Particle population as a function of density for the DD2 parameterization. The entropy and lepton fraction values chosen are typical of the interior of proto-neutron stars. Right: the same as the left panel but for the GM1L parameterization}
\label{part_pop}
\end{center}
\end{figure}

Finally, with the finite temperature extension, we are in a position to calculate the hadronic EoS both at finite temperature and at fixed entropy. It is useful to see how the hadronic model behaves with respect to what types of particles start to emerge as the density increases. Figure \ref{part_pop} shows the particle populations for the EoS used in this work.
It can be seen that the particle population strongly depends on the entropy per baryon $s=S/n$ and the lepton fraction$Y_L$. Particularly for $ \Delta $ resonances, the negative state of this particle is populated first, replacing some of the high-energy electrons. The other three states ($\Delta^0$, $\Delta^+$, y $\Delta^{++}$) are populated at densities that are slightly greater than the nuclear saturation density. Therefore, according to this model, all these states exist in the nuclei of the proto-NS. On the other hand one can see the abundance of electrons in matter where the lepton fraction is different from zero, and neutrinos are present (upper and middle panels of Figure \ref{part_pop}). Due to this, it could be speculated that the electrical conductivity of this type of matter is considerably different from the electrical conductivity of matter free of neutrinos (lower panels of the same figure), where the presence of muons leads to a lower electron density in the system, and the increasing population of the  $\Delta^-$, $\Xi^-$, and $\Sigma^-$  particles causes a reduction in the lepton numbers. Regarding the strangeness contained in the hyperons, their main contribution comes from the $ \Lambda $ and $ \Xi $ particles, whose population grows monotonically with density, dominating the composition of matter at very high densities. Other species of hyperons are present, but to a lesser degree.

\graphicspath{{Resultados/}}
\chapter{\label{ch:Resultados}Neutron stars with quark matter: results}

So far we have developed all the necessary components to model matter in the inner cores of NSs. We have described the effective hadronic and quark models and obtained the corresponding EoS. In Chapter \ref{ch:Quarks_y_contexto_astrofisico}, we have presented the hydrostatic equilibrium equations for the NSs that allow us, given an EoS, to obtain the mass, the radius and the baryonic mass of each of the stars that make up the compact objects family that correspond to that EoS. Let's see then what criteria we use to build hybrid EoS. In principle, if there were a phase transition from hadronic matter to quark matter inside the NSs, it could be both abrupt (which can be described using the Maxwell formalism), and smooth (which can be described using the Gibbs formalism), depending on the surface tension at the interface of quarks and hadrons. At this point we have a problem regarding which formalism to use, since the value of the surface tension is still indeterminate. \textit {Lattice QCD} calculations predict a surface tension value $\sim$ $0$ $\mathrm{MeV/fm^2}$ -$100$ $\mathrm{MeV/fm^2}$ \cite{KAJANTIE1991693}. According to theoretical studies, a surface tension greater than $ 70 $ $ \mathrm {MeV/fm^2} $ would favor a Maxwell-type phase transition rather than a Gibbs-type transition. Instead of analyzing the surface tension value at the hadron-quark interface, it is possible to consider both types of phase transition in the internal nuclei of the NSs and analyze if there is any observational consequence that allows us to differentiate them. Likewise, since we are interested in creating a schematic thermal evolution in these types of objects, we also need to calculate the different isentropic EoEs corresponding to the different stages of the stars that make up each family. In what follows, we will show the results obtained by applying the quark matter and hadronic matter models described in previous chapters to the context of proto-NSs and cold NSs. We will call hybrid stars to the NSs with quark matter in their inner cores.

\section{Hybrid stars with abrupt phase transitions}

To study hybrid stars that have an abrupt phase transition in their cores, we will construct these transitions according to the Maxwell formalism for different values of vector interaction in the quark model, considering the DD2 and GM1L parameterizations of the hadronic model, starting from zero temperature. Under the premise of complying with the observational restriction of 2 $ M_{\odot} $, corresponding to the PSR J1614-2230 and PSR J0348 + 0432 pulsars \cite{Demorest:2010bx,Lynch:2012vv,Antoniadis:2013pzd,Arzoumanian:2017puf}, and assuming the possibility that quark matter exists inside the NS, the maximum mass of the family of stars is calculated for different values of vector interaction, $\mathrm{\zeta_v}$. This leads to $0.331<\mathrm{\zeta_v}<0.371$ for GM1L, and $0.328<\mathrm{\zeta_v}<0.385$ for DD2. The $\mathrm{\zeta_v}$ lower limit in each case, corresponds to the minimum vector interaction value, necessary to comply with the  $2M_{\odot}$ restriction.  We define the upper limit by finding the maximum vector interaction for which quark matter exists inside the star. That is, $\mathrm{\zeta_v}$ values greater than 0.371 or 0.385 for hybrid EoS built with GM1L or DD2 parameterizations respectively, result in purely hadronic stars. Once the values of the vector interaction for quark matter at zero temperature have been fixed, different EoS are calculated at fixed temperature and/or entropy and different isentropic stages for the different families of stars are analyzed. The same analysis is also carried out with the Gibbs formalism.

\begin{figure}[ht!]
\begin{center}
\subfigure{\includegraphics[width=0.45\textwidth]{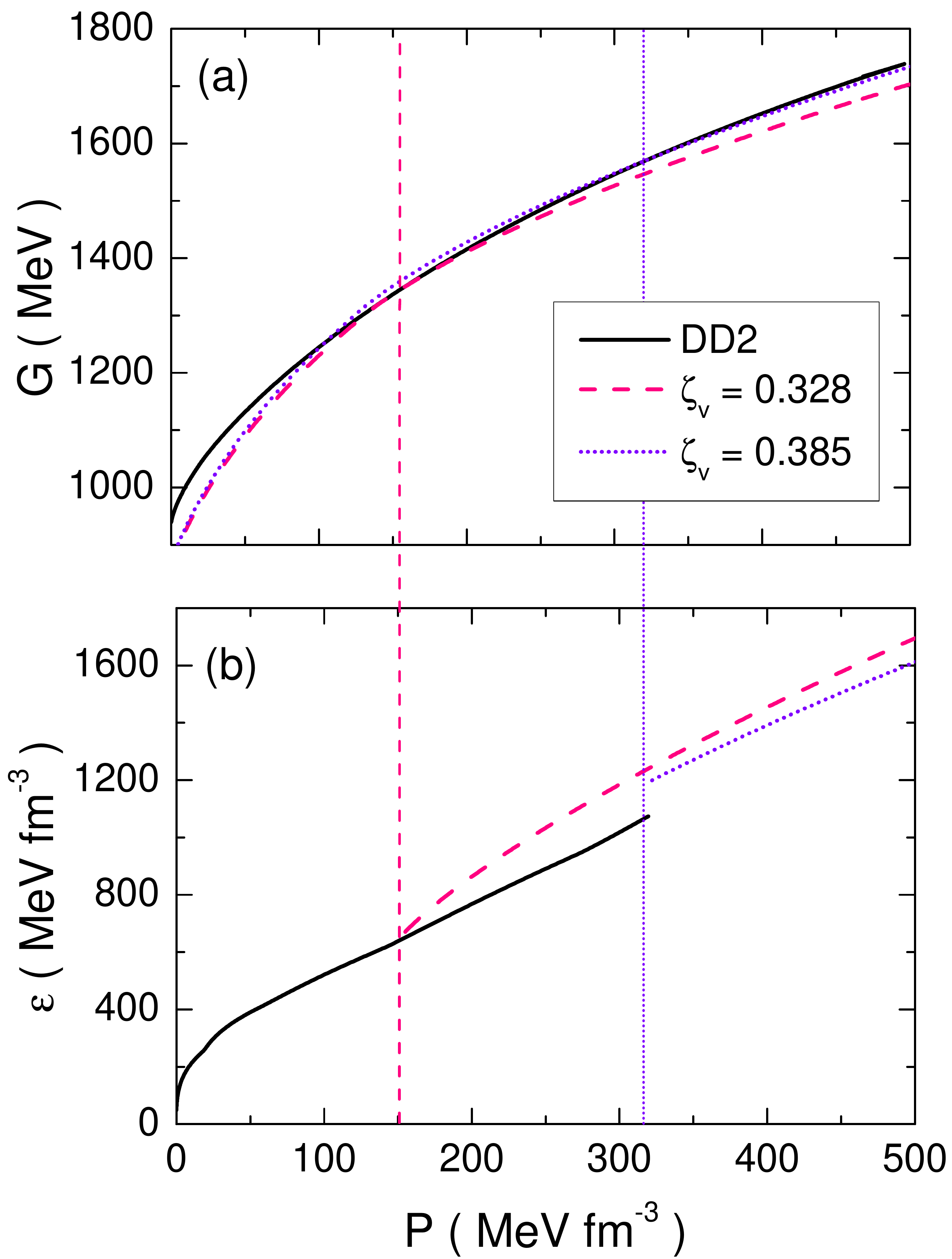}}
\hspace{0.02\textwidth}
\subfigure{\includegraphics[width=0.45\textwidth]{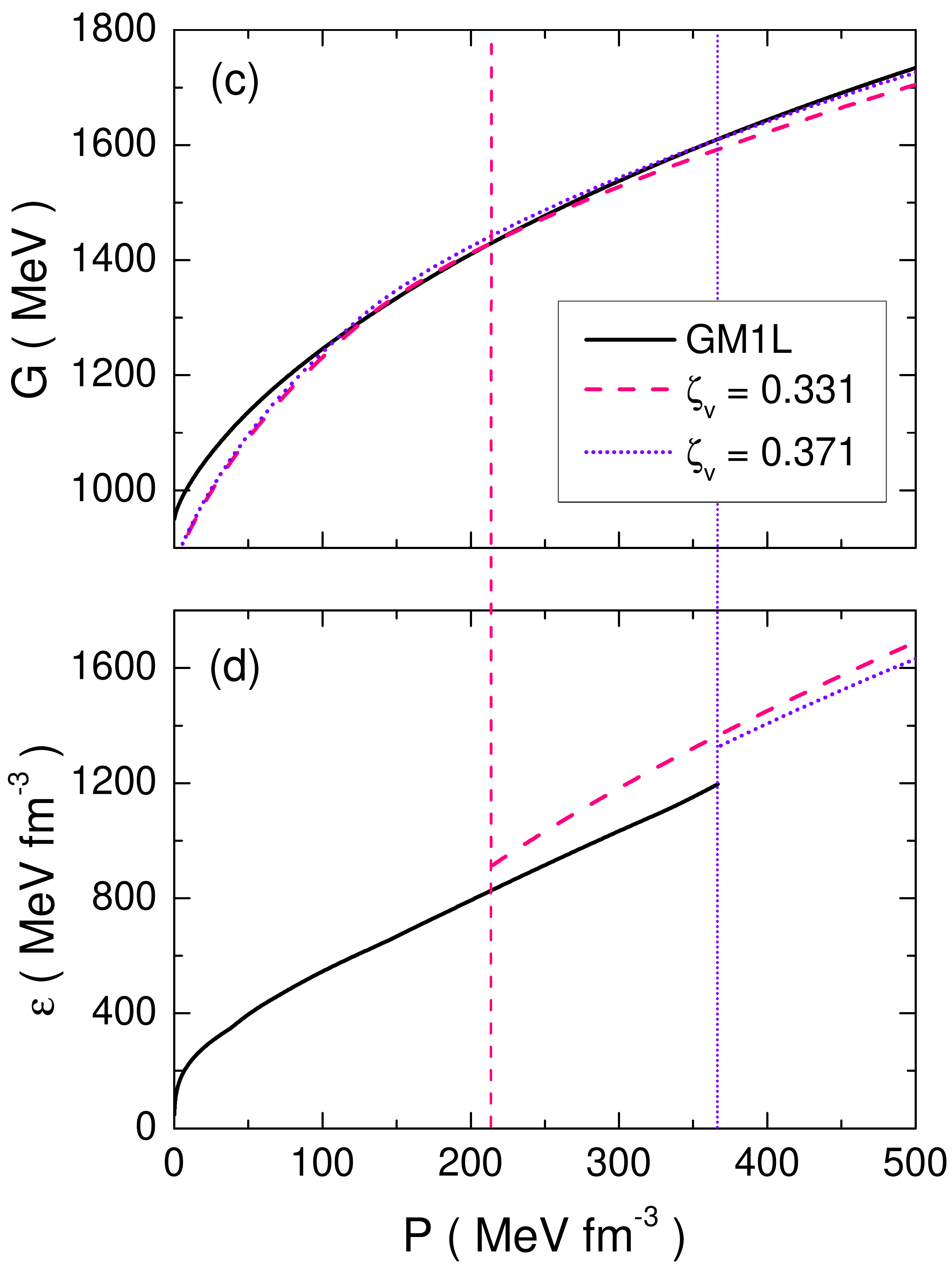}}
\caption[Construction of hybrid EoS using Maxwell formalism for the chosen parameterizations.]{Panel (a) shows the construction of the EoS (at $ T = 0 $), the Gibbs energy per baryon $ (G) $ for a hybrid parameterization DD2-nl-PNJL. The black line corresponds to the EoS for the hadronic phase and the red and blue dotted lines are the EoS for the quark phase, using the two values of vector interactions. Panel (b) shows the energy density as a function of pressure, for the mentioned values. Panels (b) and (c) show the same as panels (a) and (b) respectively, for the GM1L parameterization.}
\label{EoS_maxwell_DD2_GM1L}
\end{center}
\end{figure}

\begin{figure}[ht!]
\begin{center}
\includegraphics[width=0.49\textwidth]{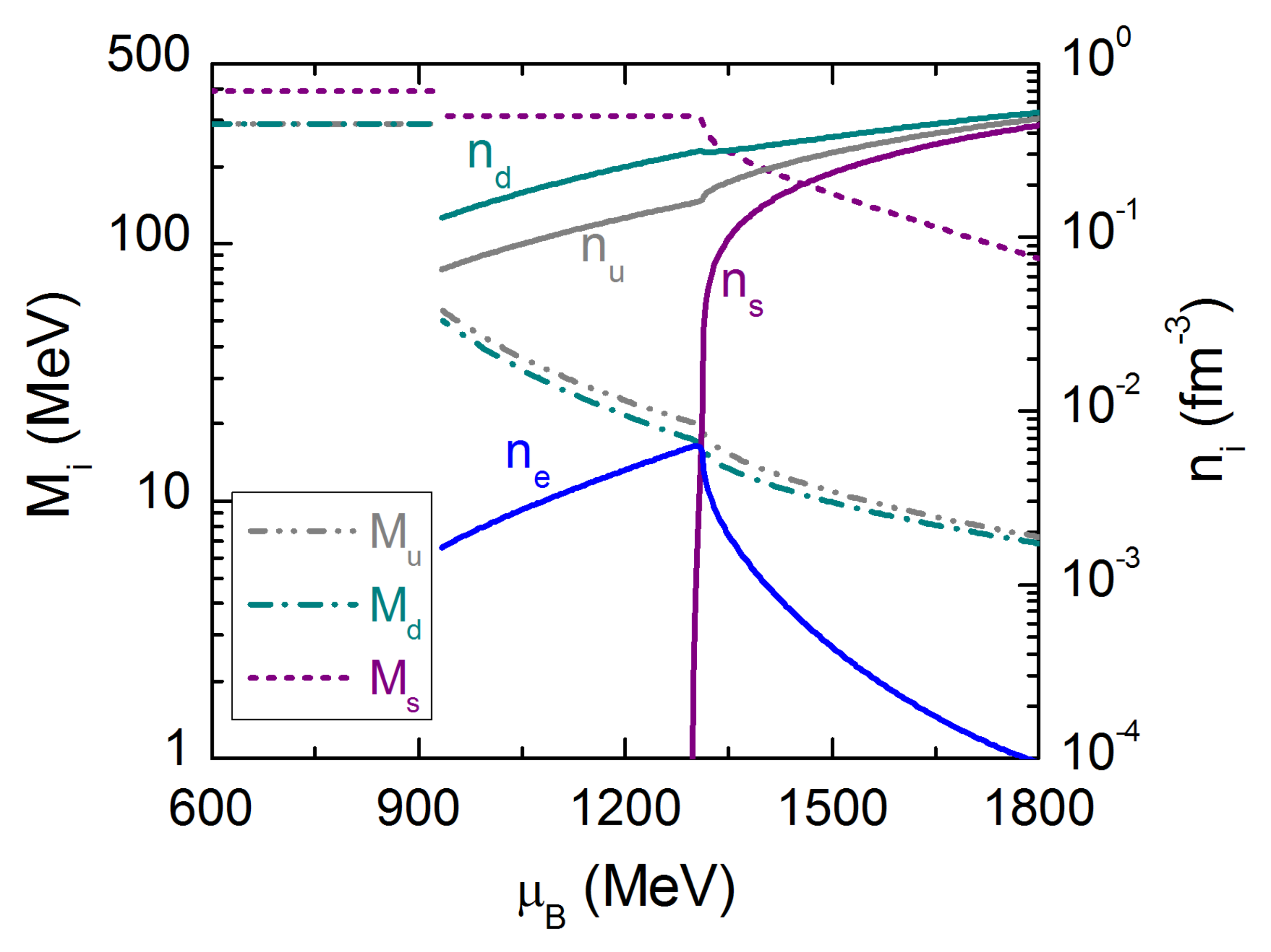}
\caption[Dynamic masses $M_i$ and densities $n_i$ fors $u,d$ and $s$ quarks as a function of baryonic chemical potential].{Dynamic masses $M_i$ and densities $n_i$ fors $u,d$ and $s$ quarks as a function of baryonic chemical potential. $n_e$ corresponds to electron density. }
  \label{masses_gibbs}
\end{center}
\end{figure}

For the construction of the phase transition, we look for the point where the hadronic and quark EoS in the ($p,E_G$) plane are equal, as explained in section \ref{construcciones}. The crosses for the curves corresponding to the mentioned vector interactions are shown in Figure \ref{EoS_maxwell_DD2_GM1L}. In that figure, it can be seen that there are two visible transitions. The first from quark matter to hadronic matter at pressure  $P \sim 100 -150$ MeV/fm$^3$, and the second a from hadronic matter to quark matter at $P \sim 350 - 400$ ~ MeV/fm$^3$. We take as the physical transition point the second mentioned crossing, since the first one is not possible to occur. The EoS are quite similar and it is difficult to distinguish between the two phases in the range of pressures  $P \sim 100 - 400$ MeV/fm$^3$, therefore, in order to find the crossing point, the precision of the graphics in that area must be increased several times..
On the other hand, it can be seen in Figure \ref{masses_gibbs}, that a first-order phase transition occurs at $\mu_B \sim$ 940 MeV, indicated by a discontinuity in particle density and dynamic masses. This corresponds to the region where the $ s $ quark  has not yet been deconfined, as mentioned in section \ref{const_diag_fases}. Therefore, for chemical potentials between 940 MeV and 1300 MeV we have a phase where the $u$ y $d$ quarks chiral condensates $\langle \bar u\, u\rangle\, =\, \langle \bar d\,d\rangle \,\sim \,0$,  while  $\langle \bar s\,s\rangle \neq 0$. This behavior could indicate the existence of a phase that has aspects of both nuclear and quark matter. (see \cite{McLerran:2007qj, Baym:2017whm} and references therein), causing a kind of masking between the two EoS \cite{Alford:2004pf} and could explain the first phase transition from quarks to hadrons, which turns out to be non-physical, since quark matter is not completely deconfined. At chemical potentials greater than $\mu_B$ $\sim 1300$ MeV and pressures  P $\sim 135$ $\mathrm{MeV/fm^3}$, the $s$ quark is deconfined, and the EoS corresponds to free quark matter, which is the part that we will use to build the hybrid EoS.

\begin{figure}[ht!]
\begin{center}
\includegraphics[scale = 0.45]{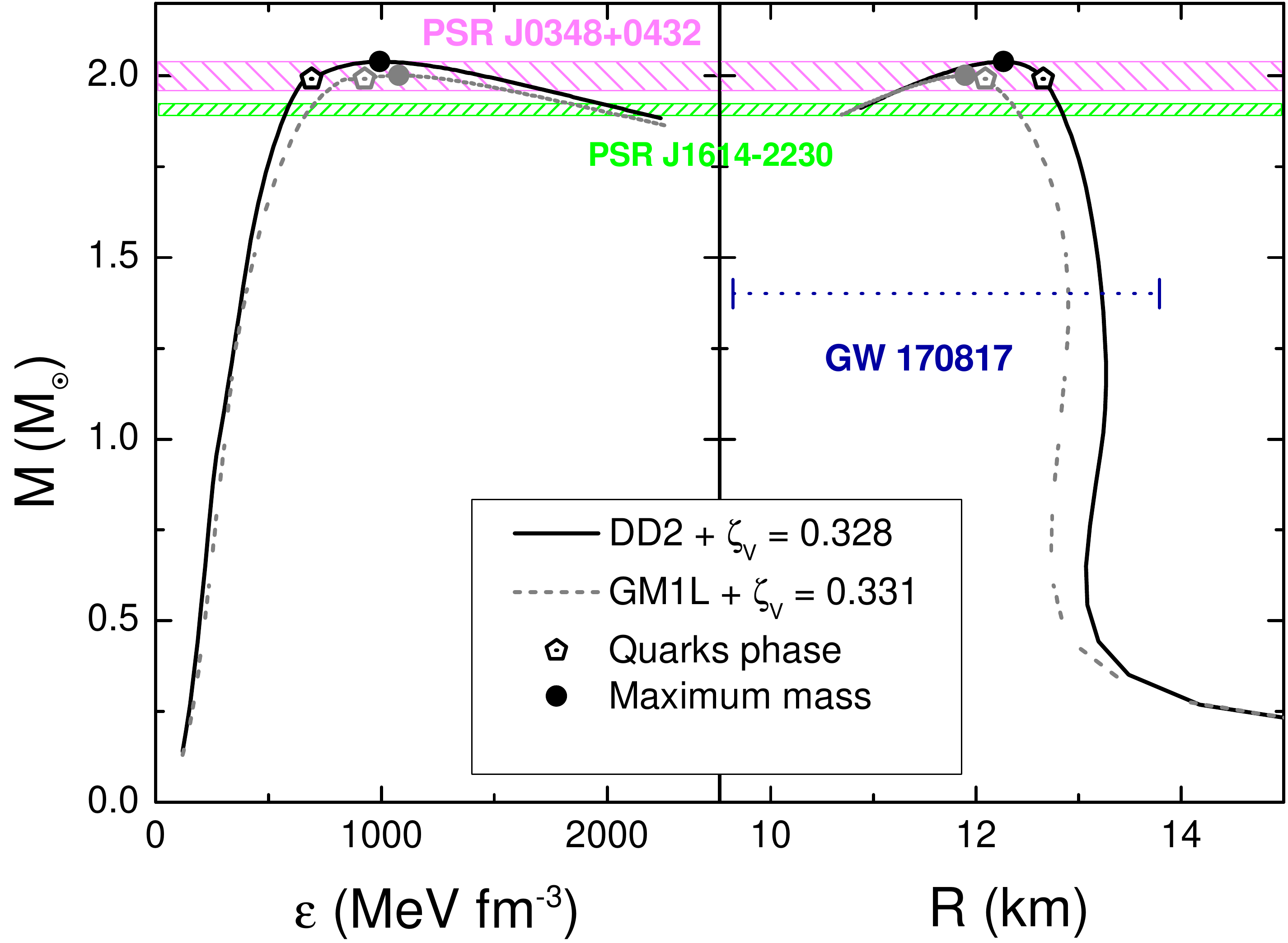}
\caption[Gravitational mass versus radius and energy density with hybrid EoS without color superconductivity.]{Gravitational mass curves as a function of energy and stellar radius for the parameterizations at $ T = 0 $ without color superconductivity.}
  \label{masas_radios_T0_NQM}
\end{center}
\end{figure}

\begin{figure}[ht!]
\begin{center}
\includegraphics[scale = 0.45]{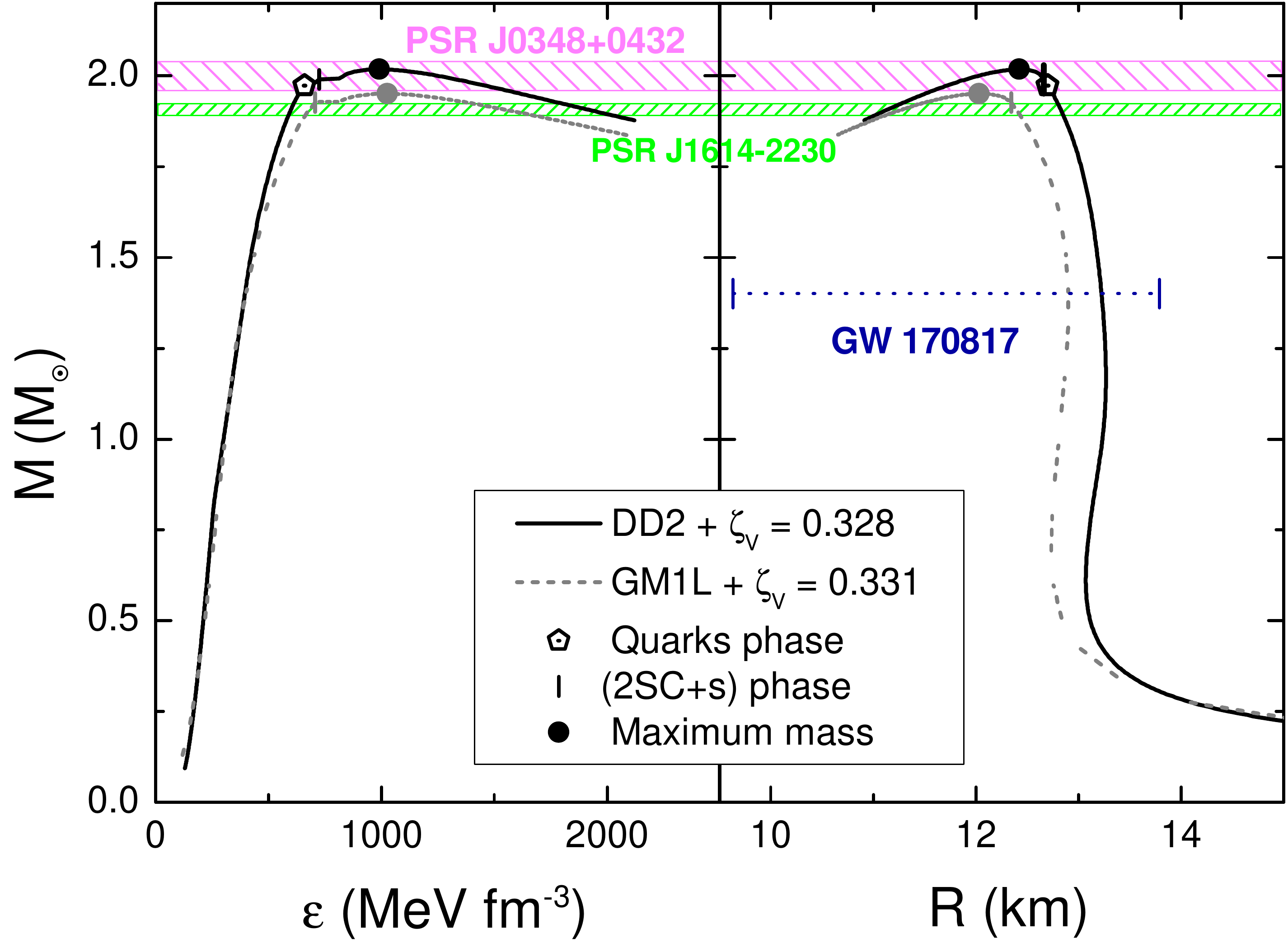}
\caption[Gravitational mass versus radius and energy density with color superconducting hybrid EoS.]{Gravitational mass curves as a function of energy and stellar radius for the parameterizations at $ T = 0 $ with color superconductivity. In the case of the GM1L parameterization and $ \zeta_V = 0.331 $ the quark phase begins at the same point as the ($ 2SC + s $) phase.}
  \label{masas_radios_T0_2SC}
\end{center}
\end{figure}

Once the hybrid EoS is constructed, we solve the TOV equations to find the mass-radius and mass-energy density curves for the NSs families corresponding to that EoS. The mass-energy density curve plays an important role in the analysis of hybrid stars: By identifying in the EoS the energy density value where quark matter appears, it is then possible to identify, in the mass-energy density results, the starting mass values from which the stars contain quark matter. In the same way, in the case of considering color superconductivity, it is possible to identify from which point this phase could be present in the NSs obtained. The results of this analysis for zero temperature, for the mentioned parameterizations, and adding the construction for the hybrid EoS at $ T = 0 $ are shown in Figure \ref{masas_radios_T0_NQM}, while in Figure \ref{masas_radios_T0_2SC}, in addition, the color superconducting phase ($2SC+s$) was included. In all the analyzes that follow, we will stick to the values corresponding to the lower limit of the calculated vector interactions, since these values favor a greater formation of quark matter inside hybrid stars.

\begin{figure}[ht]
\begin{center}
\includegraphics[scale = 0.45]{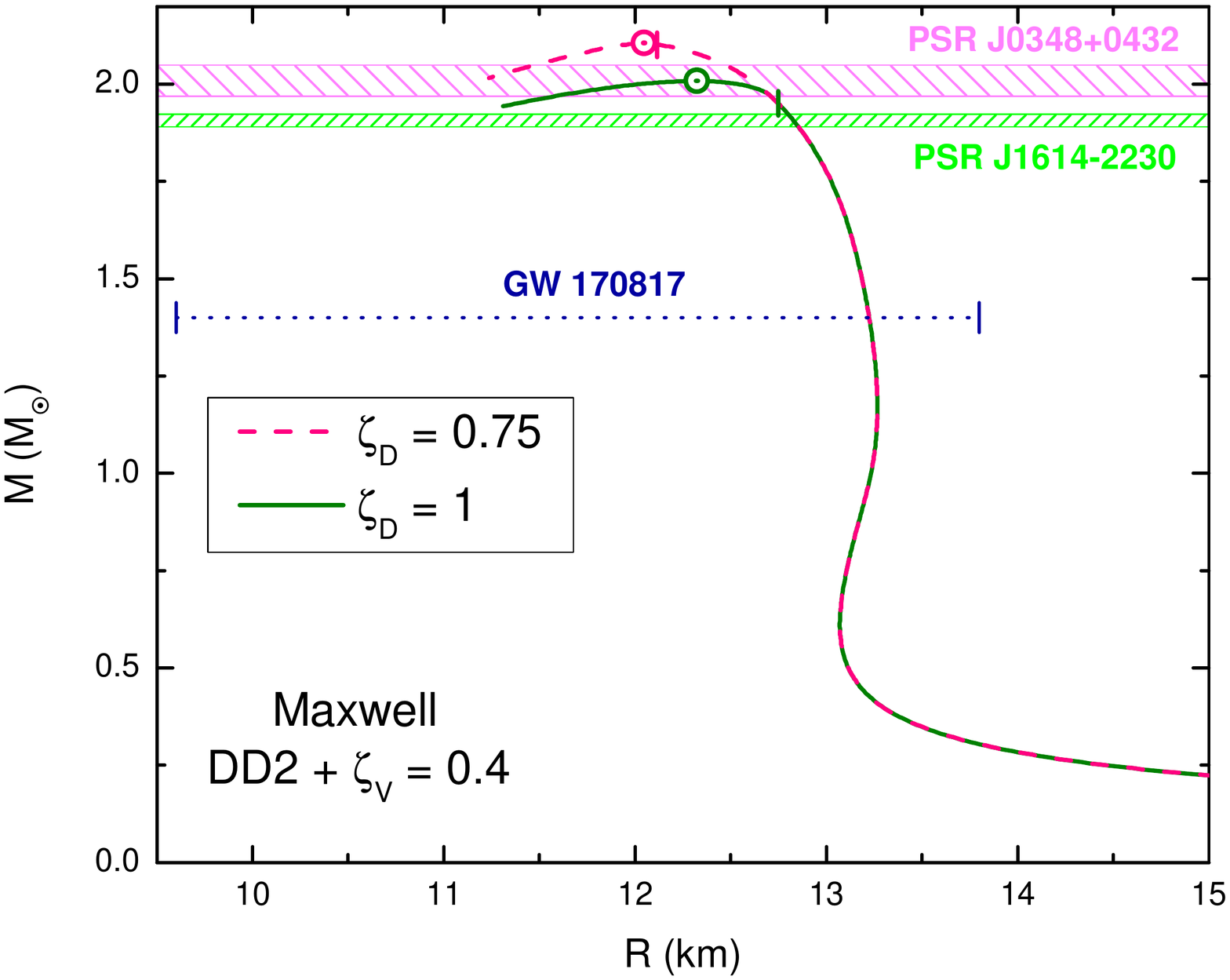}
\caption[Gravitational mass $ M_G $, as a function of the radius for the DD2 parameterization and $ \zeta_V = 0.4 $, with different coupling values for diquarks.]{Gravitational mass $ M_G $, as a function of the radius for the DD2 parameterization and $ \zeta_V = 0.4 $, with different coupling values for diquarks, the vertical lines indicate the beginning of the ($ 2SC + s $) phase, which coincides with the beginning of the quarks phase. }
  \label{Gv04Maxwell}
\end{center}
\end{figure}

It can be seen that all curves in figures \ref{masas_radios_T0_NQM} and \ref{masas_radios_T0_2SC} meet the requirement of reaching the two solar masses and also satisfy the bound established for the NSs radii. This last bound is established from the data analysis from the merger of two NSs, the gravitational waves event known as  GW170817 \cite{PhysRevLett.121.161101}. It can also be seen that the families of stars built with the hadronic parameterization corresponding to DD2 have a more extensive branch of hybrid stars than those constructed with the GM1L parameterization. This is so because DD2 hadronic EoS is stiffer, that is, matter is less compressible  than GM1L in terms of Gibbs free energy, so the transition to quark matter occurs at lower densities. We also see that in the case of the EoS with color superconductivity, first there is a transition to quark matter without superconductivity, which then transitions to the superconducting phase. This is so exclusively because we choose to stick with vector coupling values that are comparable to constructions without considering superconductivity. As we will see later, by varying that coupling it is possible to build a phase that goes from being pure hadronic to a quark color superconducting phase. In Table \ref{tabla_MR_MB} results are shown for the maximum masses obtained in each case, where it can be seen that the fact of including superconductivity slightly decreases the maximum mass obtained. On the other hand, it is necessary to mention that it is possible to find other parameterizations that meet observational requirements of two solar masses. As seen in Figure \ref{Gv04Maxwell}, for example, if the vector interaction is increased, the onset of appearance of a quark core is delayed, but if the interaction of diquarks is increased at the same time, it is possible to satisfy the two solar masses condition and that quark matter is present in the interior of the star at much lower densities. However, this slightly reduces the maximum mass obtained. In this work, as we mentioned earlier, we will consider the coupling constant value most used in the literature, $\zeta_D = G_D/G_S = 0.75$ \cite{Buballa:2005}. 

\begin{table}[ht!]
\begin{center}
\begin{tabular}{|c|c|c|c|c|}\cline{1-4}
% use packages: color,colortbl
\multicolumn{4}{|c|}{$\mathrm{GM1L}$} \\\hline
& $M_{G}~[M_{\odot}]$ & $M_{B}~[M_{\odot}]$ & $\epsilon_{c}
            ~ [\mathrm{MeV/fm^{3}}]$ \\\hline
$\mathrm{Pure\,\,hadronic}$ & $2.04$ & $2.42$ & $1194.82$ \\
$\mathrm{\zeta_v} = 0.331$ & $2.00$ & $2.36$ & $1077.02$ \\
$(2SC+s) \mathrm{\zeta_v} = 0.331$ & $1.95$  & $2.30$ & $1026.53$ \\\hline
\multicolumn{4}{|c|}{$\mathrm{DD2}$} \\\hline
& $M_{G}~ [M_{\odot}]$ & $M_{B}~ [M_{\odot}]$ & $\epsilon_{c}
            ~[\mathrm{MeV/fm^{3}}]$ \\\hline
$\mathrm{Pure\,\, hadronic}$ & $2.11$ & $2.53$ & $1110.68$ \\
$\mathrm{\zeta_v} = 0.328$ & $2.04$ & $2.43$ & $992.88$ \\
$(2SC+s)\mathrm{\zeta_v} = 0.328$ & $2.02$ & $2.40$ & $948.00$ \\\hline
\end{tabular}
\caption[Gravitational mass and baryon mass results for Maxwell transitions]{Gravitational mass $ M_G $, baryonic mass $ M_B $, and central energy density $ \epsilon_c $ for the maximum mass stars obtained for the parameterizations at $ T = 0 $.}
\label{tabla_MR_MB}
\end{center}
\end{table}

\begin{figure}[ht]
\begin{center}
\includegraphics[scale = 0.45]{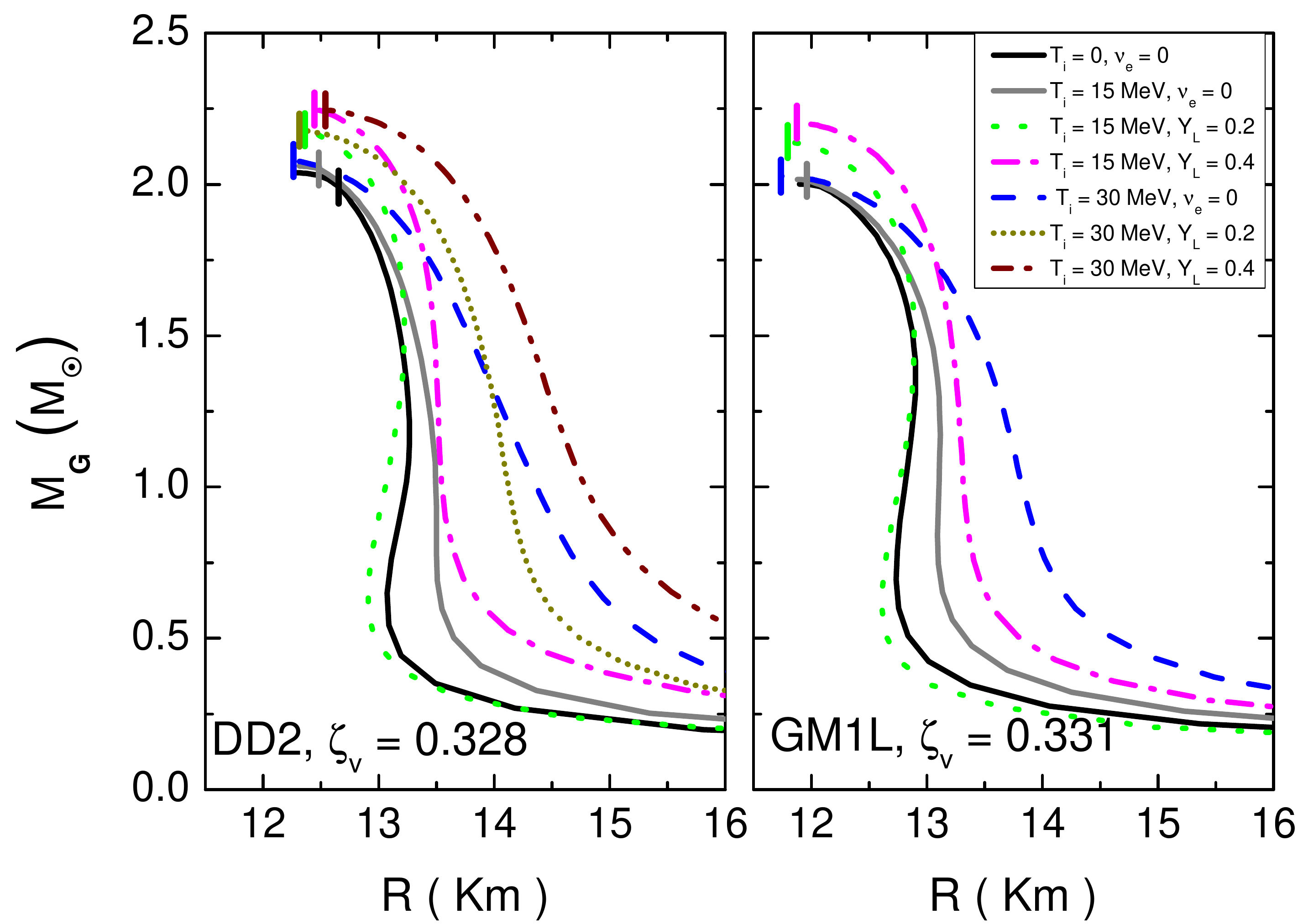}
\caption[$ M_G $ as a function of the radius for EoS at finite temperature and/or entropy.]{Gravitational mass $ M_G $, as a function of the radius for hybrid stars at different $ T_\textrm {{trans}} $ temperatures for the hadrons to quarks phase transition.  $ Y_L $ is the lepton fraction and $ Y_{\nu_e} $ the neutrino fraction. The vertical lines mark the beginning of the decofined quark matter. With the exception of stars without neutrinos ($Y_{\nu_e}=0$), the ones at  $T_{\rm trans}=0$ ($T_0$) and $T_{\rm trans}=15$ MeV ($T_{15}$), the transition occurs at the peak of maximum mass. }
  \label{MR_TEMP}
\end{center}
\end{figure}

In figures \ref{masas_radios_T0_NQM}, \ref{masas_radios_T0_2SC} and \ref{Gv04Maxwell}, we have presented the results for the cases of cold NSs. To study the proto-NSs, we need to extend the EoS to finite temperature. It is known from previous works that the evolutionary scenario of proto-ENs is based on a dynamic calculation with representative EoS (see for example \cite{Pons:1998mm}). Based on this analysis, it is inferred that proto-NSs are approximately isentropic at different stages of their evolution, and instead the temperature is variable throughout the radius of each star. To obtain an isentropic EoS for the Maxwell construction, we first calculate EoS for the hadronic and quark phase, for a fixed transition temperature (for example 15 MeV and 30 MeV). After determining the crossing point of the curves, we extend the curves of each phase isentropically for that transition temperature. In this way, we construct isentropic EoS, choosing representative transition temperatures of $T = 15$ MeV and  $T  = 30$ MeV. Depending on the stage of evolution in which we suppose the star is, we use different fractions of leptons ($Y_{\nu_e} \neq 0$), considering $Y_L = Y_e + Y_{\nu_e} =0.2$ or $Y_L = 0.4$. The radio-mass relations for stars composed of this type of matter are shown in Figure \ref{MR_TEMP}.

For the model without superconductivity and with the Polyakov loop, the results obtained lead us to the following analysis. For the maximum values of vector interactions indicated above ($\mathrm{\zeta_v} = 0.371$ and $\mathrm{\zeta_v} = 0.385$) we found that an increase in temperature (with or without the inclusion of neutrinos) prevents the formation of quark matter inside the stars. The only stars with quark matter for these vector interaction values are cold hybrid stars. For the case of minimal vector interactions, the same thing happens qualitatively. The differences come mainly from considering whether there are neutrinos trapped inside or not. For the DD2 parameterization, for example, an EoS with trapped neutrinos up to transition temperatures $ T_\textrm{{trans}} = 30 $ MeV can be built to model NSs that contain quark matter inside. However, for the GM1L parameterization, neutrinos are present only for configurations with temperatures up to $T_\textrm{{trans}} = 15$ MeV, and for higher temperatures the star becomes unstable before the hadron-quark transition occurs.
In Figure \ref{MR_TEMP} it is observed that the influence of neutrinos increases the maximum mass of the star. This influence depends significantly on the composition of the matter (see for example \cite{Prakash97}). If heavy hadrons (such as hyperons or $ \Delta $ resonances) and quarks are taken into account, the inclusion of neutrinos results in an increase in maximum mass. This is the exact opposite of the idealized EoS situation, where only nucleons and leptons are taken into account, and no other component that softens the EoS, in which case the effect of neutrinos is to reduce the maximum mass, as shown in reference \cite{Alford:2004pf}.

\begin{figure}[ht!]
\begin{center}
\includegraphics[scale = 0.3]{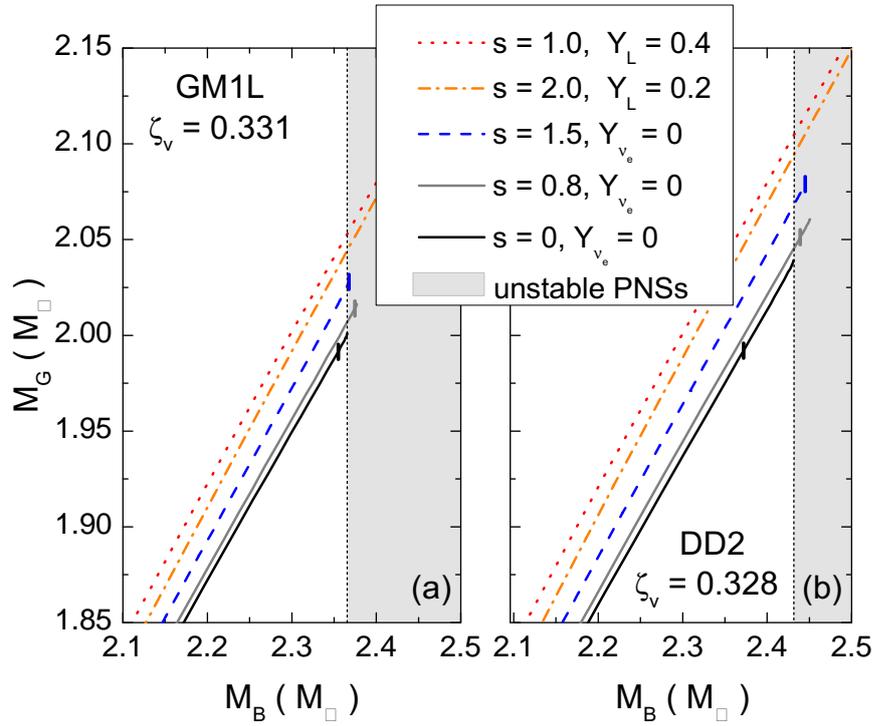}
\caption[$ M_G $ as a function of $ M_B $ for proto-NSs stages.]{Gravitational mass and baryon mass for selected evolutionary stages of proto-NSs. Each line ends at the maximum mass star for each stage. The vertical lines indicate the beginning of the deconfined quark matter. The stars in the shaded region are gravitationally unstable.}
\label{MGvsMB}
\end{center}
\end{figure}

So far we have calculated the radio-mass relationships for different EpS that would correspond to different stages of the NSs. However, even if there is quark matter in these stages, it does not mean that they are actually previous stages of the stars calculated at $ T = 0 $. To determine this we have to analyze the gravitational mass - baryon mass diagram. In Figure \ref{MGvsMB} we show these relationships for the previously calculated stages. Assuming that we are modeling isolated stars (without the possibility of accreting matter), the baryonic mass must be a constant during the sar's evolution, this condition is shown with a vertical line in the mentioned figure. he vertical bars mark the beginning of the deconfined quark matter in star's core. Proto-stars in their initial evolutionary stages (for example , $s=1$, $Y_L=0.4$ y $s=2$,$Y_L=0.2$) turn out to be purely hadronic stars no matter how massive they are. Once they are deleptonized ($Y_{\nu_e}=0$), and their entropies decrease to values of $s=1.5$ y $s=0.8$, there is presence of pure quark matter. However, this happens in our examples only for stars that are in the gravitationally unstable regions for the zero temperature star, where the proto-NSs have greater baryonic masses  than the corresponding cold star's baryonic mass. This means that while they may contain quark matter in their formation, they would eventually collapse into a black hole. The situation changes only when the stars turn into cold stars ($s=0$, $Y_{\nu_e}=0$), which contain pure quark matter in their cores. In tables \ref{mass-radius} and \ref{mass-radius2} we show the proto-stars as they evolve towards the associated cold maximum mass star.

\begin{table}[ht!]
{%
\begin{center}
\begin{tabular}{|l|c|c|l|}\cline{1-4}
% use packages: color,colortbl
\multicolumn{4}{|c|}{GM1L y $\mathrm{\zeta_v} = 0.331$} \\ \hline
\multicolumn{1}{|c|}{Stages} & $M_G~ [M_{\odot}]$ & $R~ {\rm [km]}$
& {Core composition}
\\ \hline
$s=1.0\,, ~Y_L = 0.4$ & $2.05$ & $12.75$  &Pure hadronic\\
$s=2.0\,, ~Y_L = 0.2$ & $2.04$ & $12.84$  &Pure hadronic\\
$s=1.5\,, ~Y_{\nu_e} = 0$  & $2.02$ & $ 11.94 $ &Pure hadronic\\
$s=0.8\,, ~Y_{\nu_e} = 0$  & $2.01$ & $ 11.97 $ &Pure hadronic\\
$s~\,=0.0\,, ~Y_{\nu_e} = 0$  & $2.00$ & $11.90$   &Hybrid\\ \hline
\end{tabular}
\caption[Mass, radius, and core composition for (proto-) NSs with conserved baryon mass $M_B = 2.36 M_{\odot}$ for the GM1L parameterization.]{Mass, radius, and core composition for (proto-) NSs with conserved baryon mass $M_B = 2.36 M_{\odot}$ for the GM1L parameterization.}
\label{mass-radius}
\end{center}
}
\end{table}

\begin{table}[ht!]
{%
\begin{center}
\begin{tabular}{|l|c|c|l|}\cline{1-4}
% use packages: color,colortbl
\multicolumn{4}{|c|}{DD2 y $\mathrm{\zeta_v} = 0.328$} \\ \hline
\multicolumn{1}{|c|}{Stages} & $M_G ~[M_{\odot}]$ & $R~ {\rm [km]} $
&{Core composition}
\\ \hline
$s=1.0\,, ~Y_L = 0.4$ & $2.10$ & $13.09$ &Pure hadronic \\
$s=2.0\,, ~Y_L = 0.2$ & $2.09$ & $13.15$ &Pure hadronic \\
$s=1.5\,, ~Y_{\nu_e} = 0$  & $2.07$ & $ 12.37 $ &Pure hadronic \\
$s=0.8\,, ~Y_{\nu_e} = 0$  & $2.05$ & $ 12.49 $ &Pure hadronic \\
$s~\,=0.0\,, ~Y_{\nu_e} = 0$  & $2.04$ & $12.27$  &Hybrid\\ \hline
\end{tabular}
\caption[Mass, radius, and core composition for (proto-) NSs with conserved baryon mass $M_B = 2.43 M_{\odot}$ for the DD2 parameterization.]{Mass, radius, and core composition for (proto-) NSs with conserved baryon mass $M_B = 2.43 M_{\odot}$ for the DD2 parameterization.}
\label{mass-radius2}
\end{center}
}
\end{table}

It has been proposed \cite{Prakash97,Brown:1993jz} that the unstable proto-stars described above would collapse into black holes. Furthermore, it was shown in the works of the references \cite{Prakash97,Vidana:2002rg} that the collapse could be related to the presence of $ \Delta $ hyperons and resonances, and/or quark matter, since at finite temperature and with these compositions, it is possible to reach higher mass ranges (without destabilizing the star) than at zero temperature and without trapped neutrinos.

\begin{figure}[ht!]
\begin{center}
\includegraphics[scale = 0.45]{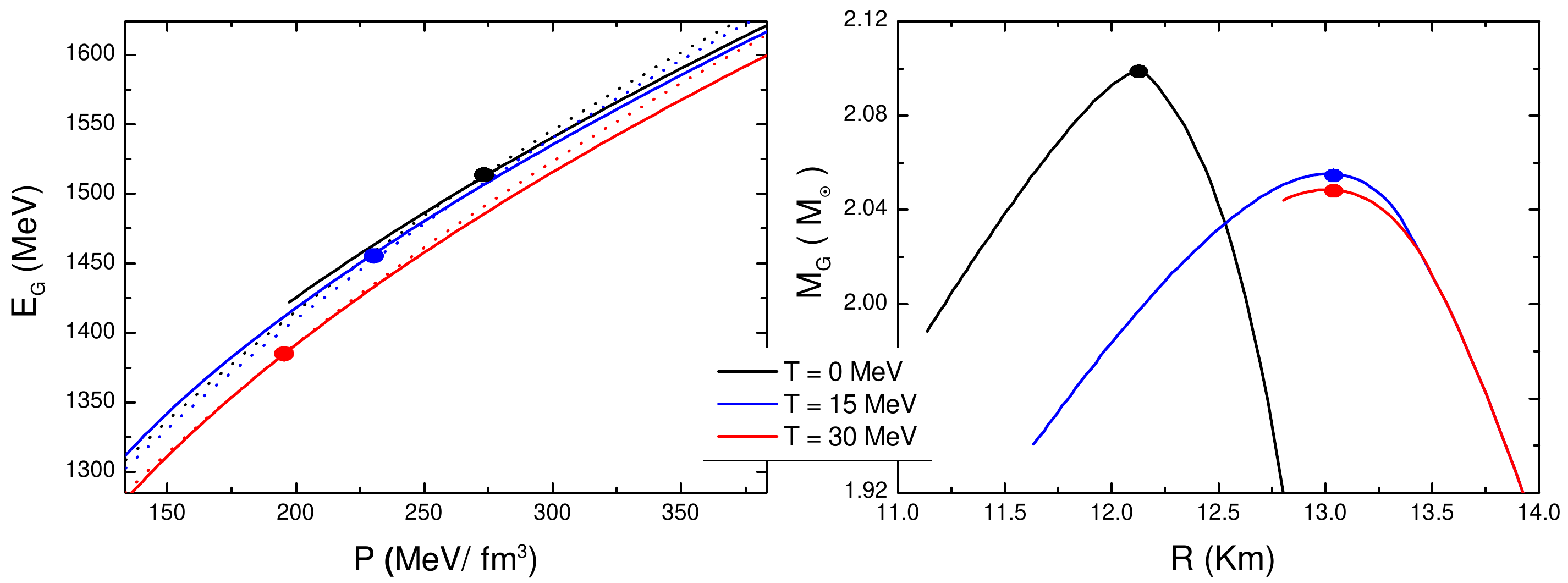}
\caption[Construction of hybrid EoS and mass-radius relationship for the ($2SC + s$) model at finite temperature.]{Left: construction of hybrid EoS for the superconductivity model at different temperatures. The dotted lines correspond to the DD2 parameterization hadronic EoS and the solid ones indicate the EoS of quarks with  $\zeta_\mathrm{V} = 0.39$ and $\zeta_\mathrm{D} = 0.75$. The solid dots indicate the ending of the hadronic phase and the start of quark. Right: mass-radius ratio for the same model, the solid dots indicate the maximum mass star.}
  \label{2SC_T_finita}
\end{center}
\end{figure}

The results with superconductivity at finite temperature are very different from those already shown. On the one hand, it is impossible to maintain the same vector interactions values with which we were working and achieve hybrid EoS. The effect of temperature, instead of moving the onset of quark matter to higher densities (as happened in the nl-PNJL model), makes the transition occur at such lower densities that they are non-physical. It is then necessary to modify the vector interaction coupling constant values, in order to see what results can be obtained in the mass-radius curves. For this, it is possible to choose (for example) a coupling such that quark matter exists at high and medium temperatures, and at the same time maintains the presence of quarks at low temperatures. As an example, these curves for a vector coupling of $\zeta_V = 0.39$ are shown in Figure \ref{2SC_T_finita}.

As can be seen, the results are completely opposite to those of the model that includes the Polyakov loop. In the ($ 2SC + s $) model, on the one hand, the effect of temperature is to move the hadron-quark transition to lower chemical potentials. On the other hand, as the temperature increases, the maximum masses of the stars decrease instead of increasing. This also in turn results in lower baryon masses, making it impossible to meet the conservation of baryonic mass for the case of maximum mass cold stars. On the other hand, the vector coupling chosen in Figure \ref {2SC_T_finita} is such that the phase transition occurs directly from hadronic matter to superconducting color matter. However, the choice of other couplings that allow a transition from hadronic matter to normal quark matter, to later transition to ($ 2SC + s $) matter, show the same behavior. That is, the effect of temperature moves the transition to lower densities, but the masses of the stars obtained as the temperature increases are lower. In Chapter \ref{ch:Conclusiones} we will explain two important observations about these behaviors. 

\section{Hybrid stars with mixed phase transitions}

To analyze the possibility of a mixed phase in stars, the treatment is the same as that explained in the section \ref{construcciones}. In the Gibbs construction, the phase transition is smoother than in the Maxwell construction, and requires global rather than local electric charge neutrality, so the hadronic and quark phases have separate non-zero electric charge and have continuous chemical potentials throughout the transition. The difference at the calculus level with the Maxwell construction is that it is not possible to build the mixed phase by simple inspection of where the curves intersect, but rather that a numerical code must be created that contains both models (of quarks and hadrons) and that it solves both EoS for each point so that the set of equations \eqref{Gibbs1}-\eqref{entropiafijagibbs} are fulfilled . This has, on the one hand the disadvantage that it is computationally very expensive, but on the other hand the advantage that it is not necessary to make isothermal constructions first and isentropic later. By having a single code that solves all the equations at the same time, it is possible to ask for the condition that the temperature is a variable, and that it is the same point to point, in such a way as to obtain a constant entropy per baryon.

\begin{figure}[ht!]
\begin{center}
\includegraphics[scale = 0.5]{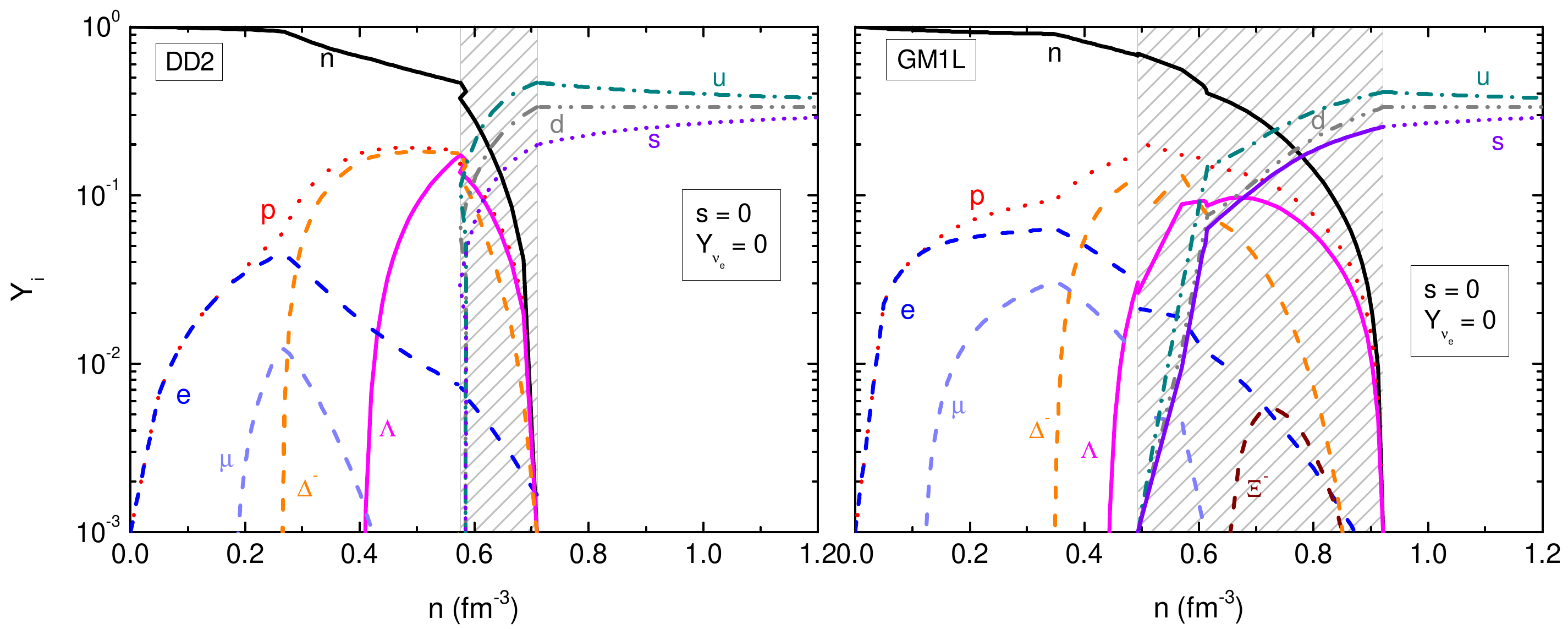}
\caption[Particle population for hybrid stars with mixed phase.]{Particle population for hybrid stars with mixed phase. Left: DD2 parameterization and $\zeta_V = 0.328$. Right: GM1L parameterization and $\zeta_V = 0.331$.}
\label{part_pop_gibbs}
\end{center}
\end{figure}

\begin{figure}[ht!]
\begin{center}
\includegraphics[scale = 0.47]{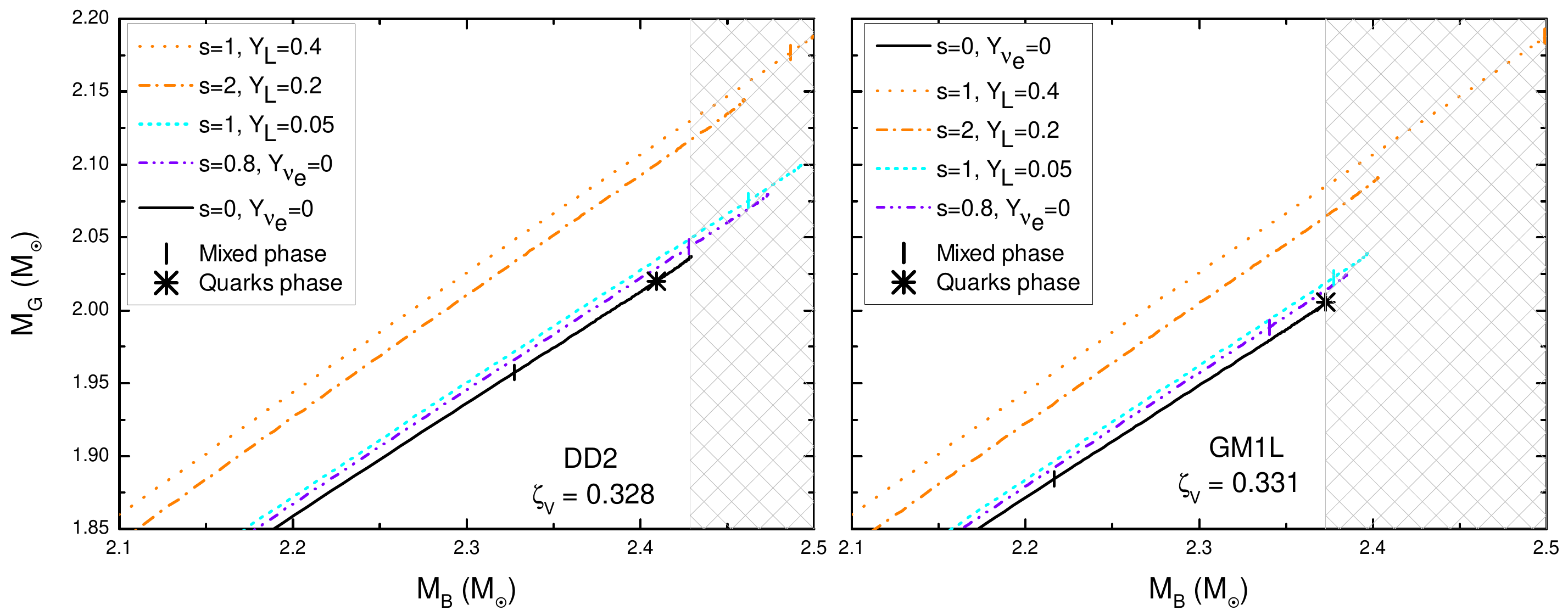}
\caption[$ M_G $ as a function of $ M_B $ for different stages of stars built with Gibbs type transition.]{Left: Gravitational mass as a function of baryon mass for the DD2 parameterization in the mixed phase construction, the shaded region corresponds to gravitationally unstable stars. The vertical lines indicate the beginning of the mixed phase, and the asterisk indicates the beginning of the pure quark phase. Right: the same curves as in the left panel but with the GM1L parameterization.}
\label{MGvsMB_Gibbs}
\end{center}
\end{figure}

Given that in the mixed phase species of hadronic particles and quarks coexist at the same time, it is interesting to study what happens to the populations of each one. In Figure \ref{part_pop_gibbs} the populations for the two parameterizations used are shown. As the mixed phase begins at the density in which the $ s $ quark  is deconfined, the effect this produces on the rest of the densities is noticeables. For the case of the GM1L parameterization, it is observed that when the mixed phase begins, the hadrons that have a negative charge have a negative jump in their population, while the neutron has a positive jump, and the $ \Lambda $ hyperon  a negative jump , but both of less magnitude than the charged particles jumps. This is due to a combined effect product of the decoupling of the $ s $ quark  at the beginning of the mixed phase and the dependence of the coupling constants with the density. Note that, for example, this effect of "breaking" in particle populations is more noticeable in the case of the GM1L parameterization, right panel of Figure \ref{part_pop_gibbs} in which only the coupling constants related to the $ \rho $ meson  depend on density. For the DD2 parameterization, left panel of Figure \ref{part_pop_gibbs}, this 'break' effect is less visible, since all coupling constants are density dependent.

n the same way as in Maxwell's construction, we now analyze the stars obtained for stages prior to the cold star, corresponding to EdE with finite values of entropy and lepton fraction. The results for hybrid stars considering this type of phase transition are shown in the gravitational mass-baryonic mass plane of Figure \ref{MGvsMB_Gibbs}. In this type of construction, two fundamental differences can be observed with respect to the Maxwell construction: first, the occurrence of quark matter in the mixed phase occurs much earlier than that of pure quark matter in the abrupt transition; second, unlike stars built considering the Maxwell transition, in this case an evolutionary stage at non-zero temperature ($s = 0.8\, ,\, Y_{\nu_e} = 0$) enters as possible evolutionary stage, preserving the cold star's baryonic mass. On the other hand, the EoS with the GM1L parameterization results in a family of stars containing mixed phase in a wider range of masses than the DD2 parameterization.

\begin{figure}[ht!]
\begin{center}
\includegraphics[scale = 0.45]{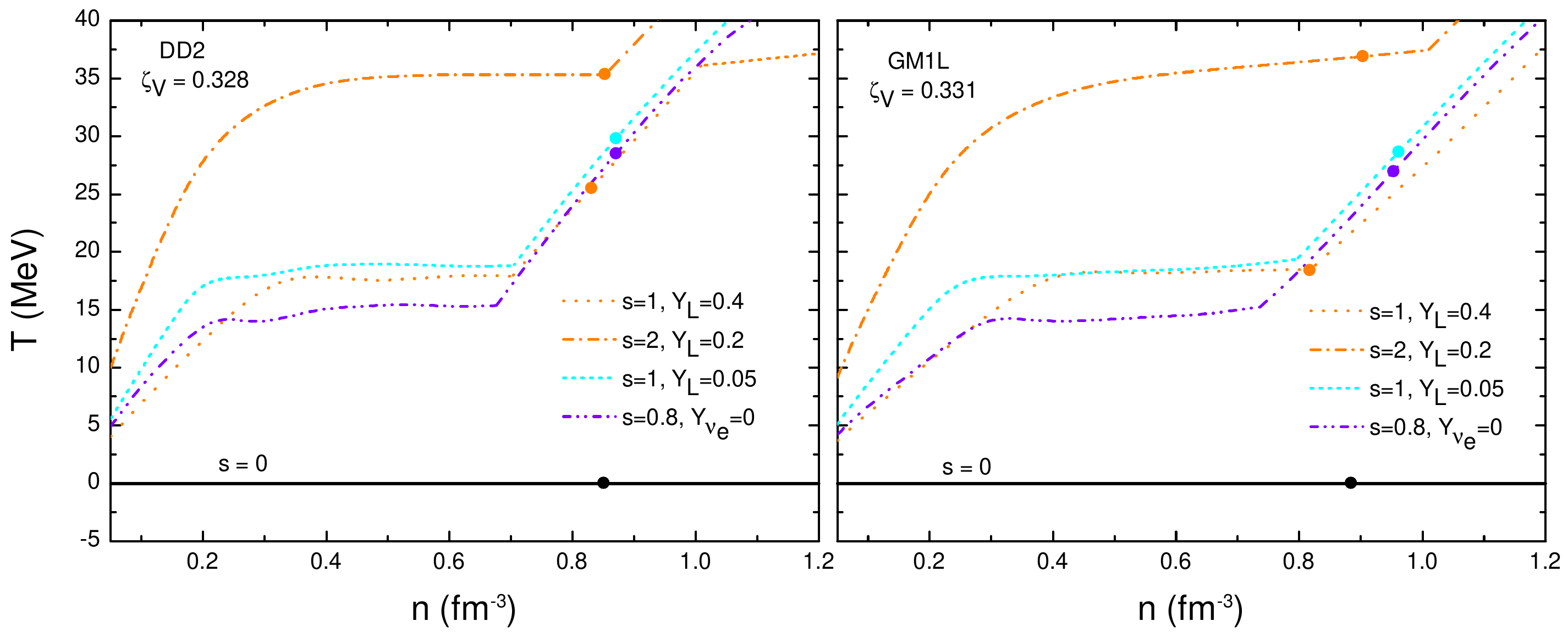}
\caption[Temperature profiles as a function of baryon density.]{Left: Temperature as a function of baryon density for the different EoS analyzed, for the DD2 parameterization. The solid points mark the density corresponding to the central density of the maximum mass star for each curve. Left: the same as the right panel but for the GM1L parameterization.}
\label{perfilesTemp}
\end{center}
\end{figure}

It is worth asking for what temperature ranges it is possible to obtain a stage prior to cold NSs. For that, we show the temperature profiles for the different EoS studied, which are shown in Figure \ref{perfilesTemp}.  When analyzing the density corresponding to the stars that meet the baryonic mass conservation condition, we find that for the stage ($s = 0.8\, ,\, Y_{\nu_e} = 0$), the temperature at the center of the star for the two parameterizations is $T_{central} \simeq 28$ MeV. On the other hand, from the same graph it can be clearly seen that the stage ($s = 2\, ,\, Y_{\nu_e} = 0.2$) corresponds to the maximum temperature of the stars, being pure hadronics.This is exactly the same as the results obtained in the case of Maxwell phase transitions, which leads us to think of two cases: or so that our results coincide with the results of the reference  \cite {Prakash:2000jr} where the maximum heating stage of the star corresponds to ($s = 2\, ,\, Y_{\nu_e} = 0$), transport phenomena must be included, or else the calculation corresponding to said work was done with less realistic EoS, and the neutrinos role with our EoS is significantly different to reach maximum heating before the neutrinos escape from the star.

\begin{figure}[ht!]
\begin{center}
\includegraphics[scale = 0.45]{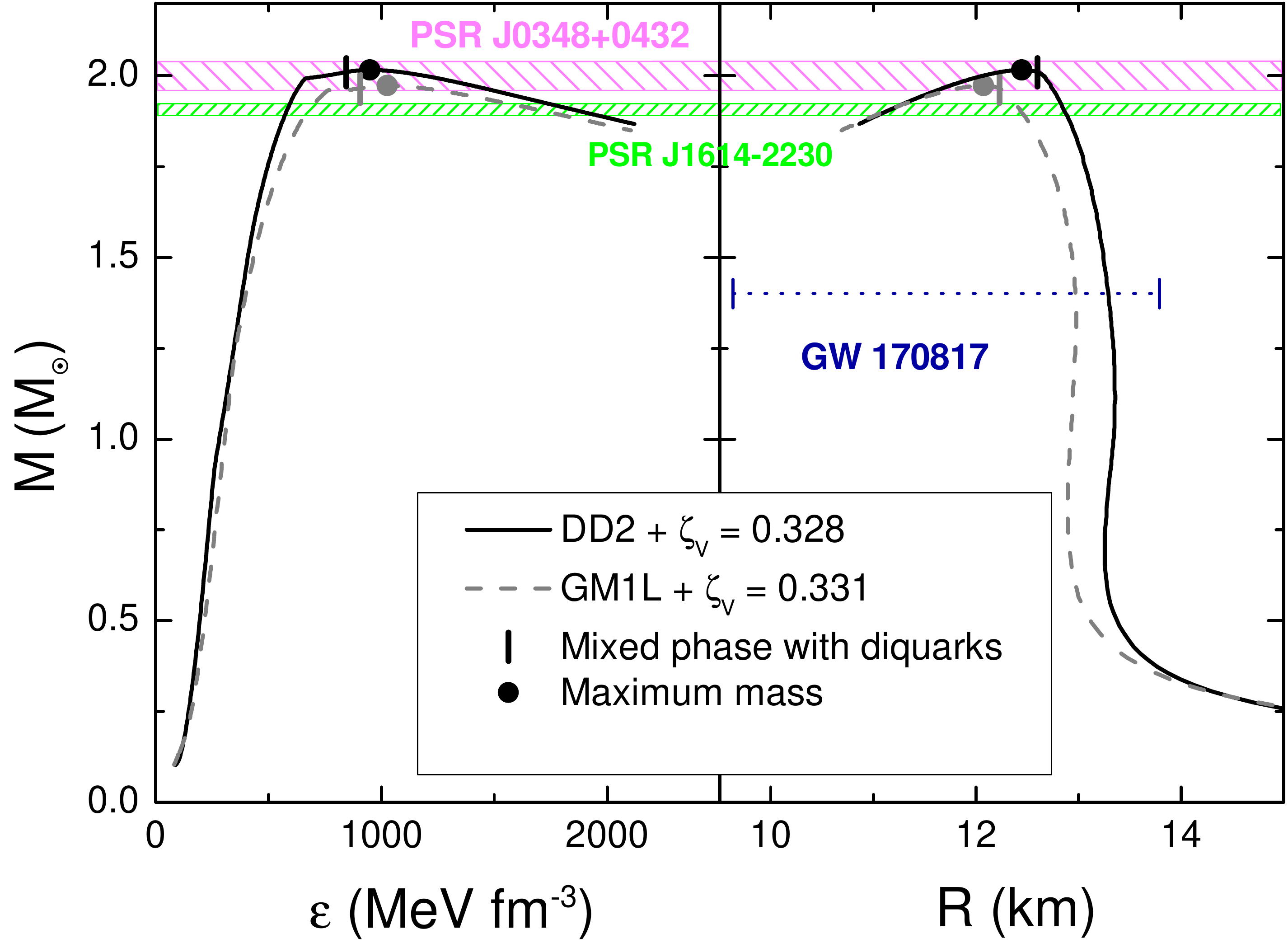}
\caption[Energy-mass and mass-radio density profiles for color superconducting hybrid stars.]{Energy-mass and mass-radio density profiles for color superconducting hybrid stars.}
\label{2sc_mixed_TOV}
\end{center}
\end{figure}

\begin{figure}[ht!]
\begin{center}
\includegraphics[width=0.6\textwidth]{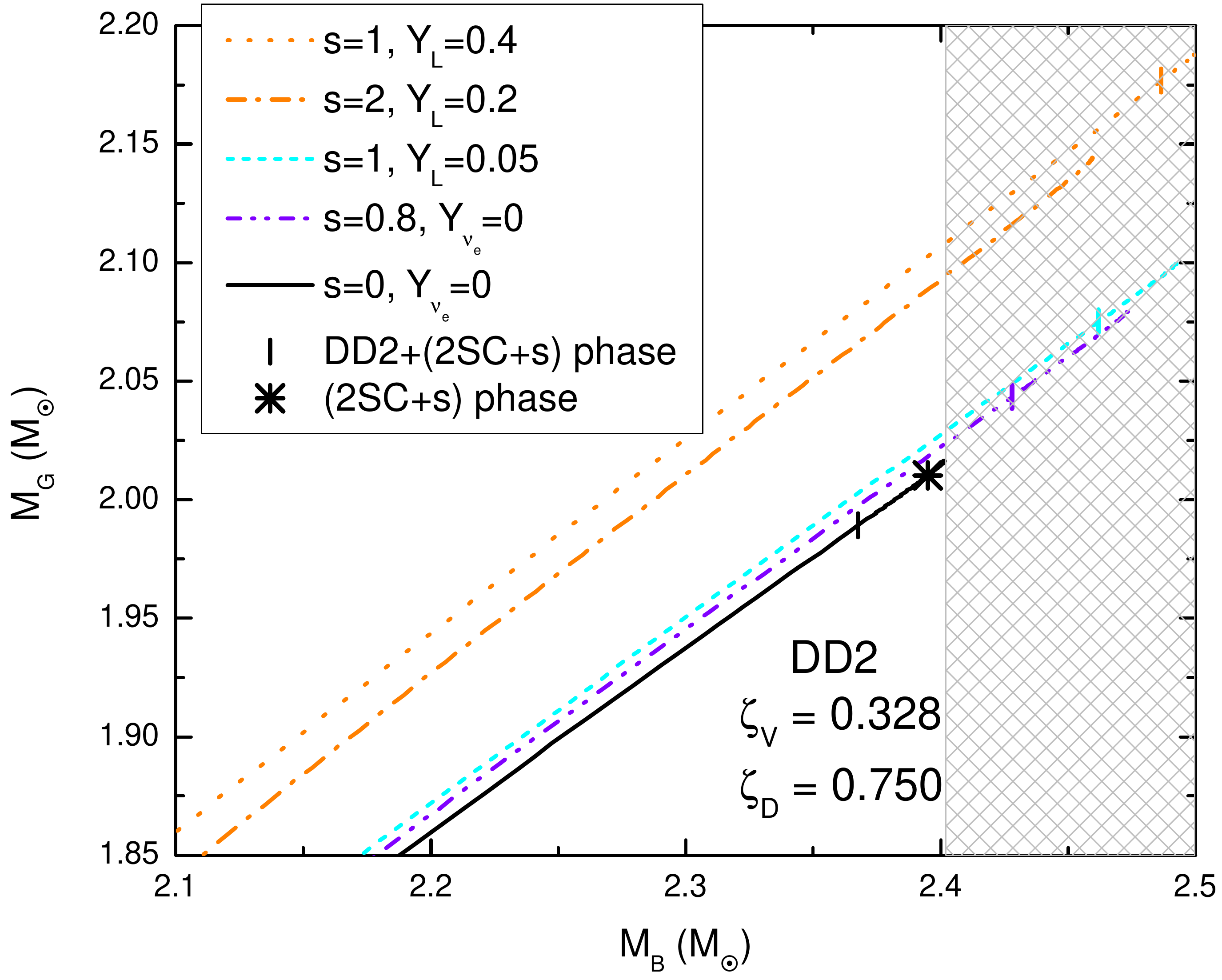}
\caption[$ M_G $ as a function of $ M_B $ for proto-stars and cold star stages with color superconductivity.]{Gravitational mass and baryon mass for the selected stages in the evolution of the proto-NSs, with the cold star corresponding to the EoS with DD2 parameterization and ($2SC + s$) phase.}
\label{MB_MG_2sc_dd2}
\end{center}
\end{figure}

Finally, to finish the analysis of the mixed phase, it remains for us to see what is the effect of the ($2SC + s$)  phase on cold stars. As expected, the inclusion of this effect reduces the maximum mass for the family of stars obtained, since the inclusion of color superconductivity makes the matter more compressible, to the point that for the GM1L parameterization the two solar masses condition is not reached, as can be seen in Figure \ref{2sc_mixed_TOV}. On the other hand, although using the parameterization DD2 the two solar masses condition is reached, the baryonic mass obtained decreases significantly compared to the case without color superconductivity. As can be seen in Figure \ref{MB_MG_2sc_dd2}, the inclusion of the ($ 2SC + s $) phase  leads to a cold star that has no previous stages containing neither pure quark matter nor a mixed phase composed of hadronic and quark matter. The same behavior is observed for both DD2 and GM1L parameterization. The results for the masses, radius and core compositions for the stars that conserve baryonic mass in the mixed phase construction, are shown in the tables \ref{mass-radius_gibbs} and \ref{mass-radius_gibbs2} for the DD2 and GM1L parameterizations respectively.
\begin{table}[ht!]
{%
\begin{center}
\begin{tabular}{|l|c|c|l|}\cline{1-4}
% use packages: color,colortbl
\multicolumn{4}{|c|}{DD2 y $\mathrm{\zeta_v} = 0.328$} \\ \hline
\multicolumn{1}{|c|}{Stages} & $M_G~ [M_{\odot}]$ & $R~ {\rm [km]}$
& {Core composition}
\\ \hline
$s=1.0\,, ~Y_L = 0.4$ & $2.13$ & $11.88$  &Pure Hadronic\\
$s=2.0\,, ~Y_L = 0.2$ & $2.12$ & $11.96$  &Pure Hadronic\\
$s=1\,, ~Y_L = 0.05$  & $2.05$ & $ 11.61 $ &Pure Hadronic \\
$s=0.8\,, ~Y_{\nu_e} = 0$  & $2.04$ & $ 11.60 $ &Pure Hadronic \\
$s~\,=0.0\,, ~Y_{\nu_e} = 0$  & $2.01$ & $12.52$   &Hybrid with $NQM$ \\
$s~\,=0.0\,, ~Y_{\nu_e} = 0$  & $2.00$ & $11.56$   &Hybrid with $(2SC+s)$ \\ \hline
\end{tabular}
\caption[Mass, radius, and core compositions for (proto-) NSs in Gibbs formalism and DD2 parameterization.]{Mass, radius, and core compositions for (proto-) NSs with conserved baryon mass $M_B = 2.42 M_{\odot}$ for DD2 parameterization and a hadron-quarks mixed phase.}
\label{mass-radius_gibbs}
\end{center}
}
\end{table}
\begin{table}[ht!]
{%
\begin{center}
\begin{tabular}{|l|c|c|l|}\cline{1-4}
% use packages: color,colortbl
\multicolumn{4}{|c|}{GM1L y $\mathrm{\zeta_v} = 0.331$} \\ \hline
\multicolumn{1}{|c|}{Stages} & $M_G~ [M_{\odot}]$ & $R~ {\rm [km]}$
& {Core composition}
\\ \hline
$s=1.0\,, ~Y_L = 0.4$ & $2.08$ & $13.10$  &Pure Hadronic\\
$s=2.0\,, ~Y_L = 0.2$ & $2.06$ & $12.51$  &Pure Hadronic\\
$s=1\,, ~Y_L = 0.05$  & $2.02$ & $ 12.01 $ &Pure Hadronic \\
$s=0.8\,, ~Y_{\nu_e} = 0$  & $2.01$ & $ 11.95 $ &Hybrid \\
$s~\,=0.0\,, ~Y_{\nu_e} = 0$  & $2.00$ & $11.85$   &Hybrid with $NQM$ \\ 
$s~\,=0.0\,, ~Y_{\nu_e} = 0$  & $1.97$ & $12.07$   &Hybrid with ($2SC+s$)\\ \hline
\end{tabular}
\caption[Mass, radius, and core compositions for (proto-) NSs in Gibbs formalism and GM1L parameterization.]{Mass, radius, and core compositions for (proto-) NSs with conserved baryon mass $M_B = 2.36 M_{\odot}$ for GM1L parameterization and a hadron-quarks mixed phase.}
\label{mass-radius_gibbs2}
\end{center}
}
\end{table}

\graphicspath{{Conclusiones/}}
\chapter{\label{ch:Conclusiones}Conclusions}

In this work we have applied non-local quark models to describe matter inside NSs. Families of proto-stars and cold hybrid stars were calculated considering two different formalisms for the study of the phase transition of hadronic matter, located in the outer core of the star, to quark matter, located in the inner core, either as a single phase or as part of a mixed phase. From the TOV equations of hydrostatic equilibrium that allow us to calculate the structure of not rotating NSs with spherical symmetry and using the hybrid EoSs obtained, different families of compact objects were built and gravitational mass-radius, gravitational mass-baryonic mass and gravitational mass - energy density diagrams were obtained. Analyzing the different stages of thermal evolution that an NS would go through from its proto-star state until it cools, we imposed different restrictions on our EoS taking into account the observational bound of 2 M $_{\odot} $ imposed by the pulsars J0348 + 0432 and J1614-2230. Likewise, the new restrictions imposed on the radii of these objects were taken into account. They come from the data analysis of the historical event known as GW170817: the first detection of gravitational waves from the two NSs merger. For the hybrid EoS construction that serve to determine the different families of stars, we describe the theoretical framework of two types of transitions: Maxwell type or abrupt and without mixed phase formation or Gibbs type, in which mixed phase formation occurs, in which components of both phases coexist and whose pressure varies monotonically with the density.

For the quark phase, we fully describe the Nambu Jona-Lasinio non-local model in $ SU (3) $, with vector interaction and Polyakov loop incorporation to model matter at zero temperature and at finite temperature. n the same way the model was described, in this case without Polyakov loop but including color superconductivity in the same two temperature regimes. This last model was developed allowing the formation of diquarks only between $u$ and $d$  quarks, and forming a ($ 2SC + s $) phase, that is, with the presence of $ s $ quark  either as a free particle, or in a chiral condensate. We obtained that different choices of vector interaction and diquarks coupling delay or advance the deconfinement of the $s$ quark, thus leading to different types of phase transitions (first order or crossover) for the restoration of the chiral symmetry of that quark. For the description of hadronic matter, we relied on the nonlinear relativistic mean field model with density-dependent coupling constants.
Within the framework of this model, we chose the parameterizations known as DD2 and GM1L because they are some of those that conform to the limits established in experiments corresponding to hadronic physics. Using DD2 and GM1L, we have calculated the population of purely hadronic particles in proto-NSs for entropies and lepton fractions typical of this type of matter in that context. According to the hadronic model used in this thesis, the results show that hyperons arise at densities of the order of the nuclear saturation density. Furthermore, the population of hyperons with strangeness increases with density, so in a phase transition from hadronic matter to quark matter, the role of the $s$ quark cannot be neglected. Both hadronic parametrizations were used in conjunction with the quark matter EoS taking into account the densities at which the quarks are deconfined and outside the meta-stable zone determined by the spinodals in the model phase diagram.

For the construction of hybrid EoS we start by considering an abrupt hadron-quark phase transition, using the Maxwell formalism, restricting ourselves to the vector interaction values for quark matter that allowed us to reach the 2 M$_{\odot}$ value. Furthermore, for each of the cases analyzed, we chose the minimum vector coupling that would allow us to achieve a phase transition from hadronic matter to deconfined quark matter at lower pressures, in order to favor the presence of the latter inside the NSs. This led us to obtain the vector interaction values in the ranges $0.331<\mathrm{\zeta_v}<0.371$ for GM1L, and $0.328<\mathrm{\zeta_v}<0.385$ for DD2. Once the different families of stars have been obtained and the condition of the maximum masses of that families for each parameterization has been established, we calculate the corresponding baryon mass in each case, to describe a schematic thermal evolution. Assuming isolated NSs, we use the gravitational mass-baryon mass diagram to determine possible stages of hot stars that can evolve to the one corresponding to the cold star, conserving its baryonic mass. Under this condition, for the Polyakov loop model we obtained that for Maxwell-type phase transitions, there are no stages prior to the NS of maximum mass that contain free quark matter: stars with quark cores are only possible for zero-entropy star configurations. That is, the effect of temperature moves the phase transition to very high chemical potentials, preventing stable star configurations with deconfined quark matter in their interior.

On the other hand, by including color superconductivity ($ 2SC + s $ phase) in the quark model, we find that the effect of increasing the temperature moves the phase transition to lower chemical potentials, thus allowing quark matter to appear much earlier in the hybrid EoS. Unfortunately, this effect instead of increasing the maximum mass in the stars, decreases it, so that the previous stages of proto-NSs with diquarks, do not conserve the baryonic mass necessary to evolve to the maximum mass cold star, collapsing into a black hole. Analyzing this behavior in more detail, we were able to detect that this effect is not a consequence of including color superconductivity in the model, but rather it occurs because it does not include the Polyakov loop, regardless of whether or not a superconducting phase is considered.

For the construction of phase transitions in which the Gibbs formalism is considered, thus allowing the possibility of a mixed phase, we obtained that for the case of GM1L parameterization and minimal vector coupling ($\zeta_{V} = 0.331$), it is possible to find an earlier evolutionary stage for the cold star of maximum mass, with entropy per baryon $s = 0.8$ and no trapped neutrinos, which corresponds to a star with a mixed-phase core composed of quarks and hadrons. The temperature obtained at the center of the star for this stage is $T_{central}^{GM1L} = 28$ MeV. The internal composition of the cold star of maximum mass product of the evolution of this stage consists of a mixed phase of quarks and hadrons, and a phase of pure deconfined quarks in the inner core of the star. For the cases in which the ($2SC + s$) phase was taken into account, we obtained that the fact of including this phase for the description of quark matter in the NSs decreases the maximum mass of the calculated stars, both for the Maxwell and Gibbs constructions. Furthermore, in the Gibbs construction, the combination of the GM1L parameterization with ($ 2SC + s $) matter does not satisfy the observational 2 M $_{\odot} $ bound. However, this same parametrization combined with quarks without considering the color superconducting phase was the only one that resulted in a non-zero entropy stage starwith quark matter in its interior, as a predecessor of the cold hybrid stars of maximum mass.

Finally, both in the abrupt or mixed phase constructions, including or not the Polyakov loop and color superconductivity, the hybrid stars obtained at zero temperature never exceeded the maximum mass that was obtained from considering pure hadronic matter only. These results were to be expected since the inclusion of the $s$ quark smoothes the quark matter EoS, and decreases the maximum mass of the calculated family of stars. However, until now there have been no works on non-local models that take into account a color superconducting phase while preserving the mixing term ('t Hooft's 6-point interaction), showing that the inclusion of that interaction does not sufficiently stiffen the EoS. For this reason, it is possible that models that include all three types of diquarks, as in the case of the $ CFL $ (\textit {Color Flavor Locked}) phase, result in stiffer EoS than in the ($2SC + s$) case. However, as we saw, if the Polyakov loop effect of the model is not considered, an increase in temperature moves the hadron-quark phase transition to lower densities. For this reason, it would be interesting to consider the possibility of incorporating new improvements to the quark model considered, for example, reviewing the role of the Polyakov loop and studying the possibility of considering a temperature-dependent coupling constant. Furthermore, although in this thesis we have developed the ($ 2SC + s $) phase, for the study of the physics of compact objects, it would also be interesting to investigate the effect of a $CFL$ phase. The problem of including both contributions (Polyakov loop and color superconductivity) simultaneously in the model due to rotation in the color space, remains unsolved. To overcome this point, in the extension of the model finite termperature including the superconducting phase of color, the temperature is introduced only through the Matsubara frequencies and without considering the Polyakov potential. Likewise, it would be interesting to study the phase diagram considering ($2SC+s$) phase. The inclusion of diquarks in the model would surely affect the phase diagram structure with the possible occurrence of new regions depending on the diquarks interaction intensity.

Regarding the hybrid star configurations calculated in this work, the results show that, if there were a phase transition inside these objects, the 2 M$_{\odot}$ NSs could contain an inner core made up of pure quarks or a color superconducting phase, or an inner core made up of a mixed phase of hadrons and quarks with or without color superconductivity. As we said at the beginning, the EoS of dense matter is still uncertain. NICER, Strobe-X, and the data provided by gravitational wave detectors like LIGO and Virgo have the potential to not only help elucidate what kind of matter makes up the interior of the NSs, but also to achieve a better understanding of the structure and stability of these types of objects, as well as the behavior of dense matter subjected to extreme conditions.

\appendix

\chapter{Fermionic integrals for the NJL model}
\label{ApendiceA}

We want to see how to calculate the integrals over the fermionic fields of equation \eqref{particionpsi}. For this, it is convenient to write the currents in terms of the Fourier transformed regulator of equation \eqref {regulador}, having then
\begin{equation}
 j_a^s (x) = \int d^4 z \frac{d^4 p}{(2\pi)^4}g(p) \overline{\psi}\left(x + \frac{z}{2}\right) \lambda_a \psi\left(x - \frac{z}{2}\right) e^{-izp},
\end{equation}
then, one of the terms that appeared in the fermionic integral is written as
\begin{equation}
 \int d^4x\, \sigma_a(x)j_a^s(x) = \int d^4x\,d^4z\frac{d^4 p}{(2\pi)^4} g(p) \sigma_a(x)\psi\left(x + \frac{z}{2}\right)e^{-izp}\label{A2}.
\end{equation}

Now it is convenient to write both the Fourier transforms for both the fermionic and bosonic fields
\begin{eqnarray}
 \overline{\psi}\left(x - \frac{z}{2}\right) &=& \int \frac{d^4q}{(2\pi)^4} \overline{\psi}(q) e^{iq\left(x+\frac{z}{2}\right)}  \label{A3}\\ 
 \psi\left(x + \frac{z}{2}\right) &=& \int \frac{d^4t}{(2\pi)^4} \psi(t) e^{-it\left(x-\frac{z}{2}\right)} \label{A4}\\ 
 \sigma_a (x) &=& \int \frac{d^4s}{(2\pi)^4} \sigma_a (s) e^{-is\left(x\right)} \label{A5},
\end{eqnarray}
where we explicitly omit the different symbols for the transformed functions to ease the notation, because in the following equation the transformed ones will appear only in momentum space. Using \eqref{A3}, \eqref{A4} and \eqref{A5} in \eqref{A2} and suitably grouping the exponents, we can write
\begin{eqnarray}
 \label{Fourierlargo}
 \int d^4x\, \sigma_a(x)j_a^s(x) &=& \int \frac{d^4x \,d^4z \, d^4p \,d^4q\,d^4s \,d^4t}{(2\pi)^{16}} g(p)\overline{\psi}(q)\sigma_a(s)\psi(t) e^{-ix(s+t-q)}e^{-iz\left(p - \frac{q + t}{2}\right)} \nonumber \\
 &=& \int \frac{d^4p \,d^4q\,d^4s \,d^4t}{(2\pi)^8} \delta^4(s+t-q)\delta^4(p-\frac{q+t}{2})g(p)\overline{\psi}(q)\lambda_a\sigma_a(s)\psi(t) \nonumber \\
 &=&\int \frac{d^4q \,d^4t}{(2\pi)^8} g\left(\frac{q+t}{2}\right)\overline{\psi}(q)\sigma_a(q-t)\psi(t) \nonumber \\
 &=& \int \frac{d^4p}{(2\pi)^4} \frac{d^4p'}{(2\pi)^4}g\left(\frac{p + p'}{2}\right)\overline{\psi}(p)\lambda_a\sigma_a(p - p') \psi (p'), 
\end{eqnarray}
where to obtain the last term of the equality, only a change of dummy variables was made within the integral. Another fermionic term from equation \eqref{particionpsi} included the interaction $\pi_a j_a^p (x)$ but in this case the procedure is the same as what was done in equation \eqref{Fourierlargo} replacing $\sigma_a$ for $\pi_a$ and $\lambda_a$ for $i\gamma_5 \lambda_a$, for which we will not develop it. The remaining term is that of the free part, which by Fourier transforming again and writing the delta in the corresponding space results
\begin{equation}
\label{Fourierlibre}
 \int d^4x ,\overline{\psi}(x)(-i\slashed{\partial} - m )\psi(x) = \int \frac{d^4p}{(2\pi)^4} \frac{d^4p'}{(2\pi)^4} \overline{\psi}(p)(\slashed{p} - m)(2\pi)^4 \delta^4(p-p')\psi(p).
\end{equation}
If we put together all the expressions already transformed and with the deltas applied, we have
\begin{eqnarray}
 &\,&\int d^4 x\left[ \overline{\psi}(x)\left(i\slashed{\partial} - m\right)\psi(x) - \sigma_a(x) j_a^s(x) - \pi_a(x)j_a^p(x)\right]  \nonumber \\
 &=&-\int \frac{d^4p}{(2\pi)^4} \frac{d^4p'}{(2\pi)^4} \overline{\psi}(p) A\left(p,p'\right)\psi(p') \nonumber \\ 
\end{eqnarray}
where the  $A(p,p')$ operator is defined as
\begin{equation}
\label{operadorA}
  A(p,p') = \left(-\slashed{p} + m \right)(2\pi)^4\delta^4(p-p') + g\left(\frac{p+p'}{2}\right)\lambda_a\left[\sigma_a(p-p') + i\gamma_5 \pi_a(p-p')\right].
\end{equation}
If we replace all this in the term of the fermionic integrals of equation \eqref{particionpsi}, we have
\begin{eqnarray}
 &\,& \int {\cal{D}}\overline{\psi} {\cal{D}} \psi \,\mathrm{exp}\left\{ \int d^4x \left[\overline{\psi}\left( i \slashed{\partial} + m \right)\psi - \sigma_a j_a^s - \pi_a j_a^p\right]\right\}  \nonumber \\
 &=& \int {\cal{D}}\overline{\psi} {\cal{D}} \psi \,\mathrm{exp}\left\{ -\int \frac{d^4p}{(2\pi)^4} \frac{d^4p'}{(2\pi)^4} \overline{\psi}(p) A\left(p,p'\right)\psi(p') \right\}  \nonumber \\
 &=& \mathrm{det}\left[A(p,p')\right],
\end{eqnarray}
with which everything that was within the functional integral of the fermionic fields in the partition function is solved.

\chapter{Thermodynamic quantities for the nl-NJL model with color superconductivity} \label{ApendiceB}

In this appendix we will use the $d\hat{p}$ notation to refer to $p^2\,dp\,dp_0$. The integrals that have limits from zero to infinity with this differential represent double integrals, one in the $ p $ variable and the other in the $ p_0 $ variable, with those limits in both cases.  The $\sumint$ symbol indicates that we must calculate the double integrals with the mentioned limits and sum over the the three colors in the $c$ variable. Any sum over different colors, or integrals over different limits is explicitly expressed.

\section{Regularized grand potential with vector interaction and diquarks at T = 0}
\begin{eqnarray}
 \Omega =  &-& \int_{0}^{+\infty}\frac{d\hat{p}}{\pi^3} \sum_c \mathrm{Real}\left\{ \mathrm{log}\left[\frac{{q_{sc}^+}^2 + M_{sc}^2}{p_{sc}^2 + m_s^2}\right]\right\}\nonumber \\ 
 &-& \int_{0}^{+\infty}\frac{d\hat{p}}{\pi^3}  \sum_c \left\{ \frac{1}{2} \mathrm{log} |A_c|^2 -  \mathrm{Real}\left[ \mathrm{log}\left({p_{uc}^+}^2 + m_u^2 \right)\right] - \mathrm{Real}\left[\mathrm{log}\left({p_{dc}^+}^2 + m_d^2\right)\right]    \right\} \nonumber \\ 
 &-& \frac{1}{2}\left[ \sum_f \left(\bar\sigma_f  \bar S_f  + \frac{G_S}{2}\bar S_f^2 +  \bar\theta_f  \bar V_f - \frac{G_V}{2}\bar V_f^2\right) + \frac{H}{2}\bar S_u \bar S_d \bar S_s + 2\bar\Delta \bar D +G_D\bar D^2\right]  \\
 &-& \sum_{f,c} \frac{\Theta\left(\mu_{fc} - m_f\right)}{24\pi^2}\left(-5m_f^2 + 2\mu_{fc}^2\right)\mu_{fc}\sqrt{\mu_{fc}^2 - m_f^2} + 3m_f^4\mathrm{log}\left(\frac{\mu_{fc} + \sqrt{\mu_{fc}^2 - m_f^2}}{m_f}\right).\nonumber
\end{eqnarray}
Where 
\begin{equation}
 A_c = \left[{q_{uc}^+}^2 + M_{uc}^2\right]\left[{q_{dc}^-}^2 + {M_{dc}^*}^2\right] + \left(1 - \delta_{bc}\right){\Delta^p}^2\left[{\Delta^p}^2 + 2 q_{uc}^+ . q_{dc}^- + 2M_{uc}M_{dc}^* \right],
\end{equation}
\begin{eqnarray}
q_{fc}^\pm &=& \left( \vec{p}, p_0 \mp i \left[\mu_{fc} - \bar{\theta_f}g\left({p_{fc}^\pm}^2 \right) \right] \right) ,\nonumber \\
q_{0fc}^\pm &=&  p_0 \mp i \left[\mu_{fc} - \bar{\theta_f}g\left({p_{fc}^\pm}^2 \right) \right] ,\nonumber \\
p_{fc}^\pm &=& \left(\vec{p}, p_0 \mp i \mu_{fc} \right) ,\nonumber \\
p_{0fc}^\pm &=& p_0 \mp i \mu_{fc}  ,\nonumber \\
M_{fc} &=& m_f + \sigma_f \,g\left( {p_{fc}^+}^2  \right) ,\nonumber \\
\Delta^p &=& \bar{\Delta}\,g\left(\frac{ \left[{p_{ur}^+} + {p_{dr}^-} \right]^2 }{4}\right).
\end{eqnarray}
\begin{eqnarray}
 \mu_{ur} &=& \mu_{ug} = \frac{\mu_B}{3} - \frac{2}{3}\mu_e + \frac{1}{3} \mu_8 ,\nonumber \\
 \mu_{ub} &=&  \frac{\mu_B}{3} - \frac{2}{3}\mu_e - \frac{2}{3} \mu_8 ,\nonumber \\
 \mu_{dr} &=& \mu_{dg} = \frac{\mu_B}{3} + \frac{1}{3}\mu_e + \frac{1}{3} \mu_8 ,\nonumber \\
 \mu_{db} &=&  \frac{\mu_B}{3} + \frac{1}{3}\mu_e - \frac{2}{3} \mu_8 ,\nonumber \\
 \mu_{sr} &=& \mu_{sg} = \mu_{dr} ,\nonumber \\
 \mu_{sb} &=& \mu_{db} .
\end{eqnarray}
\section{Auxiliary fields}
For the following derivatives we use that $\frac{\partial Log |A_c|^2}{\partial k} = \frac{\partial \mathrm{Log} |A_c^* A_c|}{\partial k} = 2\mathrm{Real}\left(\frac{1}{A_c} \frac{\partial A_c}{\partial k}  \right)$. The auxiliary fields are obtained by minimizing the grand potential with respect to each field and solving the corresponding coupled equation in the system. That is, we must consider the $ D $ field minimizing with respect to $ \Delta $, the $ S_i $ fields with respect to $ \sigma_i $ and the $ V_i $ fields with respect to $ \theta_i $. The $ g^\Delta $ symbol is defined as
\begin{equation}
 g^\Delta = \left(\frac{ \left[{p_{ur}^+} + {p_{dr}^-} \right]^2 }{4}\right),
\end{equation}
and the results for the fields are
\begin{equation}
 \bar{D} = -2 \sum_{c = r,g} \int_{0}^{+\infty} \frac{d\hat{p}}{\pi^3} \mathrm{Real}\left\{\frac{2{\Delta^p}^3 g^\Delta + {\Delta^p} g^\Delta\left(2 q_{uc}^+ . q_{dc}^- + 2M_{uc}M_{dc}^*\right)}{A_c}\right\},
\end{equation}
\begin{equation}
  S_u = -4\sumint \frac{d\hat{p}}{\pi^3} \mathrm{Real}\left\{\frac{M_{uc}g\left({p_{uc}^+}^2\right)\left[{q_{dc}^-}^2 + {M_{dc}^*}^2\right]        + g\left({p_{uc}^+}^2\right)\left(1 - \delta_{bc}\right){\Delta^p}^2M_{dc}^*     }{A_c}\right\},
\end{equation}
\begin{equation}
  S_d = -4\sumint \frac{d\hat{p}}{\pi^3} \mathrm{Real}\left\{\frac{M_{dc}^* g^*\left({p_{dc}^+}^2\right)\left[{q_{uc}^+}^2 + M_{uc}^2\right]        + g^*\left({p_{dc}^+}^2\right)\left(1 - \delta_{bc}\right){\Delta^p}^2M_{uc}     }{A_c}\right\},
\end{equation}
\begin{equation}
  S_s = -4\sum_{c = r,g,b}\int_{0}^{+\infty} \frac{d\hat{p}}{\pi^3}  \mathrm{Real}\left\{\frac{M_{sc} \,g\left({p_{sc}^+}^2 \right)    }{{q_{sc}^+}^2 + M_{sc}^2}\right\},
\end{equation}
\begin{equation}
  V_u = -4\sumint \frac{d\hat{p}}{\pi^3} \mathrm{Real}\left\{\frac{i\,q_{0uc}g\left({p_{uc}^+}^2\right)\left[{q_{dc}^-}^2 + {M_{dc}^*}^2\right]       + ig\left({p_{uc}^+}^2\right)\left(1 - \delta_{bc}\right){\Delta^p}^2 q_{0dc}^-    }{A_c}\right\},
\end{equation}
\begin{equation}
  V_d = -4\sumint \frac{d\hat{p}}{\pi^3} \mathrm{Real}\left\{\frac{-i\,q_{0dc}g\left({p_{dc}^-}^2\right)\left[{q_{uc}^+}^2 + M_{uc}^2\right]        - ig\left({p_{dc}^-}^2\right)\left(1 - \delta_{bc}\right){\Delta^p}^2 q_{0uc}^+    }{A_c}\right\},
\end{equation}
\begin{equation}
  V_s = -4\sum_{c = r,g,b}\int_{0}^{+\infty} \frac{d\hat{p}}{\pi^3} \mathrm{Real}\left\{\frac{i\,q_{0sc}\,g\left({p_{sc}^+}^2 \right)    }{q_{sc}^2 + M_{sc}^2}\right\},
\end{equation}
\section{Densities}
The densities are obtained by minimizing the grand potential with respect to each chemical potential: $\rho_{fc} = \frac{\partial \Omega}{\partial \mu_{fc}}$. For each color and flavor they result
\begin{eqnarray}
 \rho_{ur} &=& \rho_{ug} = 2\int_{0}^{+\infty} \frac{d\hat{p}}{\pi^3} \mathrm{Real}\Bigg(A_c^{-1}\Bigg\{\left[q_{0ur}^+\frac{\partial q_{0ur}^+}{\partial \mu_{ur}}  + M_{ur}\frac{\partial M_{ur}}{\partial \mu_{ur}}\right] \left[\left(q_{dr}^-\right)^2 + \left(M_{dr}^*\right)^2\right] \nonumber \\
 &+& \Delta^2 g^\Delta \frac{\partial g^\Delta}{\partial \mu_{ur}}\left[{\Delta^p}^2 + 2 q_{ur}^+ . q_{dr}^- + 2M_{ur}M_{dr}^*  \right]  \nonumber \\
 &+& {\Delta^p}^2\left[\Delta^2 g^\Delta \frac{\partial g^\Delta}{\partial \mu_{ur}} +  \frac{\partial q_{0ur}^+}{\partial \mu_{ur}} . q_{0dr}^- + \frac{\partial M_{ur}}{\partial \mu_{ur}}M_{dr}^*\right]\Bigg\} \nonumber \\
 &-& \left[\frac{  p_{0ur}^+ \frac{\partial p_{0ur}^+}{\partial \mu_{ur}} }{{p_{ur}^+}^2 + m_u^2}\right]  \Bigg) \nonumber \\
 &+& \Theta\left(\mu_{ur} - m_u\right) \frac{\left(\mu_{ur} - m_u\right)^{\frac{3}{2}}}{3\pi^2}
\end{eqnarray}
\begin{eqnarray}
 \rho_{dr} &=& \rho_{dg} = 2\int_{0}^{+\infty} \frac{d\hat{p}}{\pi^3} \mathrm{Real}\Bigg(A_c^{-1}\Bigg\{\left[{q_{uc}^+}^2 + M_{uc}^2\right] \left[q_{0dr}^-\frac{\partial q_{0dr}^-}{\partial \mu_{dr}} + M_{dr}^*\frac{\partial M_{dr}^*}{\partial \mu_{dr}}\right] \nonumber \\
 &+& \Delta^2 g^\Delta \frac{\partial g^\Delta}{\partial \mu_{dr}}\left[{\Delta^p}^2 + 2 q_{uc}^+ . q_{dr}^- + 2M_{uc}M_{dr}^*  \right]  \nonumber \\
 &+& {\Delta^p}^2\left[\Delta^2 g^\Delta \frac{\partial g^\Delta}{\partial \mu_{dr}} +  q_{0uc}^+. \frac{\partial q_{0dr}^-}{\partial \mu_{dr}} + M_{uc} \frac{ \partial M_{dr}^*}{\partial \mu_{dr}}  \right]\Bigg\} \nonumber \\
 &-& \left[\frac{  p_{0dr}^+ \frac{\partial p_{0dr}^+}{\partial \mu_{dr}} }{{p_{dr}^+}^2 + m_d^2}\right]   \Bigg) \nonumber \\
 &+& \Theta\left(\mu_{dr} - m_d\right) \frac{\left(\mu_{dr} - m_d\right)^{\frac{3}{2}}}{3\pi^2}
\end{eqnarray}
\begin{eqnarray}
 \rho_{ub} &=&  2\int_{0}^{+\infty} \frac{d\hat{p}}{\pi^3} \mathrm{Real}\Bigg(A_c^{-1}\Bigg\{\left[q_{0ub}^+\frac{\partial q_{0ub}^+}{\partial \mu_{ub}}  + M_{ub}\frac{\partial M_{ub}}{\partial \mu_{ub}}\right] \left[\left(q_{dr}^-\right)^2 + \left(M_{dr}^*\right)^2\right]\Bigg\} \nonumber \\
 &-& \left[\frac{  p_{0ub}^+ \frac{\partial p_{0ub}^+}{\partial \mu_{ub}} }{{p_{ub}^+}^2 + m_u^2}\right]  \Bigg) \nonumber \\
 &+& \Theta\left(\mu_{ub} - m_u\right) \frac{\left(\mu_{ub} - m_u\right)^{\frac{3}{2}}}{3\pi^2}
\end{eqnarray}
\begin{eqnarray}
 \rho_{db} &=& 2\int_{0}^{+\infty} \frac{d\hat{p}}{\pi^3} \mathrm{Real}\Bigg(A_c^{-1}\Bigg\{\left[{q_{uc}^+}^2 + M_{uc}^2\right] \left[q_{0db}^-\frac{\partial q_{0db}^-}{\partial \mu_{db}} + M_{db}^*\frac{\partial M_{db}^*}{\partial \mu_{db}}\right]\Bigg\} \nonumber \\
 &-& \left[\frac{  p_{0db}^+ \frac{\partial p_{0db}^+}{\partial \mu_{db}} }{{p_{db}^+}^2 + m_d^2}\right]   \Bigg) \nonumber \\
 &+& \Theta\left(\mu_{db} - m_d\right) \frac{\left(\mu_{db} - m_d\right)^{\frac{3}{2}}}{3\pi^2}
\end{eqnarray}
\begin{eqnarray}
\rho_{sr} &=&  \rho_{sg} =  2\int_{0}^{+\infty} \frac{d\hat{p}}{\pi^3} \mathrm{Real}\left\{\left[\frac{q_{0sr}^+\frac{\partial q_{0sr}^+}{\partial \mu_{sr}}  + M_{sc}\frac{\partial M_{sr}}{\partial \mu_{sr}}}{{q_{sr}^+}^2 + M_{sr}^2}\right] - \left[\frac{  p_{0sr}^+ \frac{\partial p_{0sr}^+}{\partial \mu_{sr}} }{{p_{sr}^+}^2 + m_s^2}\right]\right\} \nonumber \\
&+& \Theta\left(\mu_{sr} - m_s\right) \frac{\left(\mu_{sr} - m_s\right)^{\frac{3}{2}}}{3\pi^2}
\end{eqnarray}
\begin{eqnarray}
\rho_{sb} &=& 2 \int_{0}^{+\infty} \frac{d\hat{p}}{\pi^3} \mathrm{Real}\left\{\left[\frac{q_{0sb}^+\frac{\partial q_{0sb}^+}{\partial \mu_{sb}}  + M_{sc}\frac{\partial M_{sb}}{\partial \mu_{sb}}}{{q_{sb}^+}^2 + M_{sb}^2}\right] - \left[\frac{  p_{0sb}^+ \frac{\partial p_{0sb}^+}{\partial \mu_{sb}} }{{p_{sb}^+}^2 + m_s^2}\right]\right\} \nonumber \\
&+& \Theta\left(\mu_{sb} - m_s\right) \frac{\left(\mu_{sb} - m_s\right)^{\frac{3}{2}}}{3\pi^2}
\end{eqnarray}
The total densities are obtained by making the sum over color: $\rho_f = \sum_{c = r,g,b} \rho_{fc}$. The derivatives that appear are of the form:
\begin{eqnarray}
\frac{\partial p_{0fc}^\pm}{\partial \mu_{fc}} &=& \mp i \nonumber \\
 \frac{\partial q_{0fc}^\pm}{\partial \mu_{fc}} &=& \mp i \pm i\theta_f \frac{\partial g \left( p_{fc}^\pm \right)}{\partial \mu_{fc}} \nonumber \\
 \frac{\partial M_{fc}}{\partial \mu_{fc}} &=& \bar{\sigma}_f \frac{\partial g \left( p_{fc}^+ \right)}{\partial \mu_{fc}} \nonumber \\
 \frac{\partial M_{fc}^*}{\partial \mu_{fc}} &=& \left[\frac{\partial M_{fc}}{\partial \mu_{fc}}\right]^* \nonumber \\
 \frac{\partial g \left( p_{fc}^\pm \right)}{\partial \mu_{fc}} &=& \pm 2 i p_{0fc} \frac{g \left( p_{fc}^\pm \right)}{\Lambda^2} \nonumber \\
 \frac{\partial g^\Delta}{\partial \mu_{ur}} &=& i\frac{2p_0 - i \left(\mu_{ur} - \mu_{dr}   \right)}{2 \Lambda^2}g\left(\frac{ \left[{p_{ur}^+} + {p_{dr}^-} \right]^2 }{4}\right) \nonumber \\
  \frac{\partial g^\Delta}{\partial \mu_{dr}} &=& -i\frac{2p_0 - i \left(\mu_{ur} - \mu_{dr}   \right)}{2 \Lambda^2}g\left(\frac{ \left[{p_{ur}^+} + {p_{dr}^-} \right]^2 }{4}\right)
\end{eqnarray}
\section{Equations system to solve}
\begin{eqnarray}
 \sigma_u + G_S S_u + \frac{H}{2}S_d S_s &=& 0 \\
 \sigma_d + G_S S_d + \frac{H}{2}S_s S_u &=& 0 \\
 \sigma_s + G_S S_s + \frac{H}{2}S_u S_d &=& 0 \\
  \theta_u - G_V V_u &=& 0 \\
  \theta_d - G_V V_d &=& 0 \\
  \theta_s - G_V V_s &=& 0 \\
  \sum_{c = r,g,b}\left(\frac{2}{3}\rho_{uc} - \frac{1}{3}\rho_{dc}- \frac{1}{3}\rho_{sc}\right) - \rho_e - \rho_{\mu} &=& 0 \\
  n_B - \sum_{c = r,g,b}\frac{\rho_{uc} + \rho_{dc} + \rho_{sc}}{3} &=& 0 \\
    \Delta + G_D D &=& 0 \\
  \sum_{f = u,d,s} \left( \rho_{fr} + \rho_{fg} - 2\rho_{fb} \right) &=& 0.
\end{eqnarray}
\section{Regularized grand potential with vector interactions and diquarks at finite temperature}
\begin{eqnarray}
\label{granpotfinita}
 \Omega =  &-& 2T \sum_{n=0}^{+\infty}\int_{0}^{+\infty}\frac{p^2 \,dp}{\pi^3} \sum_c \mathrm{Real}\left\{ \mathrm{log}\left[\frac{{q_{snc}^+}^2 + M_{snc}^2}{p_{snc}^2 + m_s^2}\right]\right\}\nonumber \\ 
 &-& 2T \sum_{n=0}^{+\infty}\int_{0}^{+\infty}\frac{p^2 \,dp}{\pi^3}  \sum_c \left\{ \frac{1}{2} \mathrm{log} |A_{nc}|^2 -  \mathrm{Real}\left[ \mathrm{log}\left({p_{unc}^+}^2 + m_u^2 \right)\right] - \mathrm{Real}\left[\mathrm{log}\left({p_{dnc}^+}^2 + m_d^2\right)\right]    \right\} \nonumber \\ 
 &-& \frac{1}{2}\left[ \sum_f \left(\bar\sigma_f  \bar S_f  + \frac{G_S}{2}\bar S_f^2 +  \bar\theta_f  \bar V_f - \frac{G_V}{2}\bar V_f^2\right) + \frac{H}{2}\bar S_u \bar S_d \bar S_s + 2\bar\Delta \bar D +G_D\bar D^2\right] \nonumber \\
 &-& T \int_{0}^{+\infty}\frac{p^2 \,dp}{\pi^3} \sum_{f,c}\left[ \mathrm{log} \left( 1 + e^{-\frac{E_f - \mu_{fc}}{T}}\right) + \mathrm{log} \left( 1 + e^{-\frac{E_f + \mu_{fc}}{T}}\right)\right]
\end{eqnarray}
Where 
\begin{equation}
 A_{nc} = \left[{q_{unc}^+}^2 + M_{uc}^2\right]\left[{q_{dnc}^-}^2 + {M_{dnc}^*}^2\right] + \left(1 - \delta_{bc}\right){\Delta^p}^2\left[{\Delta^p}^2 + 2 q_{unc}^+ . q_{dnc}^- + 2M_{unc}M_{dnc}^* \right]
\end{equation}
\begin{eqnarray}
w_{n} &=& (2n + 1)\pi T \nonumber \\
q_{fnc}^\pm &=& \left( \vec{p}, w_{n} \mp i \left[\mu_{fc} - \bar{\theta_f}g\left({p_{fnc}^\pm}^2 \right) \right] \right) \nonumber \\
q_{0fnc}^\pm &=&  w_{n} \mp i \left[\mu_{fc} - \bar{\theta_f}g\left({p_{fnc}^\pm}^2 \right) \right] \nonumber \\
p_{fnc}^\pm &=& \left(\vec{p}, w_{n} \mp i \mu_{fnc} \right) \nonumber \\
p_{0fnc}^\pm &=& w_{n} \mp i \mu_{fc}  \nonumber \\
M_{fnc} &=& m_f + \sigma_f \,g\left( {p_{fnc}^+}^2  \right) \nonumber \\
\Delta^p &=& \bar{\Delta}\,g\left(\frac{ \left[{p_{unr}^+} + {p_{dnr}^-} \right]^2 }{4}\right) \nonumber \\
E_f &=& \sqrt{p^2 + m_f^2} 
\end{eqnarray}
The densities are obtained by making the same derivatives as before. However, it is easy to see that in the term of the exponentials of the grand potential in equation \eqref{granpotfinita}  the only color index comes from the chemical potentials. Therefore the densities at finite temperature will be the same as at zero temperature making the replacement
\begin{equation}
\label{recetamala2}
 \int \frac{dp_0}{(2\pi)}\int \frac{d^3p}{(2\pi)^3}f(p_0,\bm{p}) \rightarrow T \sum_{n = -\infty}^{\infty} \int \frac{d^3p}{(2\pi)^3}f(w_n - i\mu, \bm{p}).
\end{equation}
 and replacing the term $\Theta\left(\mu_{f,c} - m_f\right)$ by the corresponding derivative of the exponential term. Since it is a simple derivative, and including the complete densities again at finite temperature has a very similar writing to that shown at zero temperature, we will avoid writing them explicitly. Similarly, auxiliary fields $D$, $S_i$ y $V_i$, do not include the exponential term, which comes from the free part. Therefore, its writing is exactly the same as at zero temperature, making the mentioned replacement.

\begin{biblio}
\bibliography{Malfatti}
\end{biblio}

%\begin{thebibliography}{100}
%\end{thebibliography}
\begin{postliminary}

%\begin{seccion}{Agradecimientos}
%A todos los que se lo merecen, por merecerlo
%\end{seccion}

\end{postliminary}

\end{document}